\documentclass[accepted]{melba}

\usepackage{amsmath,amsfonts}

\usepackage{hyperref}
\usepackage{mathtools}

\usepackage{graphicx}
\usepackage{multirow}
\usepackage{subcaption}
\usepackage{booktabs}

\melbaid{2025:014}  
\doi{https://doi.org/10.59275/j.melba.2025-f3g6}
\melbaauthors{Chalcroft, Pappas, Price and Ashburner}  
\email{liam.chalcroft.20@ucl.ac.uk}
\volume{3}
\firstpageno{317}  
\melbayear{2025}  
\datesubmitted{2024-11-08}  
\datepublished{2025-08-14}  

\ShortHeadings{Synthetic Data for Robust Stroke Segmentation}{Chalcroft et al.}

\title{Synthetic Data for Robust Stroke Segmentation}

\author{\firstname Liam \surname Chalcroft \aff{1} \orcid{0000-0003-3363-6454},
        \firstname Ioannis \surname Pappas \aff{2} \orcid{0000-0002-0168-7014},
        \firstname Cathy J. \surname Price \aff{1} \orcid{0000-0001-7448-4835},
        \firstname John \surname Ashburner \aff{1} \orcid{0000-0001-7605-2518}
    }

\affiliations{
	\num 1 \addr Wellcome Centre for Human Neuroimaging, University College London \\
	\num 2 \addr University of Southern Californa
}

\abstract{
	Current deep learning-based approaches to lesion segmentation in neuroimaging often depend on high-resolution images and extensive annotated data, limiting clinical applicability. This paper introduces a novel synthetic data framework tailored for stroke lesion segmentation, expanding the SynthSeg methodology to incorporate lesion-specific augmentations that simulate diverse pathological features. Using a modified nnUNet architecture, our approach trains models with label maps from healthy and stroke datasets, facilitating segmentation across both normal and pathological tissue without reliance on specific sequence-based training. Our method achieves robust out-of-domain performance where conventional approaches fail, with in-domain performance of 48.2\% Dice compared to 57.5\% for conventional training. Crucially, even with oracle knowledge of the optimal domain adaptation method - an unrealistic scenario in practice - conventionally-trained models cannot match our synthetic approach in out-of-domain settings. The framework demonstrates that synthetic pre-training provides fundamental robustness unachievable through test-time adaptation alone. Our approach reduces reliance on domain-specific training data and helps bridge the gap between research-grade and clinical scans to improve clinical stroke neuroimaging workflows. PyTorch training code and weights are publicly available at \href{https://github.com/liamchalcroft/SynthStroke}{https://github.com/liamchalcroft/SynthStroke}, along with an SPM toolbox featuring a plug-and-play model at \href{https://github.com/liamchalcroft/SynthStrokeSPM}{https://github.com/liamchalcroft/SynthStrokeSPM}.}


\begin{document}

\twocolumn[\maketitle]

\section{Introduction} \label{intro}

    \enluminure{S}{emantic} segmentation is a critical component of neuroimaging pipelines, enabling precise quantification of anatomical structures and lesions for applications like tracking disease progression and planning treatments. In research settings, segmentation labels are typically derived from high-quality, standardised structural scans (e.g., MPRAGE) that benefit from consistent field-of-view, spacing, orientation, and minimal artifacts. In contrast, clinical scans often exhibit significant variability in these factors, which can severely impact deep learning model performance. Consequently, models trained on homogeneous research-grade data may not generalise well to the diverse, lower-quality images encountered in clinical practice.

    Both traditional probabilistic methods \citep{Ashburner2005} and modern deep discriminative methods \citep{Isensee2020} require prior information to be provided for a given sequence - in the case of a traditional method this may be an atlas or template, and in modern methods this would come in the form of training data. Atlas-based methods build a template of the anatomical structure of the brain, which may be deformed into alignment with a new subject to assign voxel-wise anatomical classes. This is proven to be robust for delineating healthy structure even with shifts in contrast \citep{Puonti2016a}, however it is non-trivial to include classes of pathology such as stroke within such a model, due to the inherent heterogeneity in location and geometric properties. In the context of generative models, lesions may be included in the form of anomaly detection as demonstrated in \cite{Seghier2008}. This method is not however directly attempting to label the pathology, and (by design) will label physiological changes such as ventricular enlargement in addition to the responsible infarct.

    Deep discriminative models trained using supervised learning have been able to reach human-level performance when tested in-domain on large datasets for a variety of brain pathologies and imaging modalities \citep{Baid2021TheClassification,delarosa2024robustensemblealgorithmischemic}. There is still however a significant gap when trying to translate these models to clinical data, where each hospital is likely to vary both in the scanning equipment used and the choice of imaging sequences \citep{Nguyen2024}. This poses a significant challenge to the adoption of deep learning for automating the labelling of clinical data, which could greatly help to accelerate the translation of modern research in stroke prognosis \citep{Loughnan2019}.
    
    To extend such methods to the open-ended domain of clinical scans, models often need to perform on sequences for which no training data may be available. To this end, domain randomisation via synthetic data has been shown to give impressive results for healthy brain parcellation in SynthSeg \citep{Billot2023}. In this method, a set of ground truth healthy tissues are used to generate synthetic images, under the assumption that each tissue class' intensity distribution should roughly follow a Gaussian. By assigning random Gaussian distributions to each class, a deep learning model can learn to extract shape information for parcellation in a way that is invariant to the input image's relative tissue contrast, hence allowing the model to be used on any sequence at test-time, without training data or prior knowledge of the sequence. This method of training with synthetic data has since been extended to tasks such as image registration \citep{hoffmann2022synthmorph,hoffmann2023anatomy,hoffmann2024anatomy}, image super-resolution \citep{iglesias_joint_2021,Iglesias2023}, surface estimation \citep{gopinath2023corticalanalysisheterogeneousclinical,gopinath2024reconallclinicalcorticalsurfacereconstruction} and vascular segmentation \citep{chollet2024neurovascularsegmentationsoctdeep}. A comprehensive overview is available in \citet{Gopinath2024}.

    An additional benefit of this method is that the 'forward model' of creating an MRI (or CT) image from tissues of different physical properties is a perfect 1:1 mapping to the 'inverse model' of labeling the tissues (i.e. segmentation) from the acquired image. In structures that exhibit a large amount of inter-rater variability, this is likely to help prevent a model from imitating under- or over- segmentation from imperfect ground truths - the images segmented are generated from the corresponding segmentation labels and so labels will always be a consistent method of segmentation.
    
    Prior work to SynthSeg demonstrated the potential of encoding anatomical priors in a neural network through pre-training with unpaired parcellation labels \citep{Dalca_2018}. Such methods face significantly larger challenges when applied to the heterogeneous shape and spatial distribution of lesions. In healthy parcellation, anatomical structures have consistent positions across individuals (e.g., the brainstem reliably appears in the same region of the brain), a regularity that has motivated atlas-based approaches to parcellation. In contrast, lesions are highly variable across individuals - not only in number and size but also in their spatial distribution. Unlike anatomical structures, lesions cannot be reliably mapped to a specific location or shape within an atlas. Although the exact site of lesion initiation is often influenced by the brain's vascular architecture - meaning that certain regions are statistically more susceptible to stroke due to the location of large blood vessels - the resulting lesion's size, shape, and spread are highly variable. Multiple sclerosis serves as an exception, with lesions that are somewhat predictable in their white matter localisation \citep{Lassmann2018}, enabling modelling through synthetic deep learning frameworks \citep{Billot2021,laso2024quantifyingwhitematterhyperintensity} and traditional probabilistic models \citep{Cerri2021}. More recently, \cite{Liu2024PEPSI:MRI} has shown promising results by training a SynthSeg-like model on lesion labels from various pathologies, providing a foundation for fine-tuning on multiple downstream datasets. 
    
    Robust open-domain stroke segmentation remains an unsolved challenge. Most domain-specific frameworks for stroke lesion segmentation are targeted towards lesion-pasting \citep{Zhang_2021,dai2022softcp,basaran2023lesionmixlesionleveldataaugmentation}, aiming to augment the anatomical variety without making any attempt to augment the variety in image contrasts. Likewise, label-conditioned generative models such as TumorGAN \citep{Li2020} can similarly only generate new lesioned brains within the learned distribution of image intensities. None of these works approach the task of robustness to shifts in image appearance, instead focusing on shape-related augmentation.
    
    In our work, we extend the SynthSeg framework to the task of stroke lesion segmentation via a novel lesion-pasting method that better simulates variety in lesion appearance. Our hybrid approach trades a statistically significant 9.3\% median Dice reduction in-domain (57.5\% vs 48.2\%, p$<$0.001) for improved out-of-domain robustness. Crucially, we demonstrate that even with oracle knowledge of the optimal domain adaptation method, conventional training cannot match our synthetic approach in out-of-domain scenarios. We validate this on a comprehensive range of lesion datasets with a wide distribution of image characteristics and lesion physiology.
    To assist in widespread evaluation of this framework, we release PyTorch training code/weights, and a MATLAB toolbox for SPM to reduce the barrier to clinical adoption.

\section{Methods} \label{sec:methods}

    \subsection*{Terminology and Notation}
    For clarity, we define the key terms used throughout this work:
    \begin{itemize}
        \item \textbf{TTA (Test-Time Augmentation):} A procedure at inference where multiple augmented versions (here generated by flips) of an input are processed and their predictions averaged to improve robustness.
        \item \textbf{DA (Domain Adaptation):} Techniques applied at test time to adapt a trained model to new data distributions.
        \item \textbf{Oracle DA:} The hypothetical best-case scenario where the optimal DA method is known a priori for each dataset/modality combination.
    \end{itemize}

    \subsection{Synthetic Data Generation Framework}
    
\begin{figure*}[hbt!]
            \centering
            \includegraphics[width=0.9\textwidth]{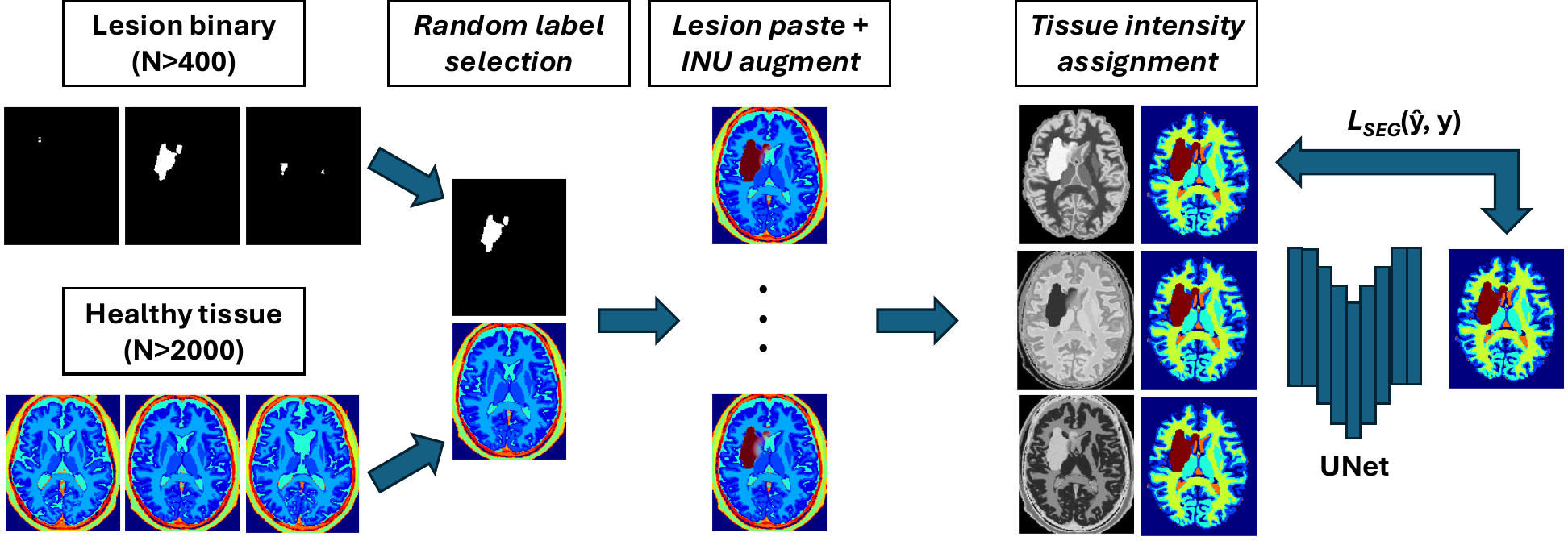}
            \caption{Schematic overview of the data generation process. Lesions are sampled from a template-normalised bank of lesion binary masks, and healthy tissue maps are sampled from a template-normalised bank of MultiBrain segmentations. Pasting of lesions onto healthy tissue maps is performed using a spatially varying lesion intensity to simulate penumbra. Tissue intensities may then be sampled from Gaussian distributions and image-label pairs used to train a segmentation model in a supervised manner.}
            \label{fig:schematic}
        \end{figure*}

        \paragraph{Rationale.}
        Our goal is to create a large, diverse and perfectly labelled training set without the labour of voxel-wise annotation. We therefore generate paired \emph{image-label} volumes by composing healthy tissue maps with realistically shaped stroke lesions, followed by intensity synthesis and heavy image-quality augmentation (Fig.~\ref{fig:schematic}).
        
\begin{figure}[hbt!]
            \centering
            \begin{subfigure}{.3\textwidth}
                \includegraphics[width=\textwidth]{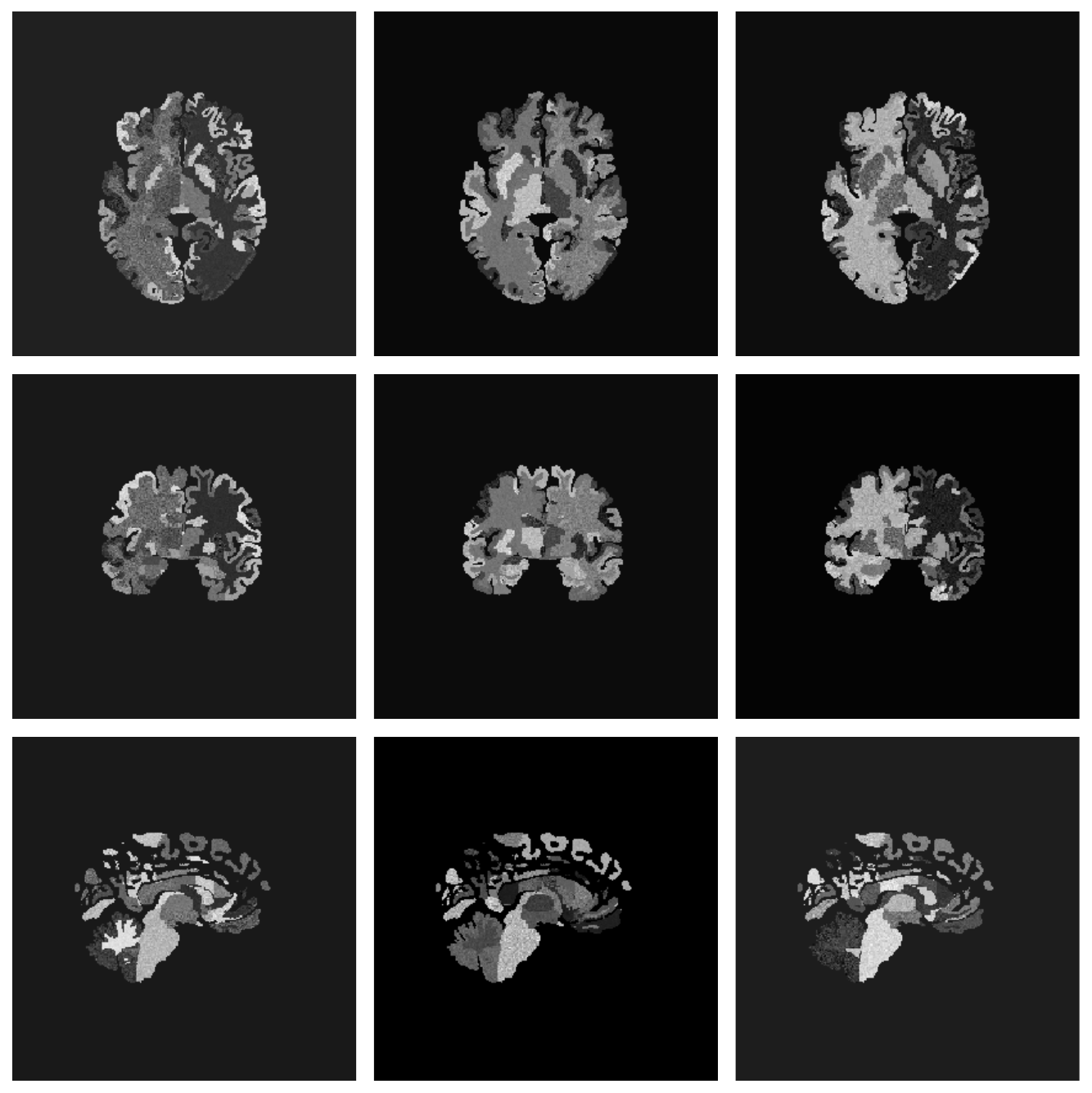}
                \caption{FreeSurfer \citep{Puonti2016a}}
                \label{fig:mb1}
            \end{subfigure}
            \begin{subfigure}{.3\textwidth}
                \includegraphics[width=\textwidth]{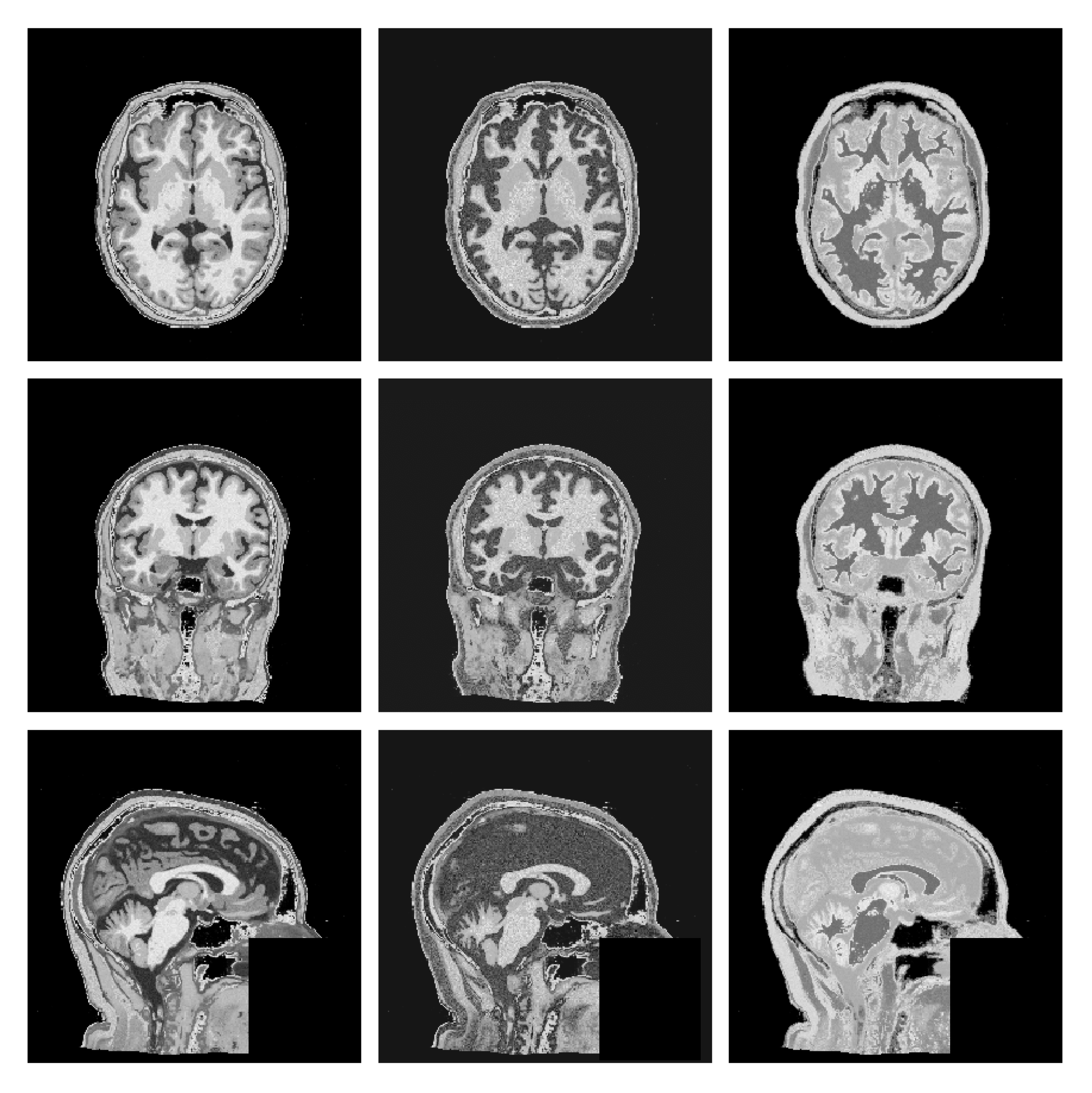}
                \caption{MultiBrain \citep{Brudfors_2020}}
                \label{fig:mb2}
            \end{subfigure}
            \begin{subfigure}{.3\textwidth}
                \includegraphics[width=\textwidth]{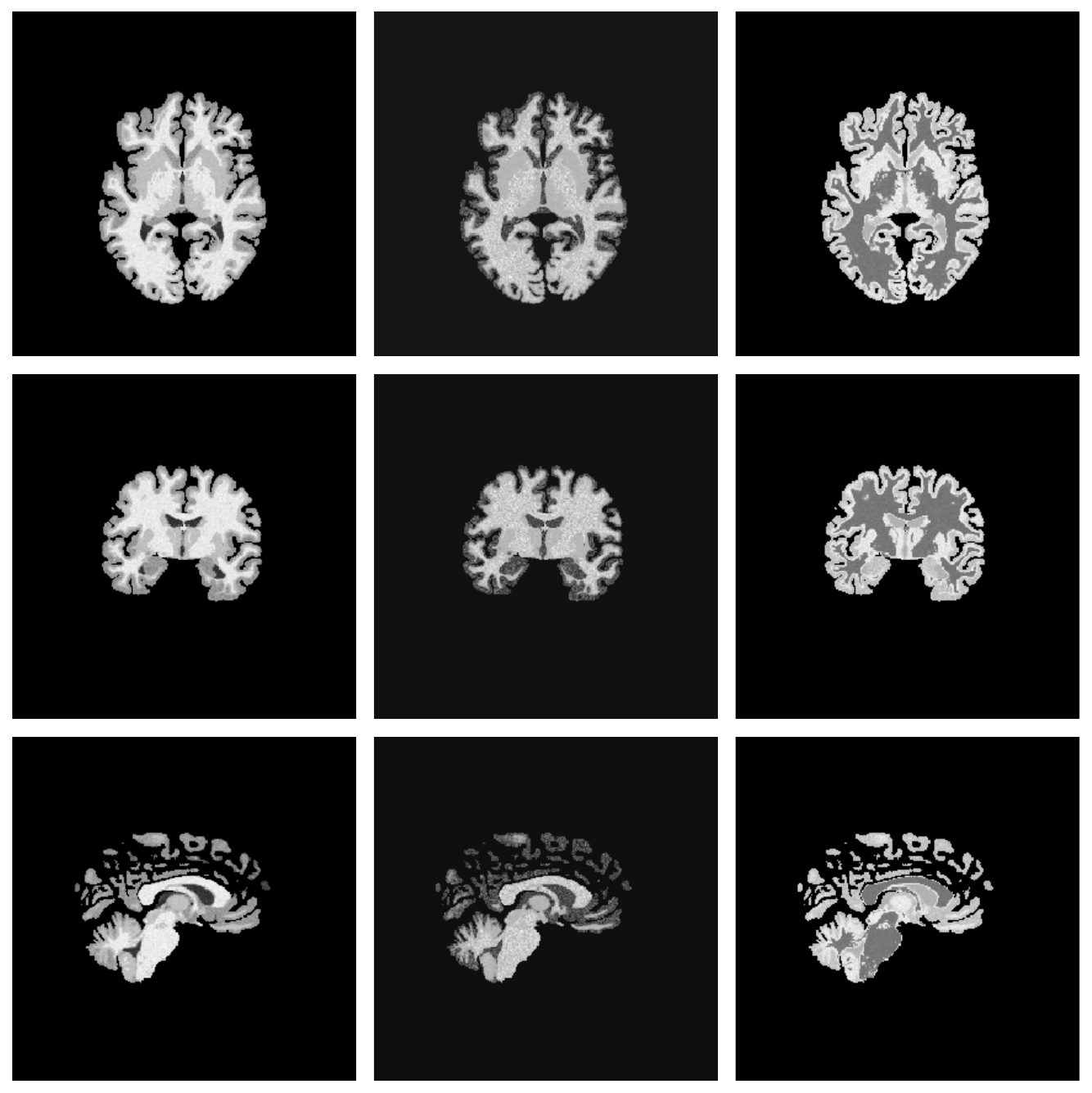}
                \caption{MultiBrain (skull-stripped)}
                \label{fig:mb3}
            \end{subfigure}
            \caption[Synthetic data samples]{Sample generated images using different labels for a single subject. \ref{fig:mb1}: FreeSurfer anatomical labels. \ref{fig:mb2}: MultiBrain tissue labels. \ref{fig:mb3}: MultiBrain tissue labels masked to simulate skull-stripping.}
            \label{fig:multibrain}
        \end{figure}

        \paragraph{(i) Healthy-tissue label bank.}
        Instead of the 100+ FreeSurfer classes used by SynthSeg, we adopt the nine posterior tissue maps produced by MultiBrain \citep{Brudfors_2020}. This reduces memory usage, speeds up sampling and still retains the GM/WM/CSF boundaries that matter for lesion realism (Fig.~\ref{fig:multibrain}).

        \paragraph{(ii) Lesion Copy-Paste (Lesion-CP).}
        We extend Soft-CP \citep{dai2022softcp} with random dilate/erode 'feathering' and a spatially varying bias-field multiplier (MONAI Random Bias Field \citep{cardoso2022monai}) to mimic penumbral intensity fall-off \citep{middleton2024localgammaaugmentationischemic}.
        
        \paragraph{(iii) Intensity sampling.}
        Each tissue class is assigned $\mu\!\sim\!\mathcal U(0,255)$, $\sigma\!\sim\!\mathcal U(0,16)$ and Gaussian blur $\text{FWHM}\!\sim\!\mathcal U(0,2)$. For stroke lesions we modulate the copy-pasted mask with the bias field to create intra-lesion heterogeneity. We sample from a single Gaussian distribution, with the implications of this choice examined via Hartigan's dip-test in Appendix \ref{app:unimodal}.
        
        \paragraph{(iv) Image-quality augmentations.}
        Table \ref{tab:aug-hyperparams} summarises every random transform (bias field, affine, elastic, skull-strip imperfection, noise, resolution, motion, contrast, etc.).
        
        Future work will explore mixture-of-Gaussians modelling \citep{Ashburner2005} to represent lesions that are simultaneously hyper- and hypo-intense.

        \begin{table*}[ht]
        \centering
        \caption{Parameter ranges for every image-quality augmentation used during synthetic training. $^\dagger$ indicates augmentations applied only to synthetic data and not to real ATLAS samples.}
        \label{tab:aug-hyperparams}
        \begin{tabular}{lll}
        \toprule
        \textbf{Category} & \textbf{Transform} & \textbf{Sampling range / notes}\\
        \midrule
        Bias field        & Multiplicative bias          & Control points $\mathcal U(2,7)$, strength $\mathcal U(0,0.5)$ \\
        Affine            & Rotation                     & $\mathcal U(-15^\circ,15^\circ)$ on each axis \\
                          & Shear                        & $\mathcal U(0,0.012)$ \\
                          & Zoom                         & $\mathcal U(0.85,1.15)$ \\
        Elastic           & Deformation grid             & Control points $\mathcal U(0,10)$, max disp.\ $\mathcal U(0,0.05)$ \\
        Skull-strip flaw$^\dagger$  & Dilation                     & $p=0.3$, radius = 2 vox. \\
                          & Erosion                      & $p=0.3$, radius = 4 vox. \\
        Noise             & Gaussian (SNR)               & SNR $\mathcal U(0,10)$, smoothed by $g$-factor $\mathcal U(2,5)$ \\
        Resolution$^\dagger$        & Slice anisotropy             & Thk.\ factor $\mathcal U(1,8)$ (base res.\ 1 mm$^{3}$) \\
        Contrast          & Gamma                        & $\gamma=10^{\mathcal N(0,0.6)}$ \\
        Motion blur$^\dagger$       & PSF width (FWHM)             & $\mathcal U(0,3)$ vox. \\
        Flip              & Mirror on each axis          & $p=0.8$ \\
        \bottomrule
        \end{tabular}
        \end{table*}

    With this synthetic data generation pipeline established, we now describe the datasets, network architecture, and training procedures used in our experiments.
    
    \subsection{Training Data and Sampling}
    
        Healthy maps: OASIS-3 (\textit{N}=2 679, 2 579/100 train/val) with MultiBrain segmentation, all warped to ICBM space.   Lesion masks: ATLAS (\textit{N}=655, 419/105/131 train/val/test) aligned to the same space. During sampling we paste one random ATLAS lesion onto one random OASIS subject and on-the-fly augment as above.
    
    \subsection{Network Architecture and Optimisation}
    
        \noindent\textbf{Backbone.}
        3D U-Net (nnUNet template) with six levels ($16\!\rightarrow\!32\!\rightarrow\!64\!\rightarrow\!128\!\rightarrow\!320$ channels), PReLU activations and one residual unit per block \citep{isensee2024nnunetrevisitedrigorousvalidation}.  
        
        \noindent\textbf{Output channels.}
        Background + Gray Matter (GM) + White Matter (WM) + GM/WM Partial Volume + Cerebrospinal Fluid (CSF) + Lesion (total six channels) for \textbf{Synth}; Background + Lesion for \textbf{Baseline}. When real images (binary GT) enter the mixed loader we mask the loss to the lesion channel only.
        
        \noindent\textbf{Training schedule.}
        $192^{3}$ crops, batch 1, 1 200 epochs × 500 iterations   (=$6\times10^{5}$ updates), AdamW ($\eta_0\!=10^{-4}$, weight-decay 0.01, poly LR decay with power 0.9), combined Dice + CE loss, dropout 0.2, gradient-norm clip 12.
        
        \noindent\textbf{Data-loader mixing.}
        Synthetic : Real ratio = 2 579 : 419 (mirrors sample counts).   Real MPRAGE images receive the same spatial/intensity transforms as synthetic batches.
        
        \noindent\textbf{Comparison model.}
        WMH-SynthSeg \citep{laso2024quantifyingwhitematterhyperintensity, Fischl2012} is included as an "off-the-shelf" robust-lesion baseline, but its training labels target small, periventricular WMHs and differ substantially from large-cortical stroke lesions.
    
    \subsection{Domain Adaptation Methods}
    
        At test time we evaluate a diverse set of \textbf{unsupervised domain-adaptation (DA)} techniques to determine:
        (i) whether the \textbf{Baseline} model can, under an \emph{oracle} choice of DA method, match the performance of the DA-free \textbf{Synth} model, and 
        (ii) whether Synth's inherent robustness offers a better starting point for DA on truly unseen data.
        
        We tested six DA methods (TTA, DAE, TENT, PL, UPL, DPL) and report the best-performing configuration for each modality/dataset combination as \textbf{Oracle DA} in Tables \ref{tab:atlas-scores}-\ref{tab:da-ploras}. This represents an upper bound on baseline performance, as in practice one cannot know \emph{a priori} which DA method will work best. Full individual results for all methods appear in Appendix \ref{app:metrics}, Tables \ref{tab:supp-atlas-scores}-\ref{tab:supp-da-ploras}.
        
        \noindent\textbf{DA techniques evaluated.}
        
        \noindent\textbf{(1) TTA - Test-Time Augmentation} (\citet{Wang_2019}; denoted \textbf{TTA}). Eight mirror-flipped volumes ($2^{3}$ axis-flip combinations) are inferred, logits averaged, and softmax/argmax yields the mask. This heuristic is cheap and rarely degrades performance.
        
        \noindent\textbf{(2) DAE - Denoising Auto-Encoder Regularisation} \citep{Karani_2021}. We train a 3-layer denoising auto-encoder to regularise noisy labels for both Baseline and Synth. At inference a three-layer normalisation network ($3{\times}3{\times}3$ kernels, 16 channels) is \emph{prepended} to the frozen segmentor; its activation is  
        
        \begin{equation}
            f(x) = \exp{-\frac{x^2}{\sigma^2}},
        \end{equation}
        
        where $\sigma$ is learned. For each test image the network is re-initialised and optimised for 100 steps to minimise Dice\,+\,L2 loss between segmentation logits and their DAE-cleaned counterpart.
        
        \noindent\textbf{(3) TENT - Test-Time Entropy Minimisation} \citep{wang2021tent}. Because 3D memory limits rule out batch-norm adaptation, we instead optimise an identical normalisation network (initialised from scratch per subject) for 100 steps to minimise Shannon entropy
        
        \begin{equation} 
        \mathcal{H}(\hat{y})=-\sum\nolimits_c p_c(\hat{y})\log p_c(\hat{y}),
        \end{equation}
        
        with $p_c$ the class-$c$ softmax probability.
        
        \noindent\textbf{(4) PL Family - Pseudo-Labelling (Self-Training)} \citep{chen2021sourcefreedomainadaptivefundus}. Three variants are tested:
        
        \begin{itemize}
          \item \textbf{PL}: threshold softmax at $\tau=\tfrac{1.5}{N_C}$ to keep only high-confidence voxels, where $N_C$ is the total number of output classes.
          \item \textbf{UPL}: PL plus uncertainty masking via 10-sample Monte-Carlo dropout (variance $>0.05$ discarded).
          \item \textbf{DPL}: full "prototype-consistency" pipeline that further removes voxels inconsistent with decoder-feature prototypes.
        \end{itemize}
        
        All PL variants fine-tune \emph{all} segmentation weights for 2000 iterations with weighted cross-entropy.
        
        \noindent\textbf{Common optimiser settings.} Every trainable DA method (DAE, TENT, PL, UPL, DPL) uses AdamW \citep{loshchilov2019decoupled} with learning-rate 0.002 and weight-decay 0.01. TTA is inference-only and therefore parameter-free.

    
    \subsection{Large-Scale Pseudo-Labelling} \label{sec:methods-pseudo}
    
        Because the synthetic pipeline decouples \emph{image realism} from \emph{label accuracy}, we can tolerate imperfect pseudo-labels. We therefore use Baseline\,+\,TTA to annotate 1159 chronic stroke MPRAGE scans from PLORAS Sample 1 \citep{Seghier2016}; this \textbf{PLORAS-MPRAGE} cohort feeds the mixed loader exactly like ATLAS.

\section{Experiments} \label{sec:experiments}

    \subsection{Datasets}
        
        Models were validated on four stroke-lesion datasets. We assessed the models' in-domain performance on the hold-out test set for the \textbf{ATLAS} dataset (131 subjects, 1~mm isotropic MPRAGE). Out-of-domain (OOD) robustness was evaluated on three additional cohorts: the \textbf{ISLES 2015} dataset \citep{Maier2017} (\textit{N}=28 subjects with skull-stripped T1w/T2w/FLAIR/DWI), the \textbf{ARC} dataset \citep{Gibson2024TheRepository,Johnson2024ProgressiveStroke} (\textit{N}=229 T2w,\;202 T1w,\;85 FLAIR; \textit{N}=84 subjects have all three), and the hospital scans from 661 acute-stroke patients in PLORAS Sample 2 (\textit{N}=106 T2w,\;300 FLAIR,\;255 CT), collectively referred to as \textbf{PLORAS} \citep{Price2010PredictingSystem}. ISLES and PLORAS also introduce an acute-versus-chronic shift. There is no overlap in patients between the PLORAS hospital scans (acute) and the PLORAS MPRAGE cohort described in Section \ref{sec:methods-pseudo}.
        
        For PLORAS, images are resampled from the original 2~mm isotropic resolution to 1~mm to maintain a single preprocessing pipeline, acknowledging that this constitutes a second resampling step.
    
    \subsection{Experimental Design}
    
        \paragraph{Pre-processing and inference.}
        All test images are re-oriented to RAS, resliced to 1 mm\,$^{3}$ voxels, histogram-normalised and $z$-scored. Inference uses a $192^{3}$ sliding window with 50 \% overlap and a Gaussian blending kernel ($\sigma=0.125$). Test-time augmentation (\textbf{TTA}) averages logits over all eight combinations of left-right, anterior-posterior and inferior-superior flips.
        
        \paragraph{Multi-modal ensembles.}
        When multiple MR sequences are available for the same subject (ISLES 2015, ARC) we average the per-modality logits before the softmax. This simple ensembling mimics a realistic clinical deployment.
        
        \paragraph{Pseudo-label training.}
        Pseudo-labels are generated for the PLORAS MPRAGE cohort with the Baseline\,+\,TTA model. A new Baseline and a new Synth model are then re-trained using the union of ATLAS and pseudo-labelled data, following exactly the same optimisation schedule as the originals.
    
    \subsection{Evaluation Metrics}
    
        Prior to metric computation, predictions and ground truth are resliced to 1 mm and zero-padded to $256^{3}$ voxels. We report Dice and Surface-Dice \citep{Seidlitz_2022} (1 mm tolerance) in the main text; HD95, absolute volume difference (AVD), absolute lesion difference (ALD), lesion-wise F1, true-positive rate (TPR) and false-positive rate (FPR) appear in Appendix \ref{app:metrics}.
        
        Dice quantifies volumetric overlap (1 = perfect, 0 = none), whereas HD95 captures boundary error while being robust to outliers. AVD reports absolute volume mismatch in $\text{cm}^3$. ALD counts mismatches in the number of connected components, and lesion-wise F1 scores per-lesion detection accuracy. TPR and FPR follow standard definitions.
        
    \subsection{Comparison Methods}

    We evaluate four primary approaches:
    
    \textbf{(i) Baseline:} A standard 3D U-Net trained solely on real MPRAGE images from the ATLAS dataset (N=419) using supervised learning. This represents the conventional approach of training on a single modality with manual annotations.
    
    \textbf{(ii) Oracle DA:} The baseline model enhanced with the single best-performing domain adaptation method (selected post-hoc from TTA, TENT, DAE, PL, UPL, DPL) for each modality. This represents the theoretical upper bound of what domain adaptation can achieve when the optimal method is known - an unrealistic scenario in practice.
    
    \textbf{(iii) WMH-SynthSeg:} The pre-trained white matter hyperintensity model from \citet{laso2024quantifyingwhitematterhyperintensity, Fischl2012}, included as an off-the-shelf robust lesion segmentation baseline. Note that this model was trained for small periventricular WMH lesions, which differ substantially from large cortical stroke lesions.
    
    \textbf{(iv) Synth (Ours):} Our proposed approach using synthetic data generation as described in Section \ref{sec:methods}. The model is trained on a mixture of synthetic data (generated from OASIS healthy maps + ATLAS lesion masks) and real ATLAS MPRAGE images, following the framework illustrated in Figure \ref{fig:schematic}.
    
    For fair comparison, TTA is \emph{only} applied when a model is explicitly labelled "+TTA" or when it is selected as the Oracle DA. All candidate maps are binarised via argmax on posterior probabilities without additional calibration or threshold tuning.

\section{Results}

\begin{figure*}[h!]
        \centering
        \begin{subfigure}[t]{\textwidth}
            \centering
            \includegraphics[width=\textwidth]{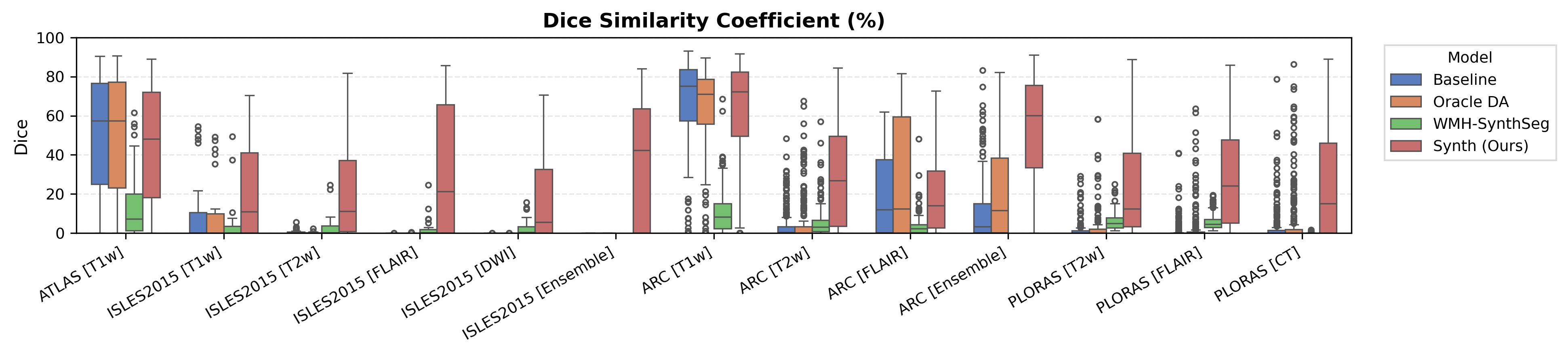}
        \end{subfigure}
        \\
        \begin{subfigure}[t]{\textwidth}
            \centering
            \includegraphics[width=\textwidth]{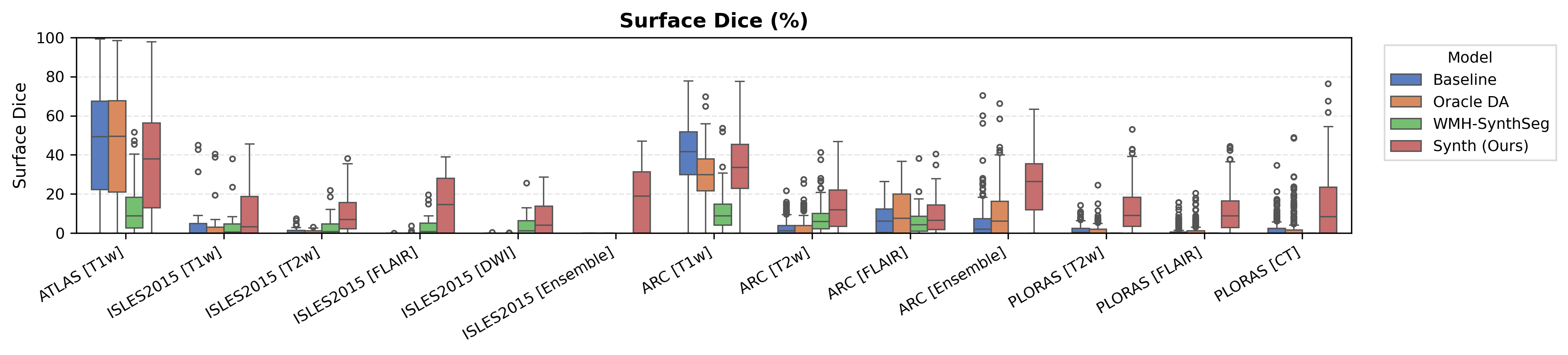}
        \end{subfigure}
        \\
        \caption{Dice and Surface Dice metrics for all reported datasets. 'Oracle DA' represents the hypothetical best-case scenario where optimal DA method is known a priori for each dataset/modality and applied to the baseline model.}
        \label{fig:combined-box}
    \end{figure*}

    Overall results are shown in Figure \ref{fig:combined-box}, which compares (i) the Baseline, (ii) the Oracle DA (best-performing domain adaptation for Baseline only), (iii) WMH-SynthSeg, and (iv) the DA-free Synth model. Oracle DA represents the hypothetical best-case scenario where the optimal DA method is known \textit{a priori} for each dataset/modality. Tables \ref{tab:atlas-scores}, \ref{tab:da-arc}, \ref{tab:da-isles2015} and \ref{tab:da-ploras} provide comprehensive results for all tested domain adaptation methods on both Baseline and Synth models.
    
    \subsection{ATLAS}

        The ATLAS dataset represents the in-domain scenario with T1-weighted images matching our training distribution. Table \ref{tab:atlas-scores} shows performance on the held-out test set. The conventional \textbf{Baseline} achieves a median Dice of \textbf{57.5 \%}, whereas our \textbf{Synth} model reaches \textbf{48.2 \%}, a \emph{9.3 \%} gap that represents the 'price of robustness' we accept for the larger out-of-domain gains reported later. Surface Dice follows the same trend (49.4 \% vs.\ 38.1 \%).
        
        Applying test-time augmentation (Baseline+TTA) changes Dice by $<$0.1 \%, indicating that TTA adds little benefit when the evaluation domain is already aligned with training.
        
        The off-the-shelf \textbf{WMH-SynthSeg}, trained for small periventricular WMH lesions, scores only 7.3 \% Dice, confirming that cortical stroke in ATLAS lies well outside its intended scope.
        
        A paired Wilcoxon test (Appendix Figure \ref{fig:wilcoxon-atlas-t1}) verifies that the Baseline–Synth Dice difference is statistically significant, underscoring that domain-invariant training still sacrifices some in-domain accuracy.  Voxel-level metrics in Appendix Table \ref{tab:supp-atlas-scores} reveal the Baseline achieves a higher recall but at the cost of more false positives, suggesting a tendency to over-segment.
        
        \textbf{Worst Scanner Syndrome.}  
        This in-domain drop is consistent with the 'Worst Scanner Syndrome' hypothesis \citep{moyer2021harmonizationworstscannersyndrome}, which posits that many domain-invariance strategies pull feature quality toward the least informative - or noisiest - domain. As the next Results sections demonstrate, that modest in-domain penalty is offset by substantial gains on heterogeneous clinical data.

        \begin{table}[h!]
        \centering
        \caption{Median results on the ATLAS hold-out set (\textit{N}=131). Best score shown in bold. Student's \textit{t} distribution 95\% confidence intervals given in brackets.}
        \resizebox{0.45\textwidth}{!}{
        \begin{tabular}{lcccccccc}
        \toprule
        \textbf{Modality} & \textbf{Model} & \textbf{Dice (\%)} & \textbf{Surface Dice (\%)} \\
        \midrule
        \multirow{4}{*}{\textbf{T1w}} & \textbf{Baseline} & \textbf{57.5 (52.3-62.7)} & 49.4 (44.5-54.3) \\
         & \textbf{Baseline+TTA} & 57.5 (52.2-62.8) & \textbf{49.5 (44.5-54.5)} \\
         & \textbf{WMH-SynthSeg} & 7.3 (4.8-9.7) & 8.9 (6.9-10.9) \\
         & \textbf{Synth (Ours)} & 48.2 (43.1-53.4) & 38.1 (33.4-42.7) \\
        \bottomrule
        \end{tabular}
        }
        \label{tab:atlas-scores}
        \end{table}
        

    \subsection{ARC}

        The ARC dataset contains research-quality chronic stroke scans, representing a moderate domain shift from our ATLAS training data. We expect T1w performance to be strongest given its proximity to the training domain.

        Table \ref{tab:da-arc} shows performance on ARC, comparing Baseline, Oracle DA (best possible domain adaptation for Baseline), WMH-SynthSeg, and our Synth approach. For T1w, both baseline and Oracle DA achieve 75.2\% Dice, with Synth at 72.3\% - all maintaining strong performance (Appendix Figure \ref{fig:wilcoxon-arc-t1} shows statistical significance). However, T2w reveals dramatic differences: baseline drops to 0.4\% and Oracle DA to 0.1\%, while Synth maintains 26.8\%. FLAIR shows intermediate performance with Oracle DA at 12.4\% versus Synth at 14.1\%. The ensemble results are particularly striking: Oracle DA achieves only 11.7\% while Synth reaches 60.2\%.

        This pattern - strong baseline performance on T1w but catastrophic failure on other modalities even with optimal DA - demonstrates that synthetic pre-training provides fundamental robustness unachievable through post-hoc adaptation.

        \begin{table}[h!]
        \centering
        \caption{Median results on the ARC dataset (\textit{N}=229). Best score shown in bold. Student's \textit{t} distribution 95\% confidence intervals given in brackets. 'Oracle DA' represents the hypothetical best-case scenario where optimal DA method is known a priori for each dataset/modality and applied to the baseline model.}
        \resizebox{0.4\textwidth}{!}{
        \begin{tabular}{lcccccccc}
        \toprule
        \textbf{Modality} & \textbf{Model} & \textbf{Dice (\%)} & \textbf{Surface Dice (\%)} \\
        \midrule
        \multirow{4}{*}{\textbf{T1w}} & \textbf{Baseline} &\textbf{75.2 (71.5-79.0)}&41.7 (39.1-44.4) \\
         & \textbf{Oracle DA} & 75.2 (71.3-79.0) & \textbf{42.1 (39.4-44.8)} \\
         & \textbf{WMH-SynthSeg} & 8.3 (6.8-9.8) & 8.9 (7.7-10.1) \\
         & \textbf{Synth (Ours)} & 72.3 (68.4-76.2) & 33.7 (31.2-36.2) \\
        \midrule
        \multirow{4}{*}{\textbf{T2w}} & \textbf{Baseline} & 0.4 (0.0-1.4) & 1.2 (0.8-1.7) \\
         & \textbf{Oracle DA} & 0.1 (0.0-0.4) & 1.0 (0.8-1.2) \\
         & \textbf{WMH-SynthSeg} & 3.2 (2.2-4.2) & 6.0 (5.2-6.8) \\
         & \textbf{Synth (Ours)} &\textbf{26.8 (23.3-30.2)}&\textbf{12.0 (10.4-13.6)} \\
        \midrule
        \multirow{4}{*}{\textbf{FLAIR}} & \textbf{Baseline} & 12.0 (7.6-16.3) & 6.3 (4.7-7.8) \\
         & \textbf{Oracle DA} & 12.4 (5.8-19.0) & \textbf{7.7 (5.2-10.2)} \\
         & \textbf{WMH-SynthSeg} & 2.4 (0.9-3.9) & 4.3 (3.0-5.5) \\
         & \textbf{Synth (Ours)} &\textbf{14.1 (9.7-18.5)}&6.6 (4.8-8.4) \\
        \midrule
        \multirow{3}{*}{\textbf{Ensemble}} & \textbf{Baseline} & 3.4 (1.3-5.4) & 2.0 (0.8-3.2) \\
         & \textbf{Oracle DA} & 11.7 (8.6-14.7) & 6.3 (4.7-7.8) \\
         & \textbf{Synth (Ours)} &\textbf{60.2 (56.6-63.8)}&\textbf{26.3 (24.3-28.4)} \\
        \bottomrule
        \end{tabular}
        }
        \label{tab:da-arc}
        \end{table}

    \subsection{ISLES 2015}

        ISLES 2015 contains skull-stripped sub-acute stroke scans, representing our most challenging domain shift with both acquisition and pathology differences from the chronic ATLAS training data.

        The results in Figure \ref{fig:combined-box} and Table \ref{tab:da-isles2015} show that the Synth model outperforms the Baseline model in all modalities in regards to both Dice and Surface Dice with high statistical significance in the Dice metric evidenced in Appendix Figures \ref{fig:wilcoxon-isles-t1}-\ref{fig:wilcoxon-isles-flair}. Baseline performance is near-zero across all modalities. Oracle DA shows minimal recovery: 10.5\% for T1w (still below Synth's 11.0\%) and 0.0\% for all other modalities (T2w, FLAIR, DWI). In contrast, Synth achieves 11.0\% (T1w), 11.1\% (T2w), 21.2\% (FLAIR), and 5.6\% (DWI). The ensemble demonstrates the starkest contrast: 0.0\% for both baseline and Oracle DA versus 42.3\% for Synth.

        The poor performance correlates with high false positive rates rather than missed detections (Appendix Table \ref{tab:supp-da-isles2015}), suggesting models struggle with tissue discrimination in this domain. When baseline predictions fail catastrophically, no DA method can recover meaningful performance.

        It is also evident from Table \ref{tab:da-isles2015} that the model performance is highly dependent on the choice of image sequence available. The ensemble improves performance over individual sequences in ISLES2015 (Table \ref{tab:da-isles2015}) but shows mixed results in ARC, suggesting dataset-specific benefits rather than universal improvement. Although we only show the upper limit of an ensemble of all four sequences, it is expected that in cases where fewer sequences are available we will still observe constructive gains from post-hoc ensembling.

        \begin{table}[h!]
        \centering
        \caption{Median results on the ISLES2015 dataset (\textit{N}=28). Best score shown in bold. Student's \textit{t} distribution 95\% confidence intervals given in brackets. 'Oracle DA' represents the hypothetical best-case scenario where optimal DA method is known a priori for each dataset/modality and applied to the baseline model.}
        \resizebox{0.4\textwidth}{!}{
        \begin{tabular}{lcccccccc}
        \toprule
        \textbf{Modality} & \textbf{Model} & \textbf{Dice (\%)} & \textbf{Surface Dice (\%)} \\
        \midrule
        \multirow{4}{*}{\textbf{T1w}} & \textbf{Baseline} & 0.0 (0.0-7.4) & 0.0 (0.0-4.8) \\
         & \textbf{Oracle DA} & 10.5 (0.0-21.7) & \textbf{3.8 (0.0-9.9)} \\
         & \textbf{WMH-SynthSeg} & 0.0 (0.0-4.5) & 0.7 (0.0-3.9) \\
         & \textbf{Synth (Ours)} &\textbf{11.0 (1.2-20.8)}&3.4 (0.0-8.5) \\
        \midrule
        \multirow{4}{*}{\textbf{T2w}} & \textbf{Baseline} & 0.0 (0.0-0.5) & 0.3 (0.0-1.0) \\
         & \textbf{Oracle DA} & 0.0 (0.0-0.2) & 0.0 (0.0-0.6) \\
         & \textbf{WMH-SynthSeg} & 0.1 (0.0-2.5) & 0.8 (0.0-3.0) \\
         & \textbf{Synth (Ours)} &\textbf{11.1 (0.7-21.6)}&\textbf{7.1 (2.7-11.5)} \\
        \midrule
        \multirow{4}{*}{\textbf{FLAIR}} & \textbf{Baseline} & 0.0 (0.0-0.0) & 0.0 (0.0-0.0) \\
         & \textbf{Oracle DA} & 0.0 (0.0-0.0) & 0.0 (0.0-0.0) \\
         & \textbf{WMH-SynthSeg} & 0.3 (0.0-2.3) & 0.8 (0.0-3.0) \\
         & \textbf{Synth (Ours)} &\textbf{21.2 (8.5-34.0)}&\textbf{14.7 (9.0-20.3)} \\
        \midrule
        \multirow{4}{*}{\textbf{DWI}} & \textbf{Baseline} & 0.0 (0.0-0.0) & 0.0 (0.0-0.0) \\
         & \textbf{Oracle DA} & 0.0 (0.0-0.0) & 0.0 (0.0-0.0) \\
         & \textbf{WMH-SynthSeg} & 0.4 (0.0-2.2) & 1.3 (0.0-3.6) \\
         & \textbf{Synth (Ours)} &\textbf{5.6 (0.0-14.4)}&\textbf{4.1 (0.8-7.4)} \\
        \midrule
        \multirow{3}{*}{\textbf{Ensemble}} & \textbf{Baseline} & 0.0 (0.0-0.0) & 0.0 (0.0-0.0) \\
         & \textbf{Oracle DA} & 0.0 (0.0-0.0) & 0.0 (0.0-0.0) \\
         & \textbf{Synth (Ours)} &\textbf{42.3 (30.2-54.5)}&\textbf{19.1 (12.7-25.4)} \\
        \bottomrule
        \end{tabular}
        }
        \label{tab:da-isles2015}
        \end{table}

    \subsection{PLORAS}

        The PLORAS dataset represents the most extreme domain shift with real clinical data exhibiting large diversity in acquisition protocols and slice thickness. We expect minimal baseline performance given these challenging conditions. For all available modalities, the Synth model outperforms the baseline with statistical significance in Dice (see Appendix Figures \ref{fig:wilcoxon-ploras-t2} - \ref{fig:wilcoxon-ploras-ct}).
        
\begin{figure*}[h!]
            \centering
            \begin{subfigure}[t]{\textwidth}
                \centering
                \includegraphics[width=\textwidth]{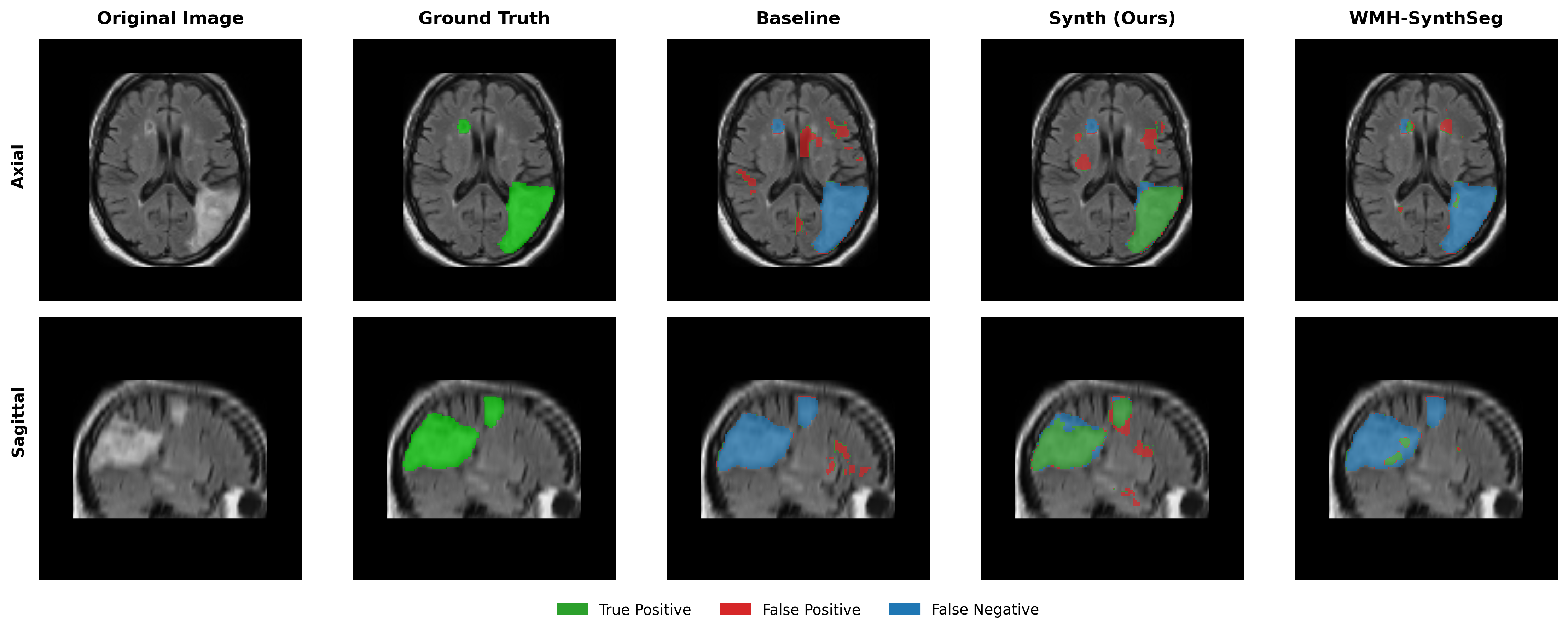}
            \end{subfigure}
            \\
            \begin{subfigure}[t]{\textwidth}
                \centering
                \includegraphics[width=\textwidth]{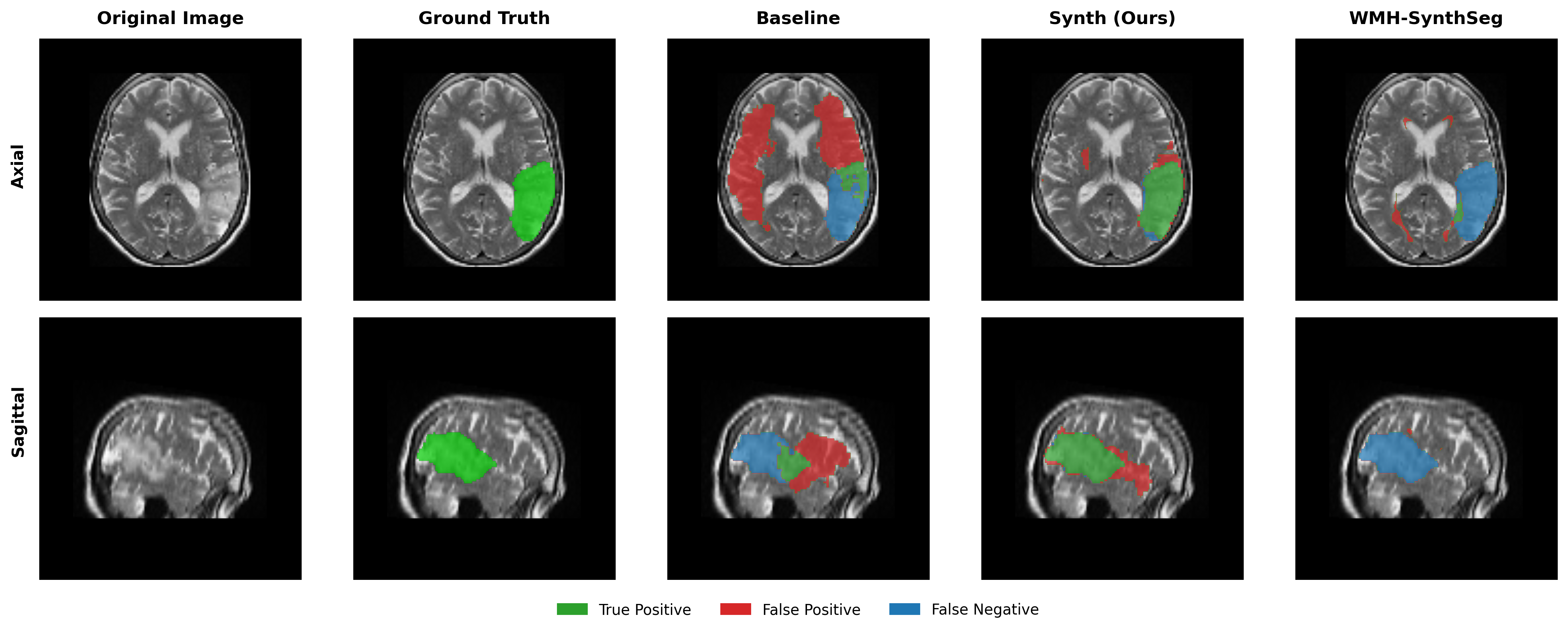}
            \end{subfigure}
            \\
            \begin{subfigure}[t]{\textwidth}
                \centering
                \includegraphics[width=\textwidth]{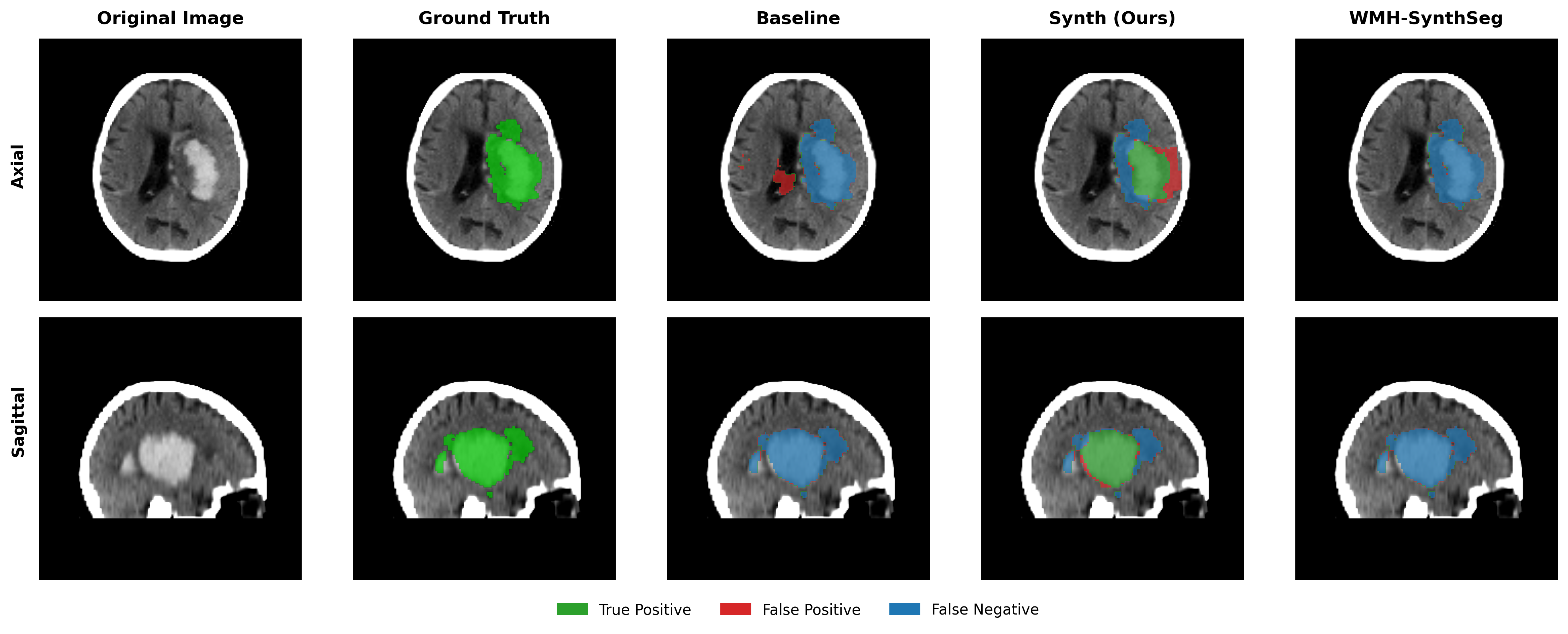}
            \end{subfigure}
            \caption{Sample visualisations in the PLORAS dataset. Green indicates a true positive prediction, red a false positive, and blue a false negative.}
            \label{fig:ploras-viz}
        \end{figure*}

        Results in Table \ref{tab:da-ploras} demonstrate the most extreme performance gap between conventional training and our synthetic approach. On this challenging clinical dataset, Oracle DA achieves at most 0.2\% Dice (CT modality), while our Synth model achieves 11.9\%, 25.4\%, and 11.3\% for T2w, FLAIR, and CT respectively. The near-total failure of both baseline and Oracle DA on real clinical data - where acquisition protocols, slice thickness, and image quality vary substantially - validates our core hypothesis: domain adaptation cannot substitute for domain-invariant training when deployment conditions diverge significantly from training data. Even WMH-SynthSeg, despite being trained for robustness, achieves only 0.0-4.9\% Dice, likely due to its focus on small periventricular lesions rather than large cortical strokes. Full results for all individual DA methods are provided in Appendix Table \ref{tab:supp-da-ploras}. A number of samples are also visualised for this dataset in Figure \ref{fig:ploras-viz}.

        \begin{table}[h!]
        \centering
        \caption{Median results on the PLORAS dataset (\textit{N}=661). Best score shown in bold. Student's \textit{t} distribution 95\% confidence intervals given in brackets. 'Oracle DA' represents the hypothetical best-case scenario where optimal DA method is known a priori for each dataset/modality and applied to the baseline model.}
        \resizebox{0.4\textwidth}{!}{
        \begin{tabular}{lcccccccc}
        \toprule
        \textbf{Modality} & \textbf{Model} & \textbf{Dice (\%)} & \textbf{Surface Dice (\%)} \\
        \midrule
        \multirow{4}{*}{\textbf{T2w}} & \textbf{Baseline} & 0.0 (0.0-1.3) & 0.1 (0.0-0.6) \\
         & \textbf{Oracle DA} & 0.1 (0.0-2.2) & 0.1 (0.0-0.8) \\
         & \textbf{WMH-SynthSeg} & 4.9 (4.0-5.8) & 0.0 (0.0-0.0) \\
         & \textbf{Synth (Ours)} &\textbf{11.9 (6.7-17.1)}&\textbf{8.4 (6.1-10.7)} \\
        \midrule
        \multirow{4}{*}{\textbf{FLAIR}} & \textbf{Baseline} & 0.0 (0.0-0.4) & 0.0 (0.0-0.2) \\
         & \textbf{Oracle DA} & 0.0 (0.0-0.4) & 0.0 (0.0-0.2) \\
         & \textbf{WMH-SynthSeg} & 4.6 (4.2-5.0) & 0.0 (0.0-0.0) \\
         & \textbf{Synth (Ours)} &\textbf{25.4 (22.5-28.3)}&\textbf{8.5 (7.4-9.6)} \\
        \midrule
        \multirow{3}{*}{\textbf{CT}} & \textbf{Baseline} & 0.0 (0.0-1.1) & 0.0 (0.0-0.5) \\
         & \textbf{Oracle DA} & 0.2 (0.0-0.5) & 0.3 (0.0-0.7) \\
         & \textbf{WMH-SynthSeg} & 0.0 (0.0-0.0) & 0.0 (0.0-0.0) \\
         & \textbf{Synth (Ours)} &\textbf{11.3 (8.0-14.6)}&\textbf{7.9 (6.0-9.8)} \\
        \bottomrule
        \end{tabular}
        }
        \label{tab:da-ploras}
        \end{table}  

    \subsection{Domain Adaptation}\label{sec:domain-adaptation}

        Tables \ref{tab:da-arc}, \ref{tab:da-isles2015} and \ref{tab:da-ploras} demonstrate that even with oracle selection of the optimal domain adaptation method, the baseline model cannot match Synth performance in out-of-domain settings. We evaluated six established DA techniques (TTA, TENT, DAE, PL, UPL, DPL) on both baseline and Synth models to determine whether domain adaptation could enable baseline generalisation. Complete results in Appendix Tables \ref{tab:supp-da-arc}-\ref{tab:supp-da-ploras} consistently show that poor initial baseline predictions prevent effective adaptation. In contrast, when applied to our Synth model, several DA methods yield substantial improvements, demonstrating the potential for compound gains when robust pre-training is combined with appropriate post-hoc adaptation.

        Analysis of complete DA results (Appendix Tables \ref{tab:supp-da-arc}-\ref{tab:supp-da-ploras}) reveals DAE as the most consistently effective method for Synth. DAE achieves substantial improvements: ARC T2w increases from 26.8\% to 54.3\% Dice, ISLES T1w from 11.0\% to 31.1\%, and PLORAS CT from 11.3\% to 23.9\%. TTA provides modest gains for PLORAS FLAIR (25.4\% to 29.4\%), while TENT benefits ARC FLAIR (14.1\% to 36.1\%). Pseudo-labeling methods (PL, UPL, DPL) show high variance - occasionally strong but often catastrophic.

        The key finding: while DA cannot rescue baseline models trained on narrow data, it can enhance robustly pre-trained models. Synth+DAE consistently outperforms both Synth alone and Baseline+Oracle DA, demonstrating complementary benefits.

        Our domain adaptation experiments serve primarily to establish an upper bound on baseline model performance. Even with oracle knowledge of the optimal DA method for each dataset/modality combination - an unrealistic scenario in practice - the baseline model cannot match Synth performance in out-of-domain settings. Full DA results for all methods appear in Appendix Tables \ref{tab:supp-da-arc}-\ref{tab:supp-da-ploras}. While individual DA methods show varied effectiveness, the key finding is that synthetic pre-training provides robustness that cannot be recovered through test-time adaptation alone. This underscores the value of appearance-invariant training, even with our current limitations in modelling stroke heterogeneity through single Gaussian distributions. Domain adaptation results varied substantially by method and modality, with no single approach providing consistent improvements. We present all tested combinations rather than cherry-picking optimal results, as real-world deployment would lack oracle knowledge of the best DA method for unseen data.

    \subsection{Ablation: To mix or not to mix?}\label{sec:ablation}

        For all experiments shown thus far, the Synth model used a mix of both synthetic data and the real ATLAS dataset. In order to evaluate whether this decision is justified, an ablation is performed where the Synth model using mixed real/synthetic data is compared to a model trained with only synthetic data. This model is trained in the exact same manner as described previously for the baseline and Synth models. Results for the test datasets ATLAS, ARC and ISLES 2015 are shown in Tables \ref{tab:atlas_combined_results}, \ref{tab:arc_combined_results} and \ref{tab:isles2015_combined_results} respectively. 
        
        The results reveal a nuanced picture. For in-domain and near-domain T1w data (ATLAS, ARC), mixing with real data provides clear benefits, with pure synthetic training achieving only 19.7\% and 46.7\% Dice respectively compared to 48.2\% and 72.3\% for mixed training. However, for several out-of-domain scenarios, pure synthetic training surprisingly often outperforms mixed training: ARC T2w (62.6\% vs 26.8\%), ISLES T1w (30.4\% vs 11.0\%), and ISLES FLAIR (37.2\% vs 21.2\%). This suggests that including real T1w data may inadvertently bias the model toward T1w-specific features, reducing generalisation to other modalities. The T2w and FLAIR improvements with pure synthetic training may reflect these sequences' different tissue contrast patterns, which are better captured by unbiased synthetic variation than by a model partially trained on T1w-specific features. Further improving the realism of the generated lesions with methods proposed in \citet{liu2025unravelingnormalanatomyfluiddriven} may aid in closing the gap for the model trained only on synthetic data, potentially achieving both in-domain performance and out-of-domain generalisation.
        
        \begin{figure*}[h!]
            \centering
            \begin{subfigure}[t]{\textwidth}
                \centering
                \includegraphics[width=0.9\textwidth]{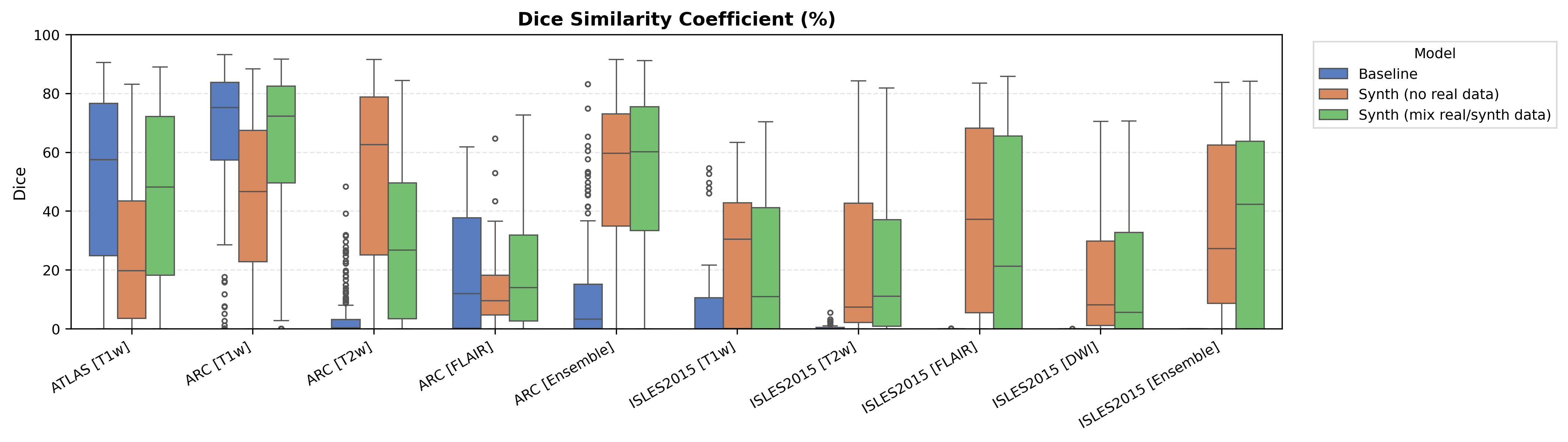}
            \end{subfigure}
            \\
            \begin{subfigure}[t]{\textwidth}
                \centering
                \includegraphics[width=0.9\textwidth]{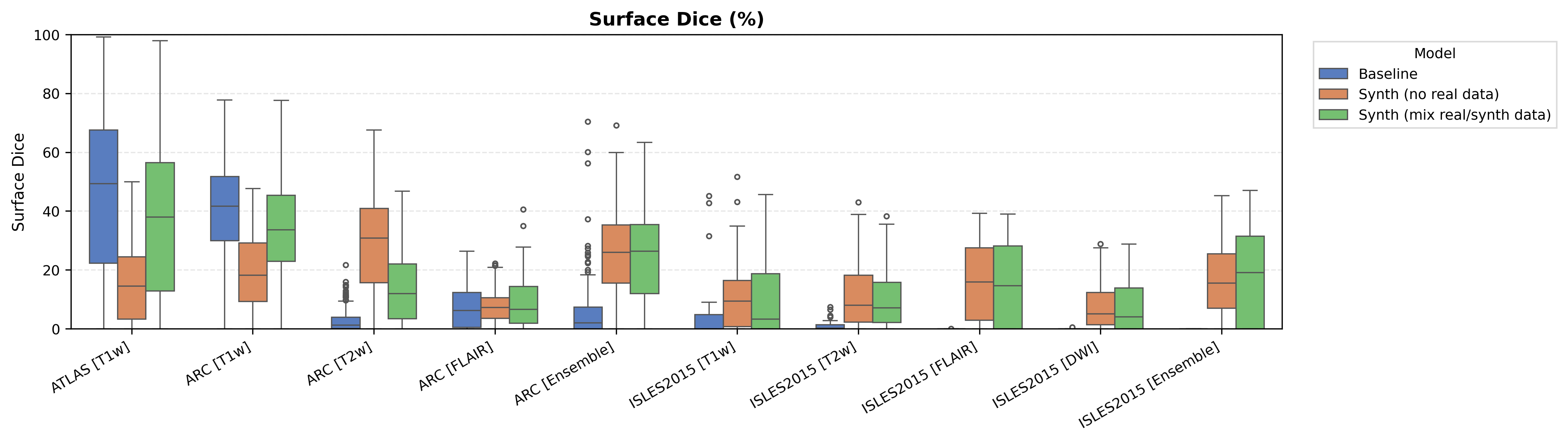}
            \end{subfigure}
            \caption{Dice and Surface Dice metrics for all reported datasets, for models trained on different combinations of real/synthetic data.}
            \label{fig:combined-ablation}
        \end{figure*}

        \begin{table}[h!]
        \centering
        \caption{Median results for ablation and pseudo-label experiments on the ATLAS test set (\textit{N}=131). Best score shown in bold. Student's \textit{t} distribution 95\% confidence intervals given in brackets.}
        \label{tab:atlas_combined_results}
        \resizebox{0.45\textwidth}{!}{
        \begin{tabular}{lcccccccc}
        \toprule
        \textbf{Modality} & \textbf{Model} & \textbf{Dice (\%)} & \textbf{Surface Dice (\%)} \\
        \midrule
        \multirow{5}{*}{\textbf{T1w}} & \textbf{Baseline} & 57.5 (52.3-62.7) & \textbf{49.4 (44.5-54.3)} \\
         & \textbf{Baseline+Pseudo} & \textbf{59.8 (54.8-64.7)} & 48.6 (44.0-53.3) \\
         & \textbf{Synth (no real data)} & 19.7 (15.5-23.9) & 14.6 (12.3-16.9) \\
         & \textbf{Synth (Ours)} & 48.2 (43.1-53.4) & 38.1 (33.4-42.7) \\
         & \textbf{Synth+Pseudo} & 49.6 (44.5-54.7) & 38.2 (33.8-42.6) \\
        \bottomrule
        \end{tabular}
        }
        \end{table}

        \begin{table}[h!]
        \centering
        \caption{Median results for ablation and pseudo-label experiments on the ARC test set (\textit{N}=229). Best score shown in bold. Student's \textit{t} distribution 95\% confidence intervals given in brackets.}
        \label{tab:arc_combined_results}
        \resizebox{0.45\textwidth}{!}{
        \begin{tabular}{lcccccccc}
        \toprule
        \textbf{Modality} & \textbf{Model} & \textbf{Dice (\%)} & \textbf{Surface Dice (\%)} \\
        \midrule
        \multirow{5}{*}{\textbf{T1w}} & \textbf{Baseline} & 75.2 (71.5-79.0) & \textbf{41.7 (39.1-44.4)} \\
         & \textbf{Baseline+Pseudo} & \textbf{75.9 (72.1-79.7)} & 38.2 (35.8-40.7) \\
         & \textbf{Synth (no real data)} & 46.7 (43.0-50.4) & 18.2 (16.5-20.0) \\
         & \textbf{Synth (Ours)} & 72.3 (68.4-76.2) & 33.7 (31.2-36.2) \\
         & \textbf{Synth+Pseudo} & 74.3 (70.4-78.2) & 36.8 (34.3-39.3) \\
        \midrule
        \multirow{5}{*}{\textbf{T2w}} & \textbf{Baseline} & 0.4 (0.0-1.4) & 1.2 (0.8-1.7) \\
         & \textbf{Baseline+Pseudo} & 1.4 (0.3-2.4) & 3.0 (2.5-3.4) \\
         & \textbf{Synth (no real data)} & \textbf{62.6 (58.8-66.5)} & \textbf{30.9 (28.7-33.0)} \\
         & \textbf{Synth (Ours)} & 26.8 (23.3-30.2) & 12.0 (10.4-13.6) \\
         & \textbf{Synth+Pseudo} & 37.2 (33.2-41.1) & 14.7 (12.7-16.7) \\
        \midrule
        \multirow{5}{*}{\textbf{FLAIR}} & \textbf{Baseline} & 12.0 (7.6-16.3) & 6.3 (4.7-7.8) \\
         & \textbf{Baseline+Pseudo} & \textbf{14.5 (10.1-18.9)} & 7.0 (5.5-8.5) \\
         & \textbf{Synth (no real data)} & 9.6 (7.0-12.3) & \textbf{7.3 (6.1-8.6)} \\
         & \textbf{Synth (Ours)} & 14.1 (9.7-18.5) & 6.6 (4.8-8.4) \\
         & \textbf{Synth+Pseudo} & 12.9 (8.4-17.3) & 6.3 (4.4-8.1) \\
        \midrule
        \multirow{5}{*}{\textbf{Ensemble}} & \textbf{Baseline} & 3.4 (1.3-5.4) & 2.0 (0.8-3.2) \\
         & \textbf{Baseline+Pseudo} & 12.4 (9.7-15.0) & 6.5 (5.1-8.0) \\
         & \textbf{Synth (no real data)} & 59.7 (56.2-63.1) & 26.0 (24.2-27.9) \\
         & \textbf{Synth (Ours)} & 60.2 (56.6-63.8) & 26.3 (24.3-28.4) \\
         & \textbf{Synth+Pseudo} & \textbf{69.4 (65.8-73.0)} & \textbf{34.0 (31.7-36.3)} \\
        \bottomrule
        \end{tabular}
        }
        \end{table}
        
        \begin{table}[h!]
        \centering
        \caption{Median results for ablation and pseudo-label experiments on the ISLES 2015 test set (\textit{N}=28). Best score shown in bold. Student's \textit{t} distribution 95\% confidence intervals given in brackets.}
        \label{tab:isles2015_combined_results}
        \resizebox{0.45\textwidth}{!}{
        \begin{tabular}{lcccccccc}
        \toprule
        \textbf{Modality} & \textbf{Model} & \textbf{Dice (\%)} & \textbf{Surface Dice (\%)} \\
        \midrule
        \multirow{5}{*}{\textbf{T1w}} & \textbf{Baseline} & 0.0 (0.0-7.4) & 0.0 (0.0-4.8) \\
         & \textbf{Baseline+Pseudo} & 0.3 (0.0-9.1) & 0.0 (0.0-4.0) \\
         & \textbf{Synth (no real data)} & \textbf{30.4 (21.3-39.6)} & \textbf{9.4 (4.1-14.8)} \\
         & \textbf{Synth (Ours)} & 11.0 (1.2-20.8) & 3.4 (0.0-8.5) \\
         & \textbf{Synth+Pseudo} & 17.9 (7.6-28.3) & 8.0 (2.1-13.8) \\
        \midrule
        \multirow{5}{*}{\textbf{T2w}} & \textbf{Baseline} & 0.0 (0.0-0.5) & 0.3 (0.0-1.0) \\
         & \textbf{Baseline+Pseudo} & 0.1 (0.0-0.7) & 0.4 (0.0-1.1) \\
         & \textbf{Synth (no real data)} & 7.4 (0.0-18.4) & 8.1 (3.0-13.1) \\
         & \textbf{Synth (Ours)} & 11.1 (0.7-21.6) & 7.1 (2.7-11.5) \\
         & \textbf{Synth+Pseudo} & \textbf{18.1 (7.1-29.1)} & \textbf{9.5 (4.4-14.5)} \\
        \midrule
        \multirow{5}{*}{\textbf{FLAIR}} & \textbf{Baseline} & 0.0 (0.0-0.0) & 0.0 (0.0-0.0) \\
         & \textbf{Baseline+Pseudo} & 0.0 (0.0-0.3) & 0.0 (0.0-0.4) \\
         & \textbf{Synth (no real data)} & \textbf{37.2 (25.5-48.9)} & \textbf{16.0 (10.8-21.1)} \\
         & \textbf{Synth (Ours)} & 21.2 (8.5-34.0) & 14.7 (9.0-20.3) \\
         & \textbf{Synth+Pseudo} & 4.2 (0.0-17.3) & 3.9 (0.0-9.8) \\
        \midrule
        \multirow{5}{*}{\textbf{DWI}} & \textbf{Baseline} & 0.0 (0.0-0.0) & 0.0 (0.0-0.0) \\
         & \textbf{Baseline+Pseudo} & 0.0 (0.0-0.4) & 0.0 (0.0-0.0) \\
         & \textbf{Synth (no real data)} & \textbf{8.2 (0.0-16.8)} & \textbf{5.1 (2.0-8.3)} \\
         & \textbf{Synth (Ours)} & 5.6 (0.0-14.4) & 4.1 (0.8-7.4) \\
         & \textbf{Synth+Pseudo} & 5.6 (0.0-14.7) & 4.3 (0.8-7.8) \\
        \midrule
        \multirow{5}{*}{\textbf{Ensemble}} & \textbf{Baseline} & 0.0 (0.0-0.0) & 0.0 (0.0-0.0) \\
         & \textbf{Baseline+Pseudo} & 0.0 (0.0-0.0) & 0.0 (0.0-0.0) \\
         & \textbf{Synth (no real data)} & 27.2 (15.7-38.8) & 15.6 (10.3-20.9) \\
         & \textbf{Synth (Ours)} & 42.3 (30.2-54.5) & 19.1 (12.7-25.4) \\
         & \textbf{Synth+Pseudo} & \textbf{50.1 (37.6-62.6)} & \textbf{24.9 (18.7-31.1)} \\
        \bottomrule
        \end{tabular}
        }
        \end{table}

    \subsection{Use of Pseudo-labelling}\label{sec:pseudo}

\begin{figure*}[h!]
            \centering
            \begin{subfigure}[t]{\textwidth}
                \centering
                \includegraphics[width=0.9\textwidth]{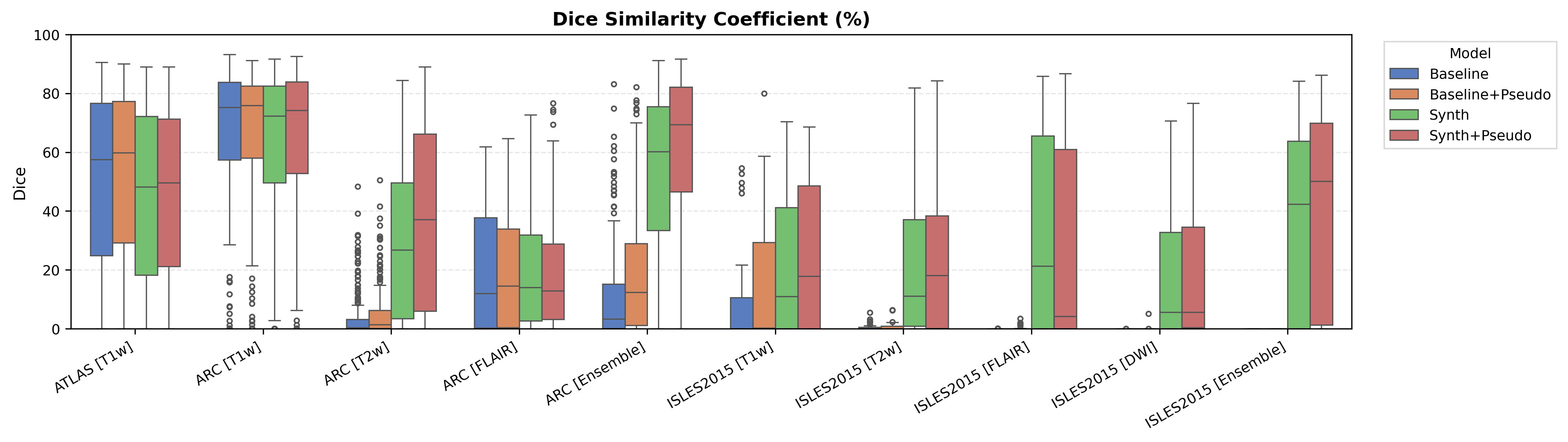}
            \end{subfigure}
            \\
            \begin{subfigure}[t]{\textwidth}
                \centering
                \includegraphics[width=0.9\textwidth]{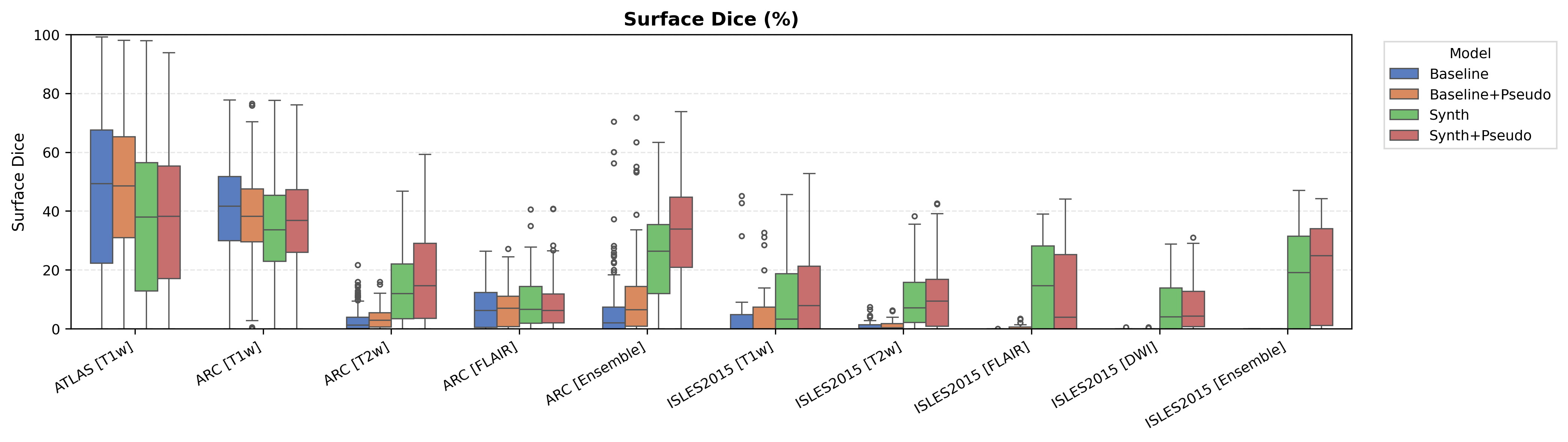}
            \end{subfigure}
            \caption{Dice and Surface Dice metrics for all reported datasets, for models trained with/without an additional training dataset of MPRAGE images and pseudo-labels.}
            \label{fig:combined-pseudo}
        \end{figure*}

        The use of synthetic data for semi-supervised pseudo-label training is also explored through the use of the PLORAS MPRAGE dataset (N=1159), distinct from the PLORAS hospital scans used in the Experiments. This approach leverages a key advantage of our framework: because synthetic training decouples image generation from label accuracy, the model can tolerate and potentially benefit from imperfect pseudo-labels. Results in Figure \ref{fig:combined-pseudo} and Tables \ref{tab:atlas_combined_results}, \ref{tab:arc_combined_results} and \ref{tab:isles2015_combined_results} demonstrate an overall positive effect of pseudo labels in the Synth model, with the exception of FLAIR images in ISLES 2015. 
        
        Notable improvements include: ARC T2w increasing from 26.8\% to 37.2\% Dice, ISLES T1w from 11.0\% to 17.9\%, and ARC ensemble from 60.2\% to 69.4\%. The baseline model shows minimal improvement with pseudo-labels in out-of-domain scenarios (e.g., ISLES ensemble remains at 0.0\%), suggesting that pseudo-labelling cannot overcome fundamental domain shift. The FLAIR degradation in ISLES 2015 (21.2\% to 4.2\%) may indicate that pseudo-labels from chronic MPRAGE data introduce biases incompatible with acute FLAIR lesion appearance. The in-domain ATLAS dataset also shows a notable improvement for both models as a result of pseudo labels, with a more marked increase in the baseline model, suggesting pseudo-labels effectively expand the training distribution when the domain gap is small.

    These results across four diverse datasets - from research-quality ATLAS to challenging clinical PLORAS scans - reveal consistent patterns that warrant deeper analysis of the fundamental principles underlying domain robustness.

\section{Discussion}

    Our results reveal fundamental principles about domain robustness in medical image segmentation. The consistent failure of domain adaptation to rescue baseline models - even with oracle selection of the optimal method - demonstrates that post-hoc adaptation cannot substitute for domain-invariant training. This finding challenges prevailing assumptions about the sufficiency of test-time adaptation for clinical deployment.
    
    The compound effect observed with Synth+DAE suggests that synthetic pre-training and test-time adaptation address different aspects of domain shift. Synthetic training provides feature representations invariant to appearance changes, while DA methods fine-tune these representations to specific test-time distributions. However, this synergy only emerges when the base model already possesses sufficient robustness - when initial predictions fail catastrophically (as with baseline models on ISLES 2015 and PLORAS), no amount of adaptation can recover performance.
    
    The variable benefit of multi-modal ensembling - substantial for ISLES 2015 but mixed for ARC - indicates that fusion strategies must consider dataset-specific characteristics rather than assuming universal improvement. Clinical deployment should balance single-modality robustness with opportunistic multi-modal fusion.

\section{Limitations and Future Directions}

    Appendix \ref{app:unimodal} confirms that 20–40\% of lesions in several cohorts are multimodal, underscoring the central limitation of our single-Gaussian sampling. While our spatially-varying approach introduces intra-lesion heterogeneity, it does not address the multi-modal nature of stroke appearance. Future implementations should explore mixture models to capture lesions that simultaneously exhibit hyper- and hypo-intense regions within the same pathology. It may also be advantageous to paste several lesions per subject, each assigned its own independently sampled intensity profile. This would allow a single synthetic brain to display chronic hypointense scars alongside acute hyperintense infarcts, better reflecting the heterogeneous lesion chronology typically observed in stroke cohorts.
    
    Additional qualitative results in Appendix Figures \ref{fig:ploras-good-t2} - \ref{fig:ploras-bad-CT} display further positive and negative results across the available image modalities within the PLORAS dataset. A number of failure modes may be observed here, such as missing cerebellar lesions or under-segmenting large hemispheric lesions.

    Although results in the paper provide strong evidence of the raw value of the segmentation metrics, it is important to further validate that the predictions made from the trained model are useful for downstream tasks. This could be validated by comparing predictivity of the segmentation masks for functional scores such as the Comprehensive Aphasia Test (CAT) \citep{Hope2024} or the NIH Stroke Score (NIHSS).

    In the context of neuroimaging data, there are many shifts in domain beyond those related to scanner sequence and resolution targeted in this work. Shifts related to anatomical shape, brought on by changes in demographics or the presence of confounding conditions such as atrophy are of equal importance to shifts in image appearance. Prior work has attempted to model such factors through the use of causal inference and counterfactual generation \citep{pawlowski2020deepstructuralcausalmodels,Pombo2023-yd,ribeiro2023highfidelityimagecounterfactuals}. It is conceivable that such an approach could be used to introduce morphological changes to the healthy tissue maps to create a model that is agnostic to shifts in domains related to both shape and appearance.

    While this study focused on stroke, the results are expected to translate to other similar domains, such as haemorrhage and glioblastoma. Future work will compare the impact of the mixing proportions of real and synthetic data. Additionally, the utility of multi-modal data in a multi-channel model will be examined versus post-hoc averaging of individual modality predictions. Lesions often appear differently across modalities, so a model trained with multi-channel inputs is expected to leverage these differences more effectively. This relationship between multi-channel inputs may better be modelled using quantitative MRI data.

\section{Conclusion}

    In this study, we introduced a novel synthetic data generation and training framework for stroke lesion segmentation, building on the success of prior works in healthy brain parcellation. A model trained using this novel framework was evaluated on a wide range of datasets covering research and clinical data, and chronic and acute stroke pathologies. Our experiments demonstrate that synthetic pre-training provides fundamental robustness unachievable through test-time adaptation alone. While accepting a 9.3\% median Dice reduction in-domain (48.2\% vs 57.5\%), our approach maintains performance where conventional methods fail entirely. Even with oracle knowledge of the optimal domain adaptation method - an unrealistic scenario in practice - conventionally-trained models cannot match our synthetic approach in out-of-domain settings.
    
    Our results demonstrate that even imperfect appearance modelling can provide substantial benefits for cross-modality segmentation. Incorporating recent contemporary work on realistic lesion simulation \citep{liu2025unravelingnormalanatomyfluiddriven} may further improve performance both in- and out-of-distribution.
    
    Qualitative results in Figure \ref{fig:ploras-viz} demonstrate the potential of the model as a starting checkpoint in an active-learning framework such as MONAI Label \citep{Diaz_Pinto_2024}, where predictions may be refined to generate training data for task-specific fine-tuning. The uncertainty of healthy-tissue segmentations via MC dropout may also provide an effective heuristic for prioritising items to refine/label in such a framework \citep{Nath2021}.

    Results also showed the potential benefit of the model for semi-supervised learning through pseudo labelling, where predicted labels may be fed back into the model with much less caution than is typically required when training with real images \citep{pham2021metapseudolabels,Xu2024}. The ablation study revealed that pure synthetic training can outperform mixed training for certain out-of-domain modalities, suggesting opportunities for further optimisation of the real/synthetic data balance. A simple solution to further reinforce this may be to train a real/fake discriminator to perform automated quality-control and threshold optimisation for the generated binary maps.

    Our approach reduces reliance on domain-specific training data and helps bridge the gap between research-grade and clinical scans to improve clinical stroke neuroimaging workflows, providing a foundation for more robust and widely applicable lesion segmentation tools.


\acks{LC is supported by the EPSRC-funded UCL Centre for Doctoral Training in Intelligent, Integrated Imaging in Healthcare (i4health) (EP/S021930/1), and the Wellcome Trust (203147/Z/16/Z and 205103/Z/16/Z). IP supported by the Alzheimer's Association grant number SG-20-678486-GAAIN2. CP is funded by Wellcome (203147/Z/16/Z, 205103/Z/16/Z and 224562/Z/21/Z). This research was supported by NVIDIA and utilised NVIDIA RTX A6000 48GB.}

%
\ethics{The work follows appropriate ethical standards in conducting research and writing the manuscript, following all applicable laws and regulations regarding treatment of animals or human subjects.}

\coi{We declare we don't have conflicts of interest.}

\data{Authors have released public code and weights to reimplement all experiments. Model weights are also made available as a toolbox for SPM. This toolbox is written in pure MATLAB and requires no external installs or compilation, to minimise the barrier for clinical evaluation. All datasets are public except the PLORAS study data, which is part of an ongoing study and will be published upon completion.}

\bibliography{citations}

\clearpage
\appendix
    \section{Lesion-intensity unimodality}\label{app:unimodal}

        \textbf{Motivation.} Most classical stroke-segmentation pipelines, including our synthetic generator, implicitly assume that the grey-level distribution of a lesion is \emph{unimodal}. That assumption underpins common appearance priors such as single-Gaussian modelling and simple intensity normalisation. However, chronic infarcts often contain a mixture of tissue constituents, and in CT the same infarct can include both hypo- and hyper-dense cores. Quantifying how often real lesions violate unimodality therefore informs whether richer mixture-based priors are warranted.
        
        \textbf{Experiment.} For every connected component in each dataset, we extracted the raw voxel intensities and applied Hartigan’s dip test \citep{Hartigan1985} to the empirical one-dimensional distribution. A lesion was labelled \emph{unimodal} when the null hypothesis of unimodality could not be rejected at $\alpha=0.05$; otherwise it was classified as \emph{multimodal}. For efficiency, we randomly subsampled at most 2048 voxels per lesion before testing. Table \ref{tab:diptest_summary} summarises the counts.
        
        \textbf{Findings.} Across MRI datasets the majority of lesions were indeed unimodal, but a non-negligible tail of multimodal cases emerged:
        
        \begin{itemize}
          \item \textbf{ATLAS} (chronic MPRAGE T1w) showed \emph{$>95\%$} unimodal lesions; only 80 of 1818 connected components exhibited multimodality, in line with mature cavities whose signal is dominated by a single CSF-like class.
          \item \textbf{ISLES 2015} exhibited 12-38\% multimodal lesions depending on modality, reflecting mixed-phase tissue in a sub-acute cohort.
          \item \textbf{ARC} (chronic research scans) retained a high unimodal rate in T1w/T2w but FLAIR contained 39\% multimodal lesions.
          \item \textbf{PLORAS} revealed the strongest departure: while T2w/FLAIR remained largely unimodal, \emph{CT} lesions were predominantly multimodal (58\%), underscoring how density heterogeneity dominates in CT.
        \end{itemize}
        
        \textbf{Implications.} A single-Gaussian appearance prior is largely adequate for \textit{T1w/T2w} MRI, where $\geq 80\%$ of lesions were unimodal; it begins to break down for modalities that emphasise subtle tissue heterogeneity - most notably \textit{FLAIR} and especially \textit{CT}. Adopting explicit mixture-based priors, or synthetically sampling lesions with mixed intensities, therefore represents a principled next step for improving model robustness across modalities and imaging protocols.
        
        \begin{table}[ht]
            \centering
            \caption{Counts of connected‐component lesions that are \textbf{unimodal} or \textbf{multimodal} in each dataset–modality pair. For every lesion larger than 10 voxels we sampled at most 2048 raw intensity values and ran Hartigan’s dip test; failure to reject the null hypothesis of unimodality at $\alpha=0.05$ yields the \textbf{Unimodal} label, otherwise \textbf{Multimodal}.  ATLAS is split into the subjects used for model development (\textit{T1w [Train]}) and the held-out evaluation set (\textit{T1w [Validation]}).}
            \label{tab:diptest_summary}
            \resizebox{0.45\textwidth}{!}{
            \begin{tabular}{llcc}
            \toprule
            \textbf{Dataset} & \textbf{Modality} & \textbf{Unimodal} & \textbf{Multimodal} \\
            \midrule
            \multirow{2}{*}{\textbf{ATLAS}} & \textbf{T1w [Train]} & 1384 & 65 \\
             & \textbf{T1w [Validation]} & 354 & 15 \\
            \midrule
            \multirow{4}{*}{\textbf{ISLES2015}} & \textbf{T1w} & 52 & 9 \\
             & \textbf{T2w} & 46 & 15 \\
             & \textbf{FLAIR} & 54 & 7 \\
             & \textbf{DWI} & 38 & 23 \\
            \midrule
            \multirow{3}{*}{\textbf{ARC}} & \textbf{T1w} & 286 & 64 \\
             & \textbf{T2w} & 335 & 48 \\
             & \textbf{FLAIR} & 90 & 57 \\
            \midrule
            \multirow{3}{*}{\textbf{PLORAS}} & \textbf{T2w} & 203 & 14 \\
             & \textbf{FLAIR} & 712 & 29 \\
             & \textbf{CT} & 248 & 342 \\
            \bottomrule
            \end{tabular}
            }
        \end{table}

    \section{Further segmentation metrics}\label{app:metrics}

        Additional metrics are reported below in Appendix Tables \ref{tab:supp-atlas-scores} - \ref{tab:supp-da-ploras} for all datasets. Here we report individual scores for each DA method, for both the baseline model and our Synth model.

        \begin{table}[h!]
        \centering
        \caption{Mean results on ATLAS hold-out set (\textit{N}=131). Best score shown in bold. FPR values are shown as a multiple of $10^{-3}.$}
        \resizebox{0.4\textwidth}{!}{
        \begin{tabular}{lcccccccc}
        \toprule
        \textbf{Modality} & \textbf{Model} & \textbf{Dice} & \textbf{HD95} & \textbf{AVD} & \textbf{ALD} & \textbf{LF1} & \textbf{TPR} & \textbf{FPR} \\
        \midrule
        \multirow{4}{*}{\textbf{T1w}} & \textbf{Baseline} & 0.500 & 44.3 & 7.60 & 3.48 & 0.480 & \textbf{0.510} & 0.251 \\
         & \textbf{Baseline+TTA} & \textbf{0.503} & \textbf{41.4} & \textbf{7.34} & 2.52 & 0.537 & 0.502 & 0.234 \\
         & \textbf{Synth (Ours)} & 0.456 & 47.4 & 8.73 & 2.26 & 0.536 & 0.426 & 0.200 \\
         & \textbf{Synth+TTA} & 0.455 & 48.3 & 8.51 & \textbf{1.76} & \textbf{0.559} & 0.424 & \textbf{0.191} \\
        \bottomrule
        \end{tabular}
        }
        \label{tab:supp-atlas-scores}
        \end{table}
        
        \begin{table}[h!]
        \centering
        \caption{Mean results on the ARC dataset (\textit{N}=229). Best score per column is shown in bold. FPR values are shown as a multiple of 1000.}
        \resizebox{0.4\textwidth}{!}{
        \begin{tabular}{lcccccccc}
        \toprule
        \textbf{Modality} & \textbf{Model} & \textbf{Dice} & \textbf{HD95} & \textbf{AVD} & \textbf{ALD} & \textbf{LF1} & \textbf{TPR} & \textbf{FPR} \\
        \midrule
        \multirow{14}{*}{\textbf{T1w}} & \textbf{Baseline} & 0.646 & 24.5 & \textbf{18.04} & 6.19 & 0.335 & 0.600 & 1.178 \\
         & \textbf{Baseline+TTA} & \textbf{0.648} & 23.7 & 18.33 & 3.42 & 0.442 & 0.598 & 1.126 \\
         & \textbf{Baseline+TENT} & 0.077 & 144.9 & 75.50 & 3.51 & 0.180 & 0.060 & \textbf{0.285} \\
         & \textbf{Baseline+DAE} & 0.614 & 26.2 & 26.53 & 3.15 & 0.507 & \textbf{0.685} & 2.238 \\
         & \textbf{Baseline+PL} & 0.325 & 67.5 & 146.72 & 175.61 & 0.017 & 0.613 & 10.342 \\
         & \textbf{Baseline+UPL} & 0.341 & 74.4 & 48.10 & 22.84 & 0.244 & 0.262 & 0.695 \\
         & \textbf{Baseline+DPL} & 0.340 & 52.5 & 52.48 & 9.53 & 0.324 & 0.247 & 0.465 \\
         & \textbf{Synth (Ours)} & 0.609 & 27.9 & 22.88 & 1.99 & 0.613 & 0.539 & 0.953 \\
         & \textbf{Synth+TTA} & 0.611 & 27.0 & 24.08 & \textbf{1.84} & \textbf{0.640} & 0.534 & 0.885 \\
         & \textbf{Synth+TENT} & 0.560 & 33.4 & 26.71 & 2.42 & 0.528 & 0.604 & 2.385 \\
         & \textbf{Synth+DAE} & 0.613 & \textbf{23.4} & 20.60 & 2.26 & 0.581 & 0.640 & 1.903 \\
         & \textbf{Synth+PL} & 0.311 & 158.6 & 48.41 & 260.54 & 0.017 & 0.310 & 2.303 \\
         & \textbf{Synth+UPL} & 0.340 & 56.3 & 54.57 & 20.35 & 0.334 & 0.243 & 0.457 \\
         & \textbf{Synth+DPL} & 0.255 & 58.3 & 63.36 & 7.44 & 0.530 & 0.169 & 0.339 \\
        \midrule
        \multirow{14}{*}{\textbf{T2w}} & \textbf{Baseline} & 0.040 & 76.8 & 73.80 & 47.12 & 0.044 & 0.063 & 6.778 \\
         & \textbf{Baseline+TTA} & 0.023 & 77.0 & 71.83 & 26.00 & 0.054 & 0.036 & 6.819 \\
         & \textbf{Baseline+TENT} & 0.000 & 148.5 & 85.40 & 5.11 & 0.047 & 0.000 & 0.206 \\
         & \textbf{Baseline+DAE} & 0.056 & 75.8 & 123.49 & 10.09 & 0.113 & 0.134 & 10.938 \\
         & \textbf{Baseline+PL} & 0.006 & 78.8 & 127.78 & 353.31 & 0.007 & 0.014 & 12.731 \\
         & \textbf{Baseline+UPL} & 0.000 & 256.0 & 85.74 & 2.57 & 0.013 & 0.000 & 0.185 \\
         & \textbf{Baseline+DPL} & 0.000 & 256.0 & 85.74 & \textbf{2.57} & 0.013 & 0.000 & \textbf{0.185} \\
         & \textbf{Synth (Ours)} & 0.299 & 66.2 & 40.79 & 38.81 & 0.060 & 0.304 & 2.282 \\
         & \textbf{Synth+TTA} & 0.284 & 66.2 & 43.70 & 16.00 & 0.117 & 0.274 & 1.893 \\
         & \textbf{Synth+TENT} & 0.321 & 55.5 & 43.43 & 15.09 & 0.140 & 0.326 & 1.663 \\
         & \textbf{Synth+DAE} & \textbf{0.437} & \textbf{40.8} & \textbf{40.19} & 11.43 & \textbf{0.196} & 0.410 & 1.344 \\
         & \textbf{Synth+PL} & 0.384 & 65.5 & 40.26 & 22.77 & 0.095 & \textbf{0.475} & 3.881 \\
         & \textbf{Synth+UPL} & 0.120 & 50.7 & 73.30 & 22.71 & 0.172 & 0.074 & 0.484 \\
         & \textbf{Synth+DPL} & 0.055 & 52.3 & 78.43 & 47.28 & 0.108 & 0.032 & 0.447 \\
        \midrule
        \multirow{14}{*}{\textbf{FLAIR}} & \textbf{Baseline} & 0.193 & 66.2 & 52.15 & 49.49 & 0.049 & 0.205 & 4.532 \\
         & \textbf{Baseline+TTA} & 0.202 & 65.7 & 47.98 & 25.98 & 0.090 & 0.197 & 3.719 \\
         & \textbf{Baseline+TENT} & 0.011 & 165.6 & 79.40 & 4.94 & 0.070 & 0.008 & 0.628 \\
         & \textbf{Baseline+DAE} & 0.287 & 54.8 & 75.47 & 11.09 & 0.232 & \textbf{0.327} & 5.981 \\
         & \textbf{Baseline+PL} & 0.126 & 81.9 & 162.13 & 382.66 & 0.007 & 0.240 & 13.548 \\
         & \textbf{Baseline+UPL} & 0.000 & 256.0 & 82.36 & 2.62 & 0.035 & 0.000 & 0.497 \\
         & \textbf{Baseline+DPL} & 0.000 & 256.0 & 82.36 & \textbf{2.62} & 0.035 & 0.000 & \textbf{0.497} \\
         & \textbf{Synth (Ours)} & 0.199 & 59.3 & 60.36 & 17.69 & 0.150 & 0.157 & 1.349 \\
         & \textbf{Synth+TTA} & 0.190 & 56.4 & 62.98 & 8.36 & 0.256 & 0.143 & 1.185 \\
         & \textbf{Synth+TENT} & \textbf{0.340} & \textbf{49.7} & 46.05 & 5.01 & 0.247 & 0.286 & 1.582 \\
         & \textbf{Synth+DAE} & 0.263 & 51.5 & 53.14 & 5.88 & 0.248 & 0.215 & 1.581 \\
         & \textbf{Synth+PL} & 0.276 & 96.8 & \textbf{45.97} & 93.18 & 0.027 & 0.255 & 2.133 \\
         & \textbf{Synth+UPL} & 0.029 & 99.1 & 80.44 & 4.79 & \textbf{0.272} & 0.016 & 0.520 \\
         & \textbf{Synth+DPL} & 0.010 & 146.9 & 81.38 & 7.11 & 0.189 & 0.005 & 0.512 \\
        \midrule
        \multirow{14}{*}{\textbf{Ensemble}} & \textbf{Baseline} & 0.107 & 53.3 & 77.04 & 22.20 & 0.309 & 0.071 & 1.137 \\
         & \textbf{Baseline+TTA} & 0.087 & 65.8 & 78.91 & 9.10 & 0.429 & 0.057 & 1.126 \\
         & \textbf{Baseline+TENT} & 0.003 & 238.0 & 85.73 & 2.60 & 0.054 & 0.002 & 0.185 \\
         & \textbf{Baseline+DAE} & 0.212 & 51.3 & 73.84 & 39.25 & 0.168 & 0.212 & 2.350 \\
         & \textbf{Baseline+PL} & 0.016 & 76.1 & 70.76 & 512.92 & 0.004 & 0.014 & 3.869 \\
         & \textbf{Baseline+UPL} & 0.000 & 256.0 & 85.74 & 2.57 & 0.013 & 0.000 & 0.185 \\
         & \textbf{Baseline+DPL} & 0.000 & 256.0 & 85.74 & \textbf{2.57} & 0.013 & 0.000 & \textbf{0.185} \\
         & \textbf{Synth (Ours)} & 0.525 & 34.9 & 35.60 & 10.09 & 0.356 & 0.451 & 0.770 \\
         & \textbf{Synth+TTA} & 0.530 & 33.0 & 36.51 & 4.50 & 0.470 & 0.452 & 0.696 \\
         & \textbf{Synth+TENT} & 0.451 & 39.1 & 38.50 & 11.12 & 0.370 & 0.412 & 0.960 \\
         & \textbf{Synth+DAE} & \textbf{0.579} & \textbf{27.6} & \textbf{27.52} & 6.77 & 0.432 & \textbf{0.531} & 0.923 \\
         & \textbf{Synth+PL} & 0.529 & 50.2 & 33.21 & 29.02 & 0.106 & 0.515 & 1.551 \\
         & \textbf{Synth+UPL} & 0.244 & 53.0 & 65.77 & 9.80 & \textbf{0.470} & 0.159 & 0.310 \\
         & \textbf{Synth+DPL} & 0.172 & 71.4 & 71.62 & 13.97 & 0.414 & 0.107 & 0.284 \\
        \bottomrule
        \end{tabular}
        }
        \label{tab:supp-da-arc}
        \end{table}

        \begin{table}[h!]
        \centering
        \caption{Mean results on the ISLES2015 dataset (\textit{N}=28). Best score per column is shown in bold. FPR values are shown as a multiple of 1000.}
        \resizebox{0.4\textwidth}{!}{
        \begin{tabular}{lcccccccc}
        \toprule
        \textbf{Modality} & \textbf{Model} & \textbf{Dice} & \textbf{HD95} & \textbf{AVD} & \textbf{ALD} & \textbf{LF1} & \textbf{TPR} & \textbf{FPR} \\
        \midrule
        \multirow{14}{*}{\textbf{T1w}} & \textbf{Baseline} & 0.115 & 84.2 & 38.14 & 3.32 & 0.269 & 0.076 & 0.034 \\
         & \textbf{Baseline+TTA} & 0.091 & 113.2 & 40.11 & \textbf{1.96} & 0.311 & 0.058 & 0.012 \\
         & \textbf{Baseline+TENT} & 0.002 & 178.2 & 43.77 & 2.14 & 0.161 & 0.001 & 0.001 \\
         & \textbf{Baseline+DAE} & 0.246 & 68.2 & 48.07 & 7.18 & 0.285 & 0.191 & 1.604 \\
         & \textbf{Baseline+PL} & 0.040 & 81.6 & 42.79 & 643.29 & 0.003 & 0.039 & 1.979 \\
         & \textbf{Baseline+UPL} & 0.000 & 256.0 & 43.79 & 2.11 & 0.000 & 0.000 & 0.000 \\
         & \textbf{Baseline+DPL} & 0.000 & 256.0 & 43.79 & 2.11 & 0.000 & 0.000 & \textbf{0.000} \\
         & \textbf{Synth (Ours)} & 0.222 & 70.5 & 30.49 & 7.14 & 0.208 & 0.168 & 0.162 \\
         & \textbf{Synth+TTA} & 0.230 & 78.5 & 31.41 & 2.18 & \textbf{0.349} & 0.163 & 0.085 \\
         & \textbf{Synth+TENT} & 0.076 & 108.2 & 33.79 & 40.86 & 0.103 & 0.058 & 0.265 \\
         & \textbf{Synth+DAE} & \textbf{0.268} & \textbf{58.1} & \textbf{29.18} & 3.64 & 0.316 & \textbf{0.223} & 0.293 \\
         & \textbf{Synth+PL} & 0.050 & 76.9 & 42.55 & 3.93 & 0.321 & 0.028 & 0.008 \\
         & \textbf{Synth+UPL} & 0.000 & 241.7 & 43.79 & 2.04 & 0.050 & 0.000 & 0.000 \\
         & \textbf{Synth+DPL} & 0.001 & 221.7 & 43.77 & 2.46 & 0.127 & 0.000 & 0.000 \\
        \midrule
        \multirow{14}{*}{\textbf{T2w}} & \textbf{Baseline} & 0.007 & 65.4 & 50.81 & 15.00 & 0.078 & 0.044 & 2.786 \\
         & \textbf{Baseline+TTA} & 0.003 & 65.5 & 48.97 & 6.86 & 0.095 & 0.025 & 2.260 \\
         & \textbf{Baseline+TENT} & 0.001 & 142.6 & 43.38 & 3.75 & 0.050 & 0.001 & 0.085 \\
         & \textbf{Baseline+DAE} & 0.006 & 71.5 & 88.17 & 9.43 & 0.071 & 0.069 & 5.805 \\
         & \textbf{Baseline+PL} & 0.002 & 74.2 & 71.28 & 155.21 & 0.013 & 0.005 & 5.055 \\
         & \textbf{Baseline+UPL} & 0.000 & 256.0 & 43.79 & 2.11 & 0.000 & 0.000 & 0.000 \\
         & \textbf{Baseline+DPL} & 0.000 & 256.0 & 43.79 & \textbf{2.11} & 0.000 & 0.000 & \textbf{0.000} \\
         & \textbf{Synth (Ours)} & 0.231 & 74.8 & 52.73 & 11.82 & 0.166 & 0.443 & 3.534 \\
         & \textbf{Synth+TTA} & 0.245 & 74.6 & 50.99 & 4.68 & 0.279 & \textbf{0.454} & 3.379 \\
         & \textbf{Synth+TENT} & 0.215 & 77.4 & 33.44 & 20.11 & 0.109 & 0.276 & 1.755 \\
         & \textbf{Synth+DAE} & \textbf{0.255} & 63.1 & 31.10 & 4.61 & \textbf{0.280} & 0.328 & 1.437 \\
         & \textbf{Synth+PL} & 0.245 & 73.3 & 37.57 & 13.71 & 0.161 & 0.288 & 2.280 \\
         & \textbf{Synth+UPL} & 0.152 & 63.3 & \textbf{31.06} & 12.43 & 0.138 & 0.116 & 0.631 \\
         & \textbf{Synth+DPL} & 0.111 & \textbf{58.6} & 35.25 & 8.11 & 0.172 & 0.073 & 0.143 \\
        \midrule
        \multirow{14}{*}{\textbf{FLAIR}} & \textbf{Baseline} & 0.000 & 71.4 & 40.06 & 10.79 & 0.018 & 0.000 & 0.515 \\
         & \textbf{Baseline+TTA} & 0.000 & 112.9 & 41.24 & 3.43 & 0.000 & 0.000 & 0.238 \\
         & \textbf{Baseline+TENT} & 0.000 & 119.0 & 43.47 & 4.32 & 0.000 & 0.000 & 0.021 \\
         & \textbf{Baseline+DAE} & 0.000 & 80.1 & 61.65 & 15.29 & 0.015 & 0.004 & 3.679 \\
         & \textbf{Baseline+PL} & 0.000 & 77.1 & 43.43 & 70.43 & 0.009 & 0.000 & 0.490 \\
         & \textbf{Baseline+UPL} & 0.000 & 256.0 & 43.79 & 2.11 & 0.000 & 0.000 & 0.000 \\
         & \textbf{Baseline+DPL} & 0.000 & 256.0 & 43.79 & \textbf{2.11} & 0.000 & 0.000 & \textbf{0.000} \\
         & \textbf{Synth (Ours)} & 0.314 & 62.9 & 18.84 & 10.61 & 0.154 & 0.346 & 0.975 \\
         & \textbf{Synth+TTA} & \textbf{0.329} & 82.1 & \textbf{18.37} & 3.50 & \textbf{0.275} & \textbf{0.359} & 0.891 \\
         & \textbf{Synth+TENT} & 0.127 & 79.0 & 28.03 & 34.54 & 0.031 & 0.107 & 0.460 \\
         & \textbf{Synth+DAE} & 0.274 & \textbf{55.7} & 20.25 & 7.86 & 0.152 & 0.305 & 0.989 \\
         & \textbf{Synth+PL} & 0.252 & 75.7 & 50.59 & 111.43 & 0.036 & 0.330 & 3.176 \\
         & \textbf{Synth+UPL} & 0.232 & 137.9 & 24.88 & 3.04 & 0.271 & 0.187 & 0.111 \\
         & \textbf{Synth+DPL} & 0.205 & 145.6 & 25.16 & 7.36 & 0.236 & 0.168 & 0.191 \\
        \midrule
        \multirow{14}{*}{\textbf{DWI}} & \textbf{Baseline} & 0.000 & 80.1 & 42.63 & 5.54 & 0.023 & 0.000 & 0.138 \\
         & \textbf{Baseline+TTA} & 0.000 & 102.8 & 43.83 & 2.43 & 0.000 & 0.000 & 0.133 \\
         & \textbf{Baseline+TENT} & 0.000 & 103.9 & 43.52 & \textbf{1.89} & 0.000 & 0.000 & 0.021 \\
         & \textbf{Baseline+DAE} & 0.000 & 76.2 & 40.04 & 11.21 & 0.025 & 0.000 & 2.123 \\
         & \textbf{Baseline+PL} & 0.004 & 77.5 & 40.33 & 835.36 & 0.001 & 0.004 & 0.726 \\
         & \textbf{Baseline+UPL} & 0.000 & 256.0 & 43.79 & 2.11 & 0.000 & 0.000 & 0.000 \\
         & \textbf{Baseline+DPL} & 0.000 & 256.0 & 43.79 & 2.11 & 0.000 & 0.000 & \textbf{0.000} \\
         & \textbf{Synth (Ours)} & 0.193 & 82.5 & 47.64 & 16.71 & 0.100 & 0.352 & 2.741 \\
         & \textbf{Synth+TTA} & \textbf{0.204} & 81.1 & 44.32 & 8.39 & 0.187 & 0.336 & 2.344 \\
         & \textbf{Synth+TENT} & 0.104 & 86.6 & \textbf{30.71} & 38.50 & 0.051 & 0.108 & 0.882 \\
         & \textbf{Synth+DAE} & 0.185 & 80.4 & 39.10 & 12.36 & 0.106 & 0.262 & 2.139 \\
         & \textbf{Synth+PL} & 0.186 & 85.2 & 81.64 & 17.43 & 0.120 & \textbf{0.397} & 5.759 \\
         & \textbf{Synth+UPL} & 0.179 & \textbf{60.8} & 32.42 & 19.43 & 0.154 & 0.155 & 0.268 \\
         & \textbf{Synth+DPL} & 0.089 & 117.4 & 39.27 & 3.75 & \textbf{0.300} & 0.060 & 0.031 \\
        \midrule
        \multirow{14}{*}{\textbf{Ensemble}} & \textbf{Baseline} & 0.000 & 242.8 & 43.79 & \textbf{2.00} & 0.000 & 0.000 & 0.000 \\
         & \textbf{Baseline+TTA} & 0.000 & 256.0 & 43.79 & 2.11 & 0.000 & 0.000 & 0.000 \\
         & \textbf{Baseline+TENT} & 0.000 & 256.0 & 43.79 & 2.11 & 0.000 & 0.000 & 0.000 \\
         & \textbf{Baseline+DAE} & 0.000 & 231.0 & 43.68 & 2.00 & 0.000 & 0.000 & 0.007 \\
         & \textbf{Baseline+PL} & 0.000 & 198.9 & 43.79 & 2.11 & 0.024 & 0.000 & 0.000 \\
         & \textbf{Baseline+UPL} & 0.000 & 256.0 & 43.79 & 2.11 & 0.000 & 0.000 & 0.000 \\
         & \textbf{Baseline+DPL} & 0.000 & 256.0 & 43.79 & 2.11 & 0.000 & 0.000 & \textbf{0.000} \\
         & \textbf{Synth (Ours)} & 0.370 & 60.4 & \textbf{19.68} & 7.50 & 0.234 & \textbf{0.339} & 0.256 \\
         & \textbf{Synth+TTA} & \textbf{0.378} & 69.7 & 20.01 & 2.75 & 0.348 & 0.334 & 0.223 \\
         & \textbf{Synth+TENT} & 0.111 & 114.8 & 32.23 & 3.21 & 0.235 & 0.089 & 0.047 \\
         & \textbf{Synth+DAE} & 0.365 & \textbf{51.4} & 19.84 & 3.75 & \textbf{0.354} & 0.331 & 0.204 \\
         & \textbf{Synth+PL} & 0.246 & 60.7 & 29.01 & 21.18 & 0.125 & 0.188 & 0.150 \\
         & \textbf{Synth+UPL} & 0.016 & 218.2 & 42.52 & 5.18 & 0.131 & 0.009 & 0.000 \\
         & \textbf{Synth+DPL} & 0.033 & 187.3 & 41.09 & 6.04 & 0.223 & 0.020 & 0.000 \\
        \bottomrule
        \end{tabular}
        }
        \label{tab:supp-da-isles2015}
        \end{table}
    
        \begin{table}[h!]
        \centering
        \caption{Mean results on the PLORAS dataset (\textit{N}=661). Best score per column is shown in bold. FPR values are shown as a multiple of 1000.}
        \resizebox{0.4\textwidth}{!}{
        \begin{tabular}{lcccccccc}
        \toprule
        \textbf{Modality} & \textbf{Model} & \textbf{Dice} & \textbf{HD95} & \textbf{AVD} & \textbf{ALD} & \textbf{LF1} & \textbf{TPR} & \textbf{FPR} \\
        \midrule
        \multirow{14}{*}{\textbf{T2w}} & \textbf{Baseline} & 0.028 & 80.3 & 112.82 & 12.22 & 0.101 & 0.064 & 9.638 \\
         & \textbf{Baseline+TTA} & 0.020 & 79.1 & 92.69 & 11.02 & 0.103 & 0.045 & 8.409 \\
         & \textbf{Baseline+TENT} & 0.005 & 101.5 & 30.79 & 5.12 & 0.020 & 0.007 & 1.849 \\
         & \textbf{Baseline+DAE} & 0.044 & 82.7 & 204.49 & 2.64 & 0.276 & 0.123 & 14.983 \\
         & \textbf{Baseline+PL} & 0.000 & 235.8 & 32.46 & 2.13 & 0.009 & 0.004 & 1.696 \\
         & \textbf{Baseline+UPL} & 0.000 & 256.0 & 32.47 & 2.12 & 0.009 & 0.004 & 1.695 \\
         & \textbf{Baseline+DPL} & 0.000 & 256.0 & 32.47 & \textbf{2.12} & 0.009 & 0.004 & \textbf{1.695} \\
         & \textbf{Synth (Ours)} & 0.250 & 71.9 & 40.20 & 8.42 & 0.243 & \textbf{0.371} & 4.156 \\
         & \textbf{Synth+TTA} & 0.253 & 70.5 & 37.00 & 4.75 & \textbf{0.329} & 0.361 & 3.917 \\
         & \textbf{Synth+TENT} & 0.202 & 77.0 & 61.99 & 21.23 & 0.118 & 0.330 & 5.612 \\
         & \textbf{Synth+DAE} & \textbf{0.291} & \textbf{58.7} & \textbf{23.68} & 4.28 & 0.307 & 0.312 & 2.969 \\
         & \textbf{Synth+PL} & 0.162 & 81.0 & 33.28 & 8.58 & 0.179 & 0.186 & 3.737 \\
         & \textbf{Synth+UPL} & 0.101 & 66.0 & 25.80 & 11.07 & 0.186 & 0.073 & 2.014 \\
         & \textbf{Synth+DPL} & 0.053 & 72.1 & 28.44 & 3.42 & 0.212 & 0.036 & 1.832 \\
        \midrule
        \multirow{14}{*}{\textbf{FLAIR}} & \textbf{Baseline} & 0.010 & 74.2 & 42.12 & 18.69 & 0.033 & 0.013 & 2.839 \\
         & \textbf{Baseline+TTA} & 0.006 & 74.6 & 38.04 & 12.57 & 0.026 & 0.008 & 2.119 \\
         & \textbf{Baseline+TENT} & 0.000 & 106.3 & 37.65 & 4.99 & 0.014 & 0.003 & 0.718 \\
         & \textbf{Baseline+DAE} & 0.027 & 76.2 & 98.18 & 3.81 & 0.155 & 0.055 & 7.805 \\
         & \textbf{Baseline+PL} & 0.000 & 249.9 & 39.25 & \textbf{2.46} & 0.000 & 0.002 & 0.582 \\
         & \textbf{Baseline+UPL} & 0.000 & 256.0 & 39.25 & 2.53 & 0.000 & 0.002 & 0.582 \\
         & \textbf{Baseline+DPL} & 0.000 & 256.0 & 39.25 & 2.53 & 0.000 & 0.002 & \textbf{0.582} \\
         & \textbf{Synth (Ours)} & 0.296 & 72.1 & 48.04 & 17.93 & 0.137 & 0.415 & 3.443 \\
         & \textbf{Synth+TTA} & \textbf{0.314} & 71.4 & 41.15 & 9.07 & \textbf{0.213} & 0.412 & 2.967 \\
         & \textbf{Synth+TENT} & 0.221 & 79.4 & 48.21 & 9.87 & 0.140 & 0.280 & 3.362 \\
         & \textbf{Synth+DAE} & 0.304 & \textbf{63.6} & 43.34 & 8.82 & 0.199 & 0.359 & 3.258 \\
         & \textbf{Synth+PL} & 0.279 & 84.3 & 81.32 & 20.01 & 0.156 & \textbf{0.519} & 5.573 \\
         & \textbf{Synth+UPL} & 0.145 & 78.5 & 31.51 & 15.50 & 0.115 & 0.108 & 1.028 \\
         & \textbf{Synth+DPL} & 0.243 & 80.1 & \textbf{25.71} & 13.82 & 0.146 & 0.224 & 1.514 \\
        \midrule
        \multirow{14}{*}{\textbf{CT}} & \textbf{Baseline} & 0.029 & 70.9 & 28.01 & 8.54 & 0.084 & 0.032 & 1.492 \\
         & \textbf{Baseline+TTA} & 0.017 & 125.0 & 27.63 & 4.11 & 0.076 & 0.011 & 0.593 \\
         & \textbf{Baseline+TENT} & 0.000 & 207.7 & 28.74 & \textbf{2.59} & 0.018 & 0.000 & 0.550 \\
         & \textbf{Baseline+DAE} & 0.064 & 78.2 & 78.65 & 2.98 & 0.198 & 0.106 & 5.426 \\
         & \textbf{Baseline+PL} & 0.011 & 78.3 & 79.93 & 2010.46 & 0.002 & 0.032 & 6.333 \\
         & \textbf{Baseline+UPL} & 0.000 & 256.0 & 28.87 & 2.84 & 0.000 & 0.000 & 0.542 \\
         & \textbf{Baseline+DPL} & 0.000 & 256.0 & 28.87 & 2.84 & 0.000 & 0.000 & \textbf{0.542} \\
         & \textbf{Synth (Ours)} & 0.234 & 66.8 & \textbf{19.71} & 5.43 & 0.233 & 0.204 & 0.965 \\
         & \textbf{Synth+TTA} & 0.229 & 85.2 & 20.34 & 2.65 & \textbf{0.317} & 0.195 & 0.913 \\
         & \textbf{Synth+TENT} & 0.222 & 58.6 & 20.82 & 9.17 & 0.139 & 0.210 & 1.245 \\
         & \textbf{Synth+DAE} & \textbf{0.307} & \textbf{51.6} & 23.81 & 4.13 & 0.258 & \textbf{0.298} & 1.657 \\
         & \textbf{Synth+PL} & 0.000 & 110.0 & 26.69 & 34.56 & 0.002 & 0.001 & 0.714 \\
         & \textbf{Synth+UPL} & 0.000 & 255.0 & 28.86 & 2.84 & 0.000 & 0.000 & 0.543 \\
         & \textbf{Synth+DPL} & 0.000 & 256.0 & 28.87 & 2.84 & 0.000 & 0.000 & 0.542 \\
        \bottomrule
        \end{tabular}
        }
        \label{tab:supp-da-ploras}
        \end{table}

    \newpage
    \section{Wilcoxon Significance Tests}

        Significance measures are provided below in Figures \ref{fig:wilcoxon-atlas-t1} - \ref{fig:wilcoxon-ploras-ct} as Wilcoxon signed-rank test values for pairwise comparisons between models on each dataset. 'Oracle DA' represents the hypothetical best-case scenario where optimal DA method is known a priori for each dataset/modality and applied to the baseline model. Median Dice values are shown along the diagonal.

        \begin{figure*}[h!]
            \centering
            \includegraphics[width=0.8\textwidth]{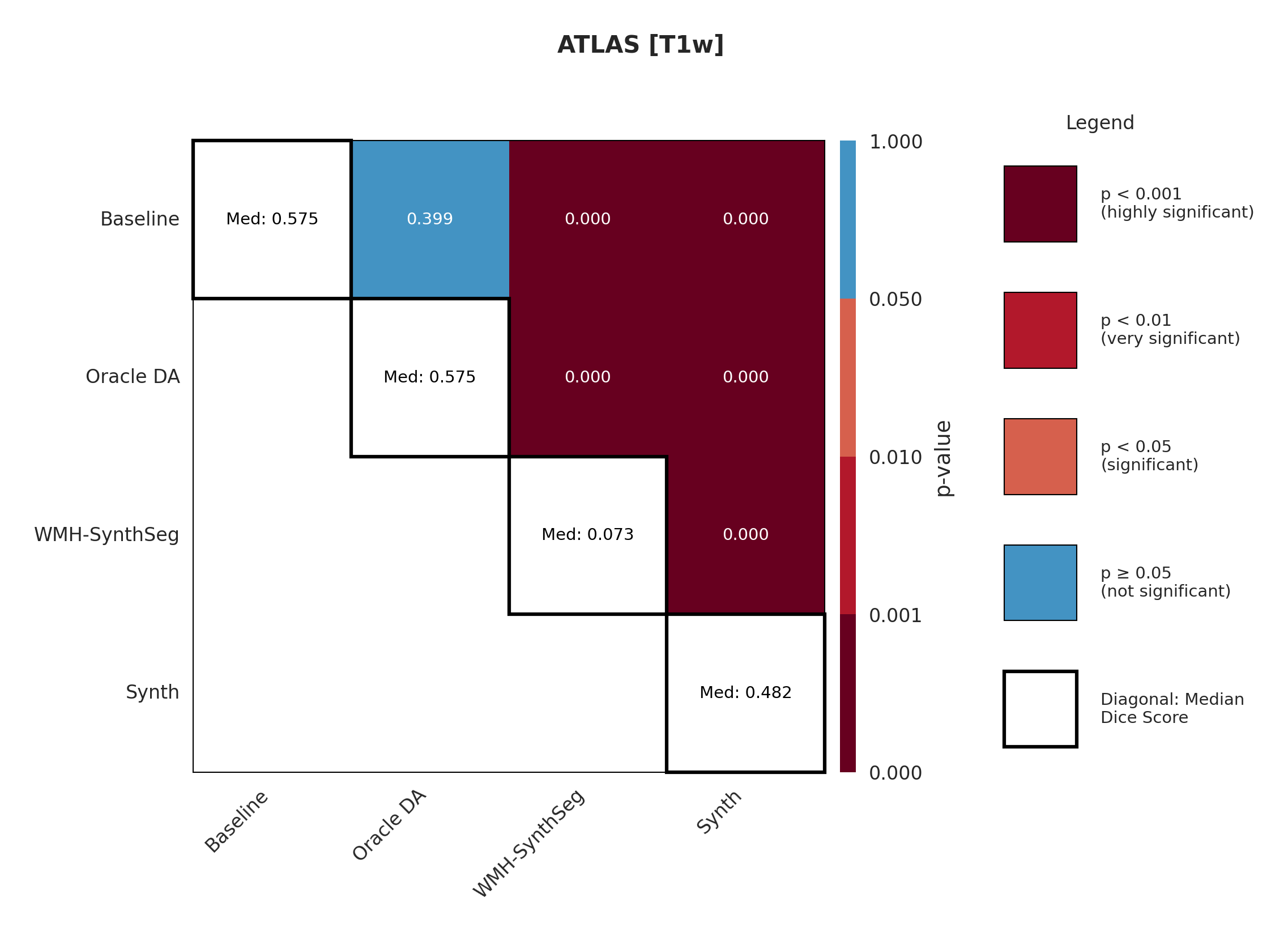}
            \caption{Wilcoxon signed-rank test values for Dice metric measurements in the ATLAS T1w dataset.}
            \label{fig:wilcoxon-atlas-t1}
        \end{figure*}

        \begin{figure*}[h!]
            \centering
            \includegraphics[width=0.8\textwidth]{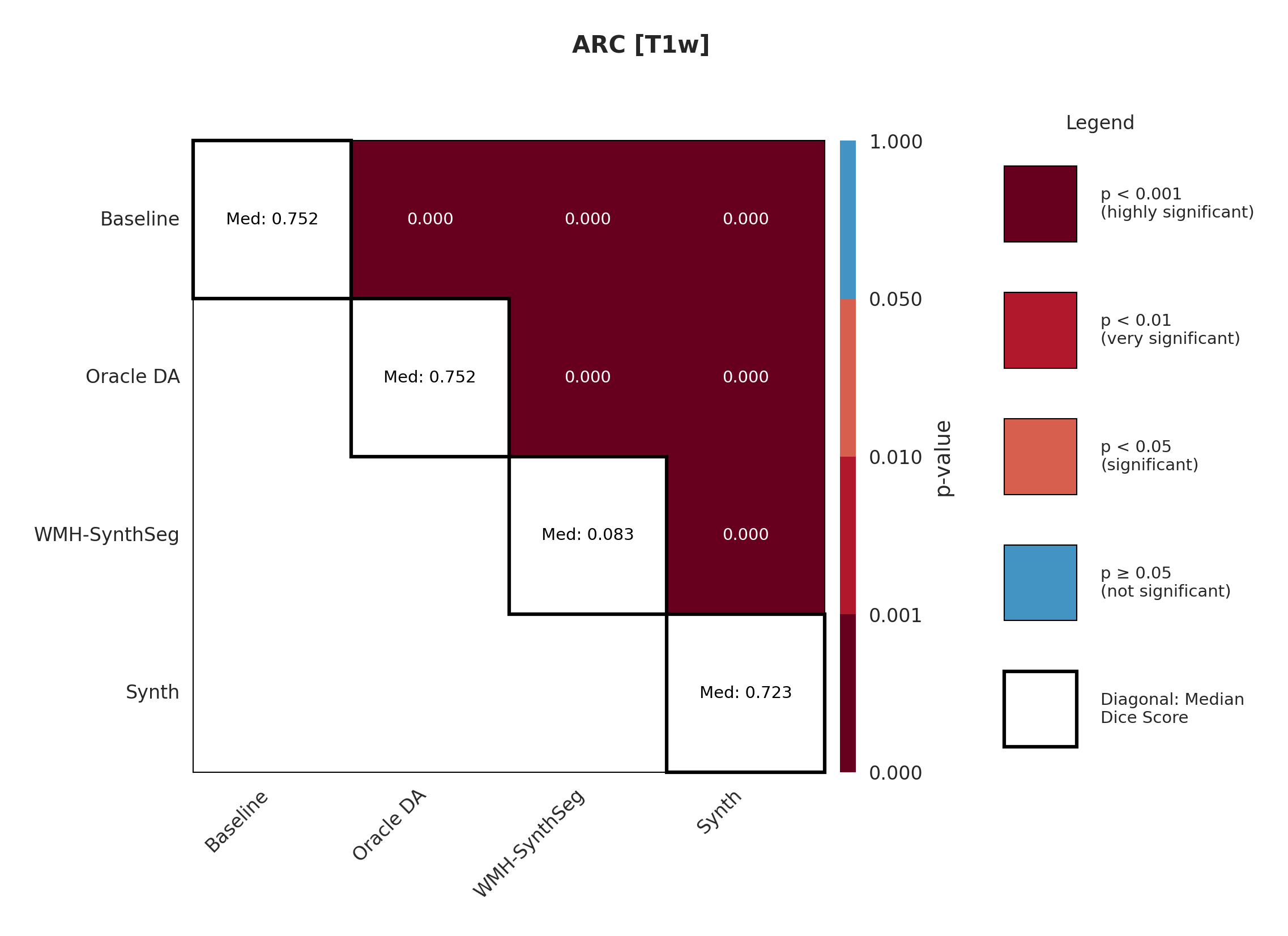}
            \caption{Wilcoxon signed-rank test values for Dice metric measurements in the ARC T1w dataset.}
            \label{fig:wilcoxon-arc-t1}
        \end{figure*}

        \begin{figure*}[h!]
            \centering
            \includegraphics[width=0.8\textwidth]{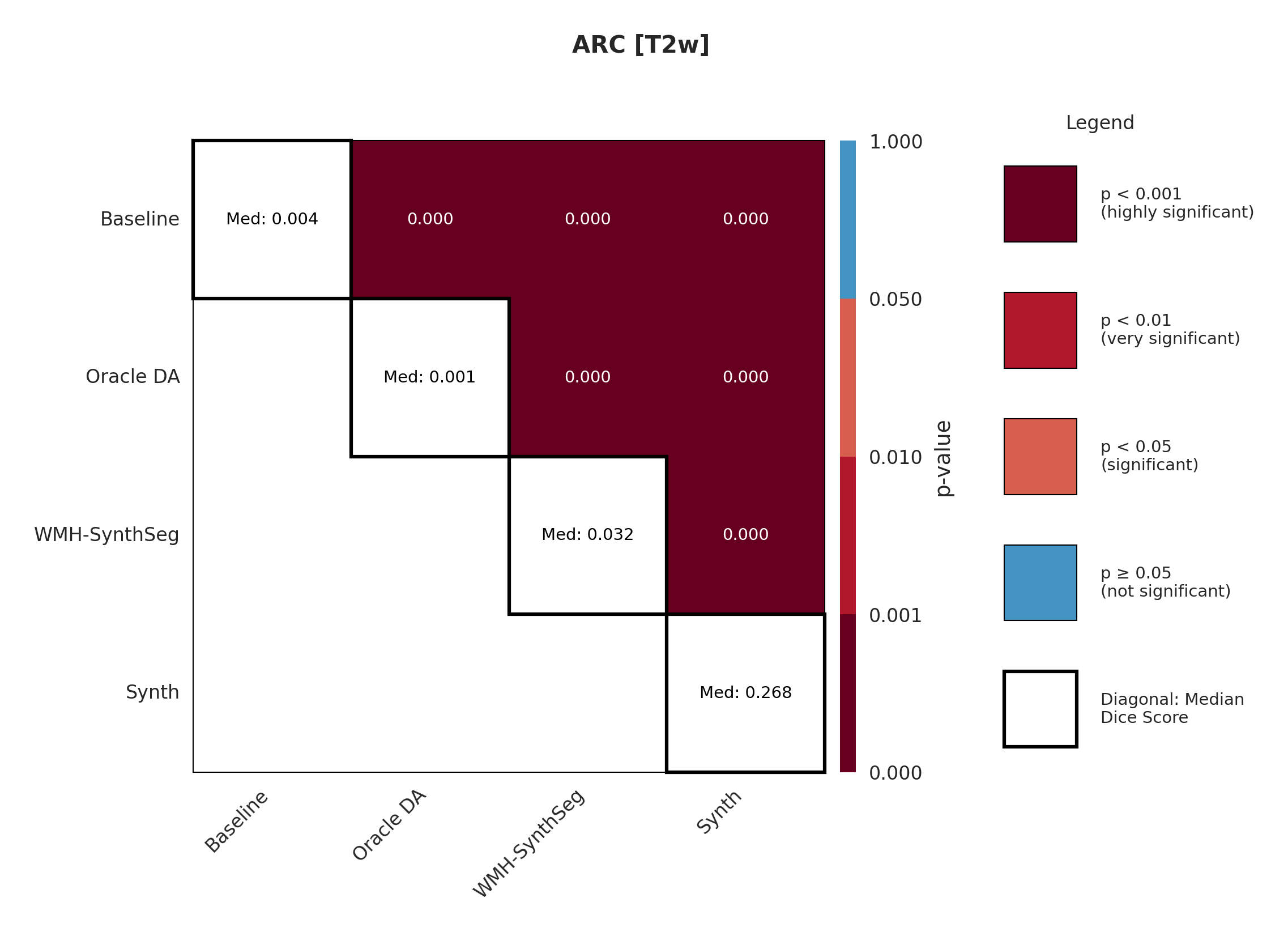}
            \caption{Wilcoxon signed-rank test values for Dice metric measurements in the ARC T2w dataset.}
            \label{fig:wilcoxon-arc-t2}
        \end{figure*}

        \begin{figure*}[h!]
            \centering
            \includegraphics[width=0.8\textwidth]{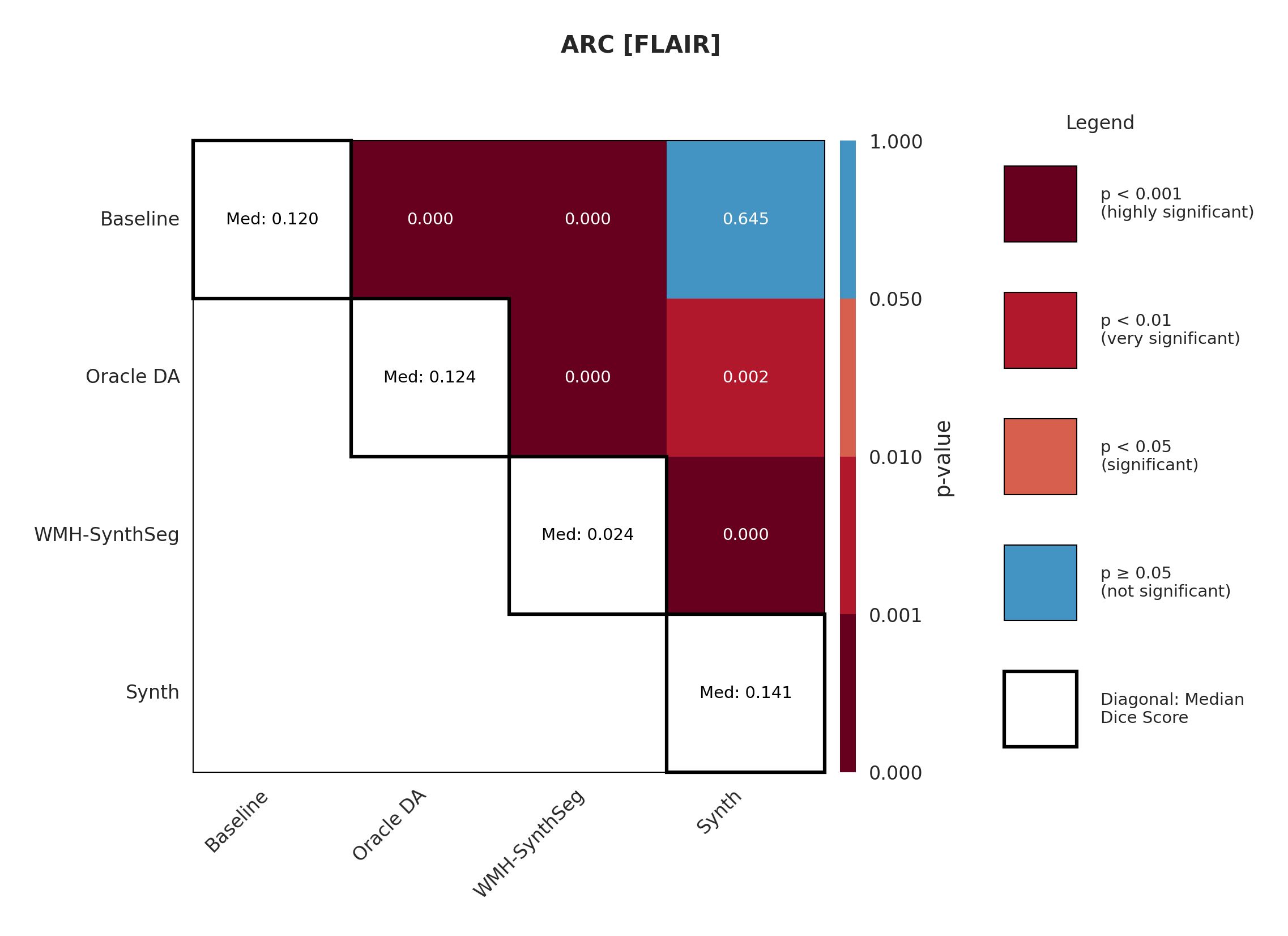}
            \caption{Wilcoxon signed-rank test values for Dice metric measurements in the ARC FLAIR dataset.}
            \label{fig:wilcoxon-arc-flair}
        \end{figure*}

        \begin{figure*}[h!]
            \centering
            \includegraphics[width=0.8\textwidth]{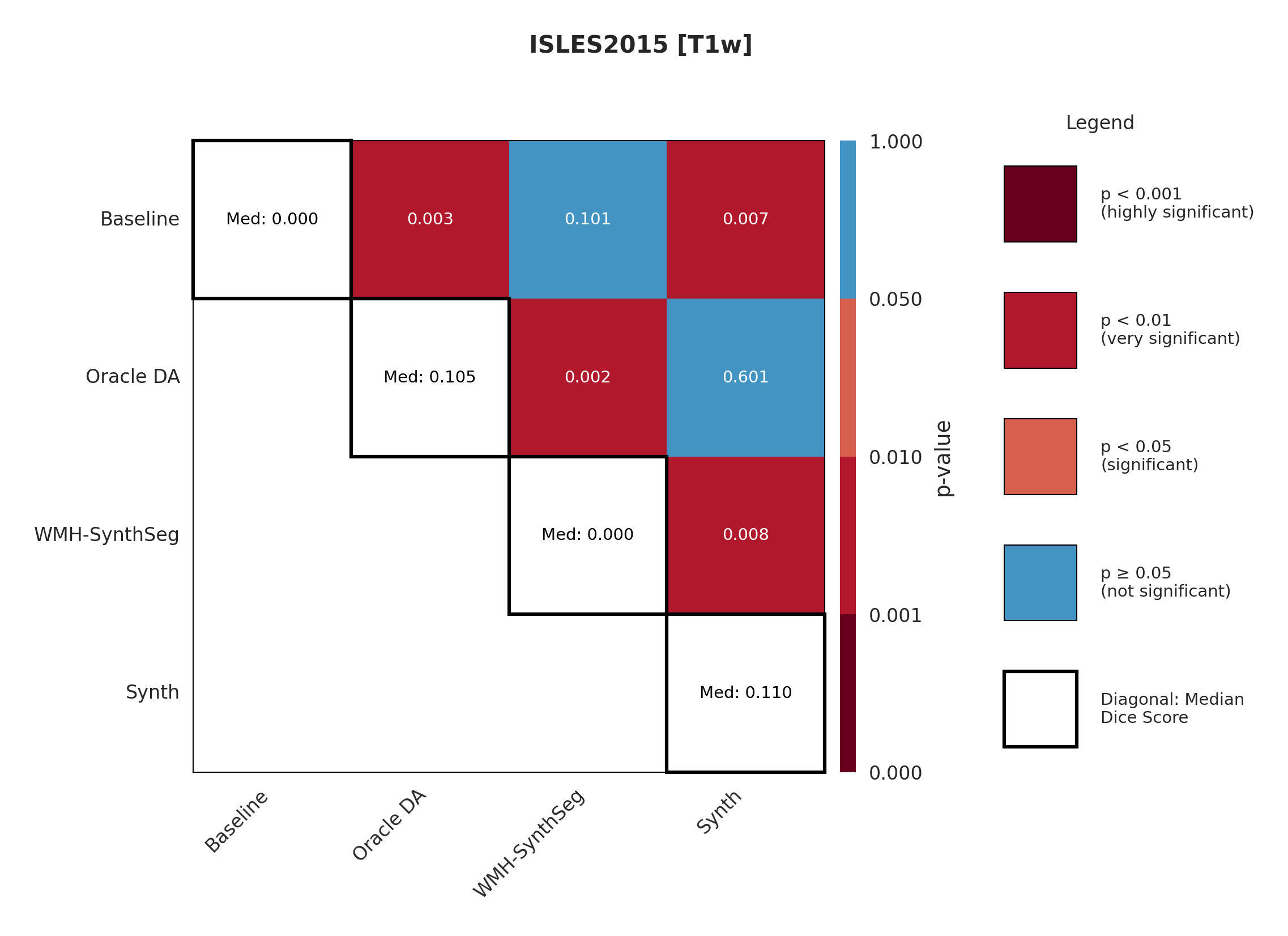}
            \caption{Wilcoxon signed-rank test values for Dice metric measurements in the ISLES 2015 T1w dataset.}
            \label{fig:wilcoxon-isles-t1}
        \end{figure*}

        \begin{figure*}[h!]
            \centering
            \includegraphics[width=0.8\textwidth]{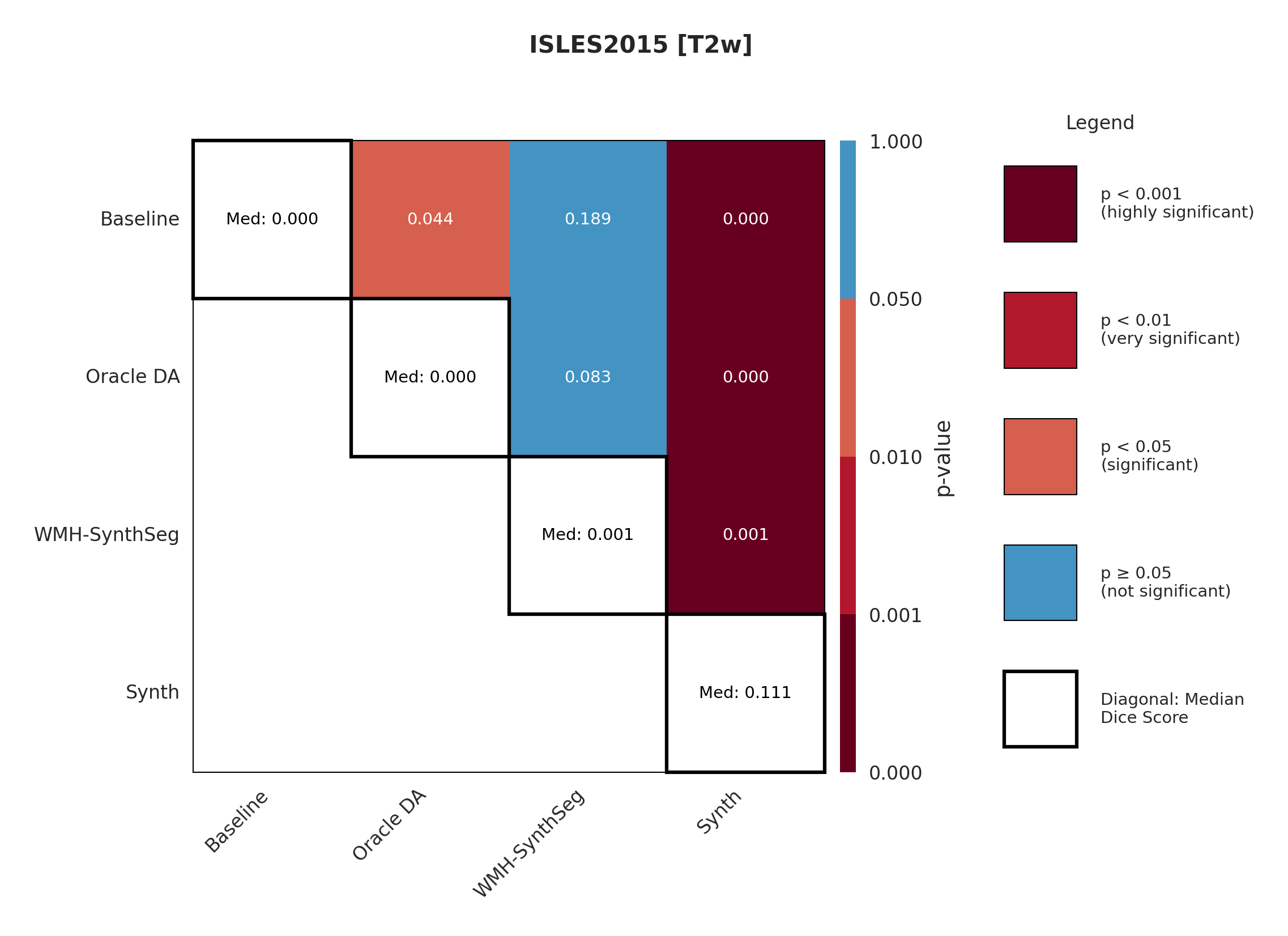}
            \caption{Wilcoxon signed-rank test values for Dice metric measurements in the ISLES 2015 T2w dataset.}
            \label{fig:wilcoxon-isles-t2}
        \end{figure*}

        \begin{figure*}[h!]
            \centering
            \includegraphics[width=0.8\textwidth]{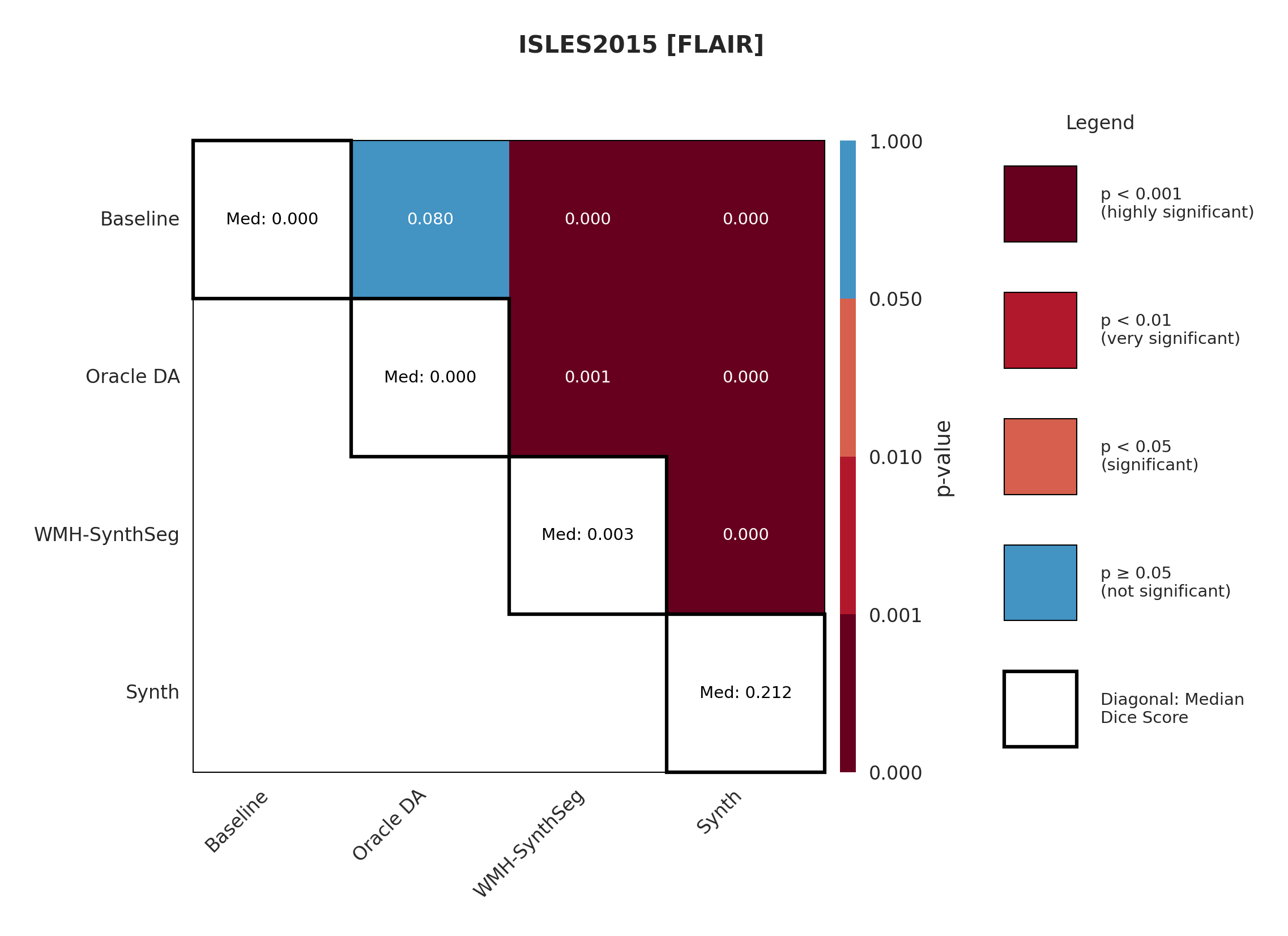}
            \caption{Wilcoxon signed-rank test values for Dice metric measurements in the ISLES 2015 FLAIR dataset.}
            \label{fig:wilcoxon-isles-flair}
        \end{figure*}

        \begin{figure*}[h!]
            \centering
            \includegraphics[width=0.8\textwidth]{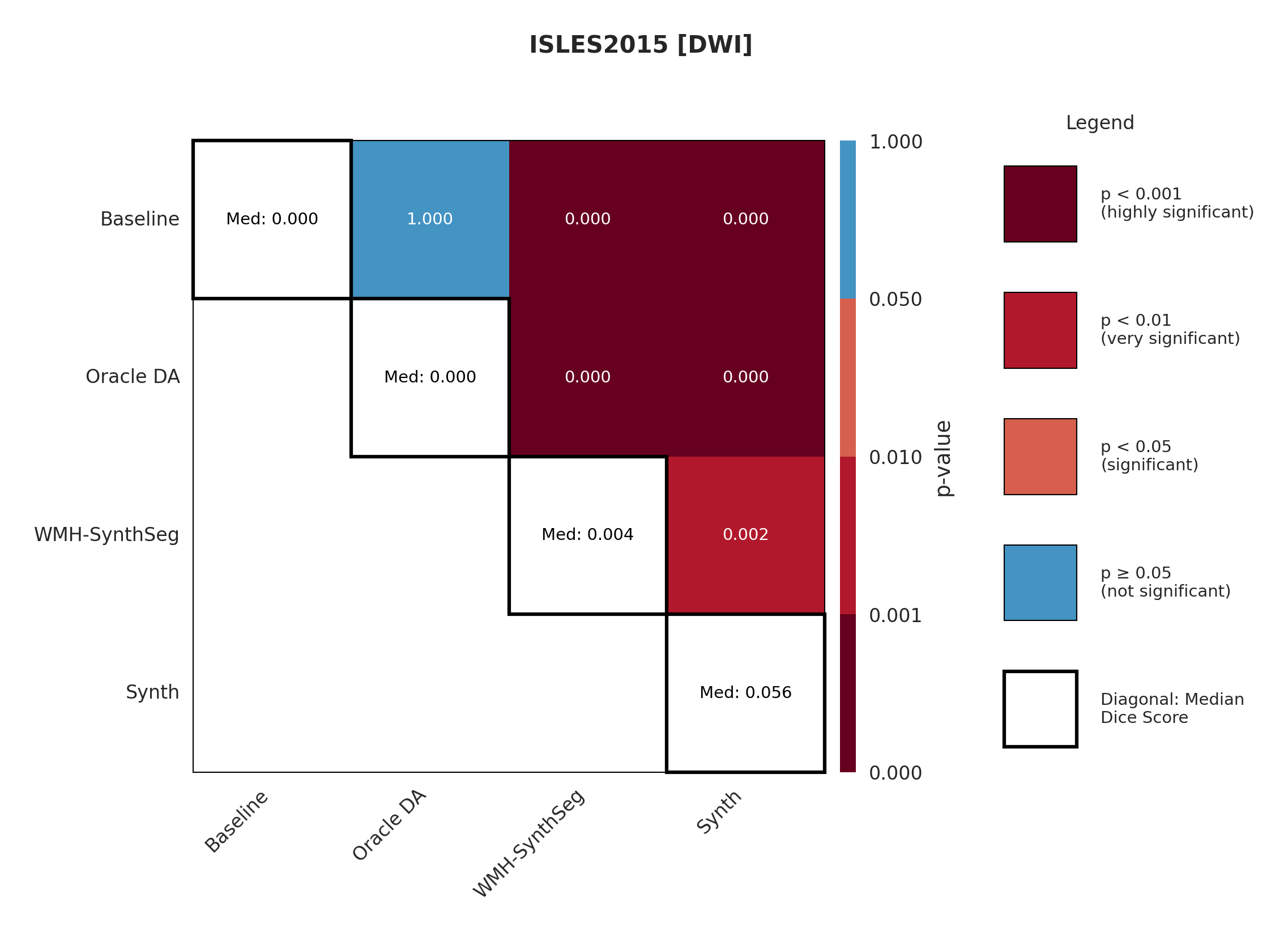}
            \caption{Wilcoxon signed-rank test values for Dice metric measurements in the ISLES 2015 DWI dataset.}
            \label{fig:wilcoxon-isles-dwi}
        \end{figure*}

        \begin{figure*}[h!]
            \centering
            \includegraphics[width=0.8\textwidth]{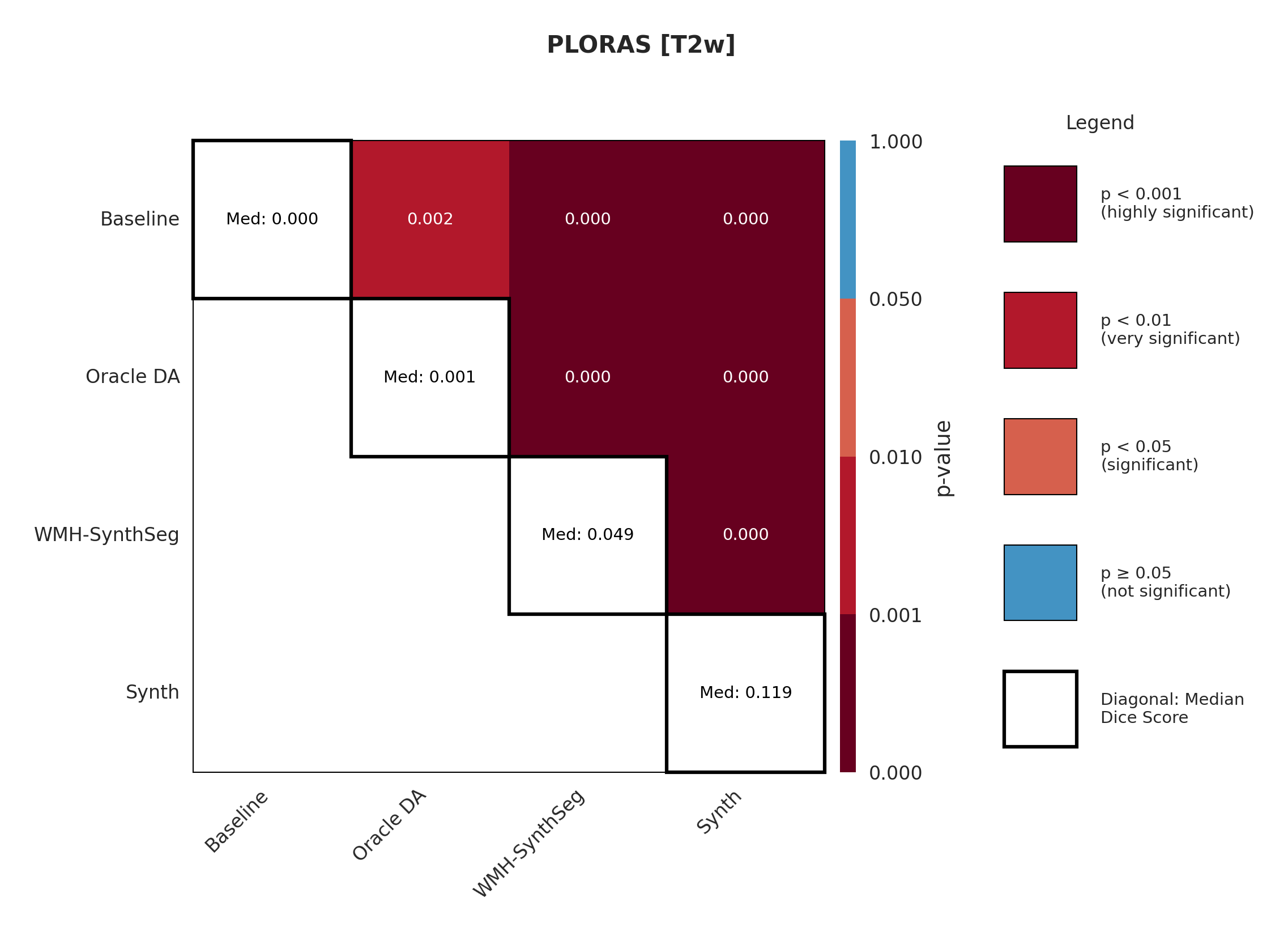}
            \caption{Wilcoxon signed-rank test values for Dice metric measurements in the PLORAS T2w dataset.}
            \label{fig:wilcoxon-ploras-t2}
        \end{figure*}

        \begin{figure*}[h!]
            \centering
            \includegraphics[width=0.8\textwidth]{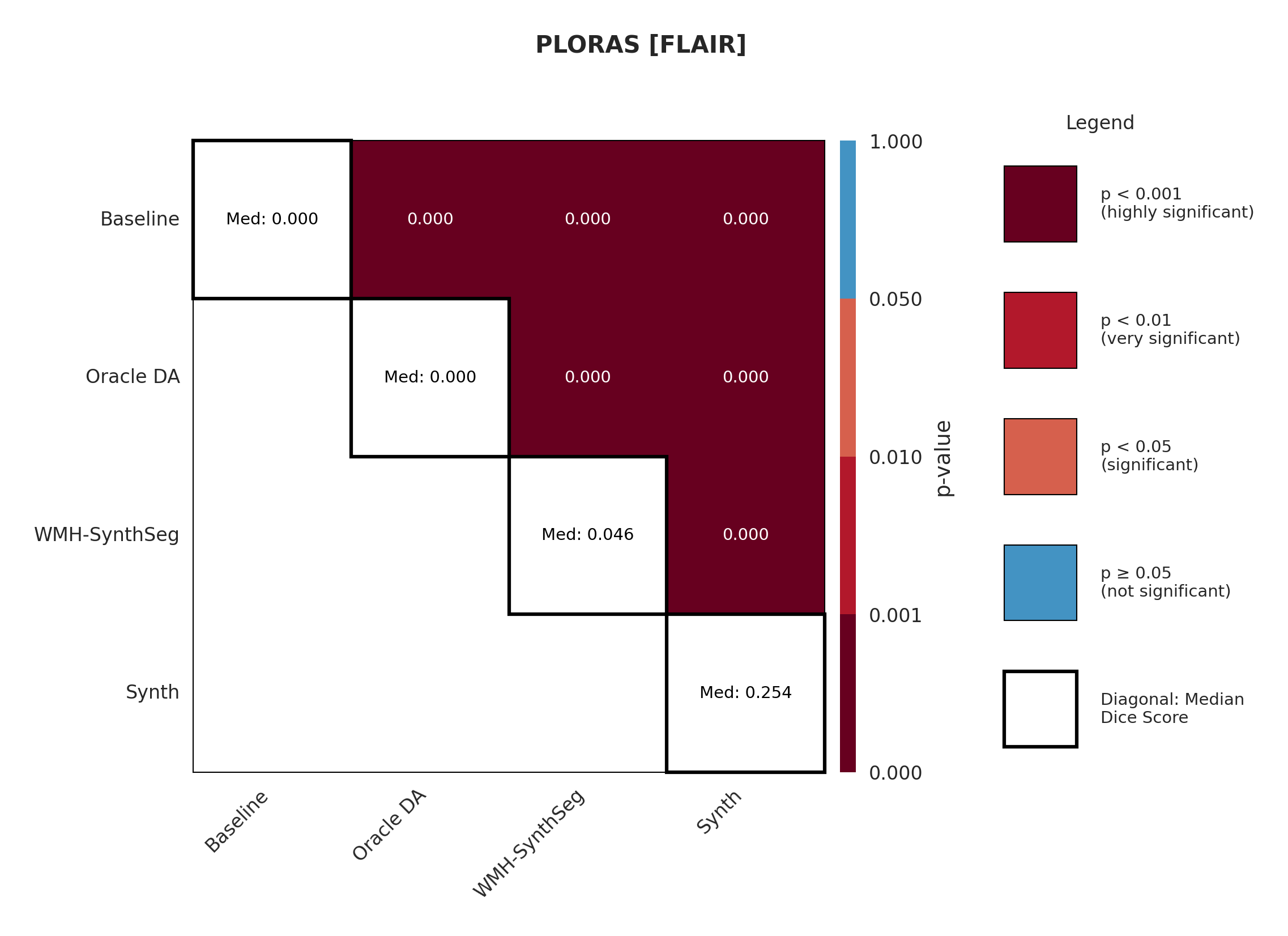}
            \caption{Wilcoxon signed-rank test values for Dice metric measurements in the PLORAS FLAIR dataset.}
            \label{fig:wilcoxon-ploras-flair}
        \end{figure*}

        \begin{figure*}[h!]
            \centering
            \includegraphics[width=0.8\textwidth]{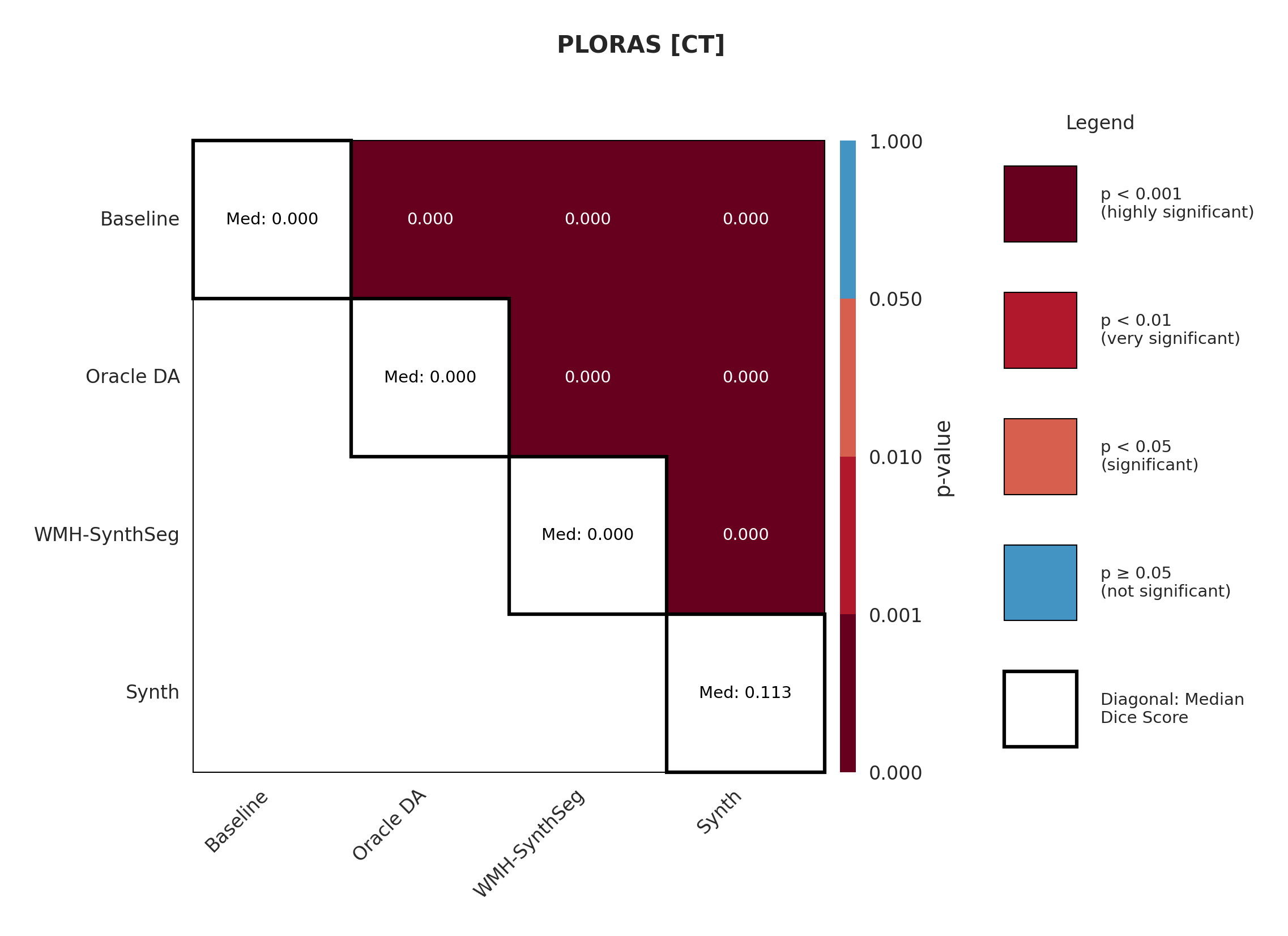}
            \caption{Wilcoxon signed-rank test values for Dice metric measurements in the PLORAS CT dataset.}
            \label{fig:wilcoxon-ploras-ct}
        \end{figure*}

    \newpage
    \section{Additional Qualitative Results}

        Results are shown below in Figures \ref{fig:ploras-good-t2} - \ref{fig:ploras-bad-CT} for failure and success cases for the three different sequences in the PLORAS dataset.

        \begin{figure*}[h!]
            \centering
            \begin{subfigure}[t]{\textwidth}
                \centering
                \includegraphics[width=\textwidth]{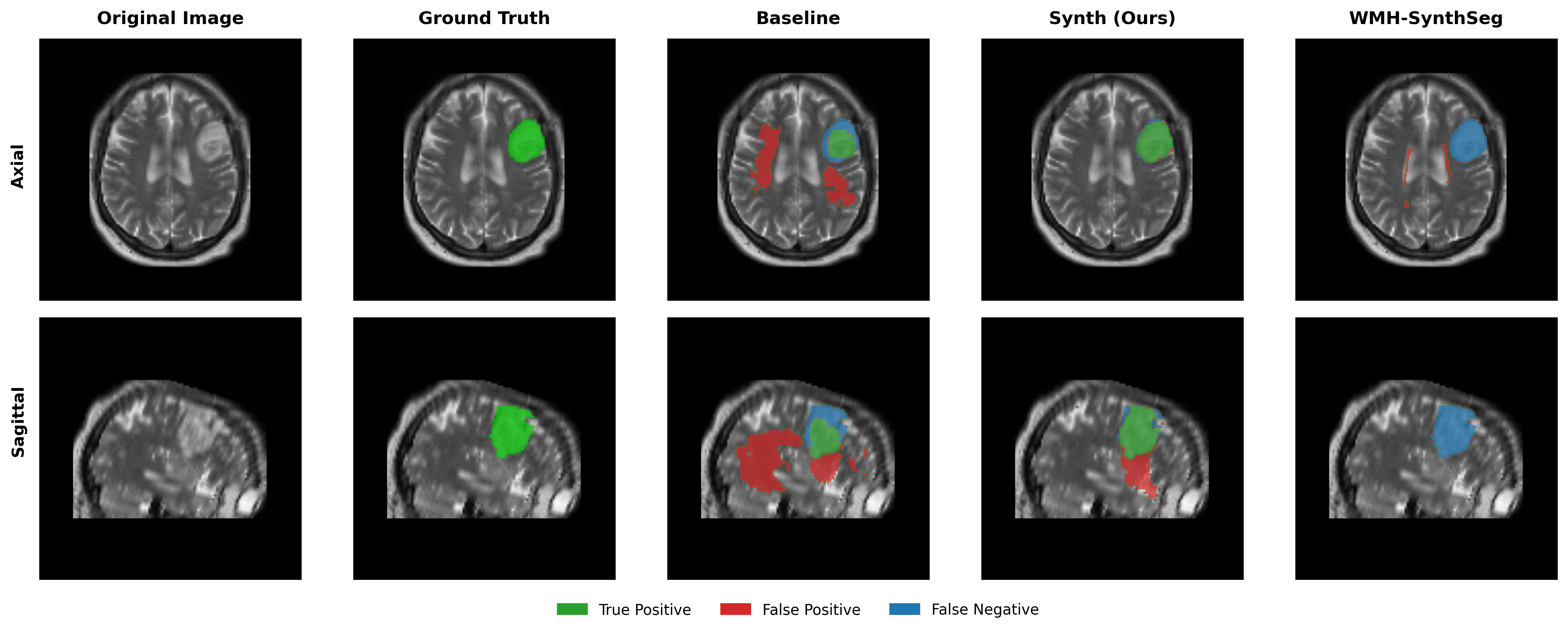}
            \end{subfigure}
            \\
            \begin{subfigure}[t]{\textwidth}
                \centering
                \includegraphics[width=\textwidth]{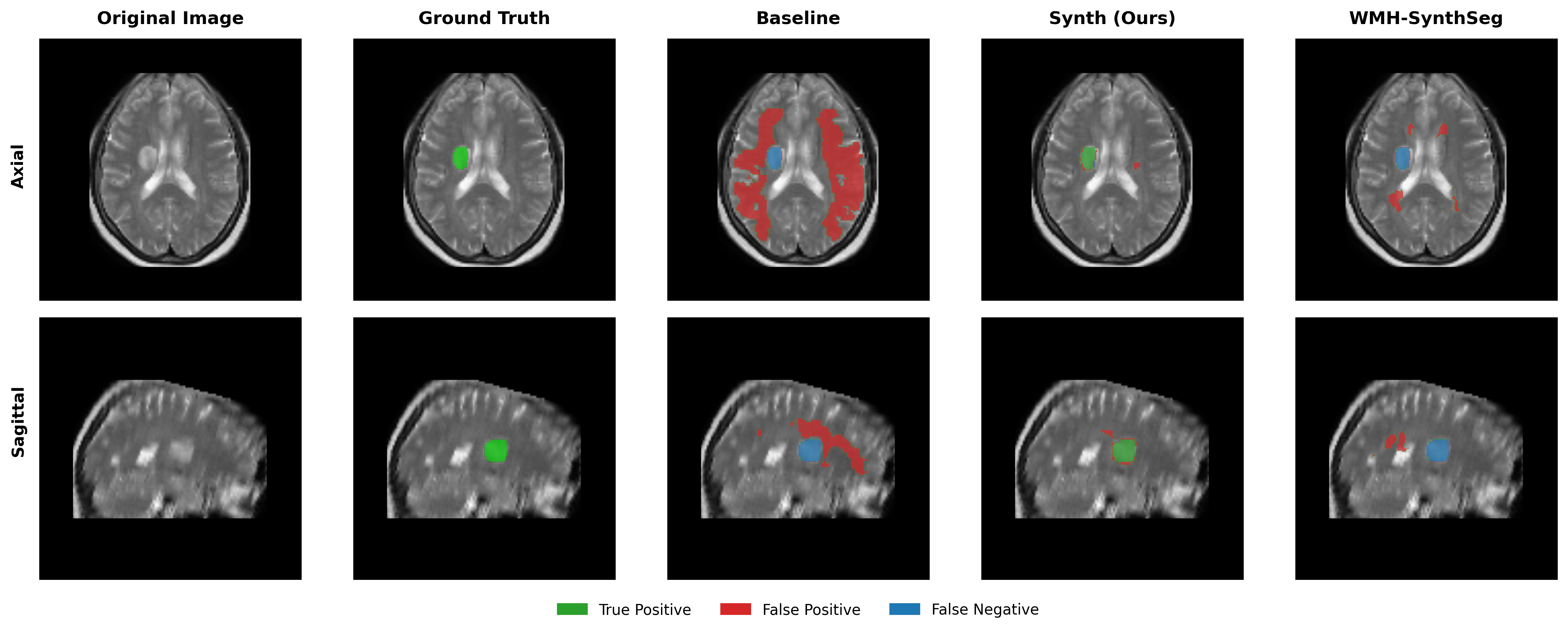}
            \end{subfigure}
            \\
            \begin{subfigure}[t]{\textwidth}
                \centering
                \includegraphics[width=\textwidth]{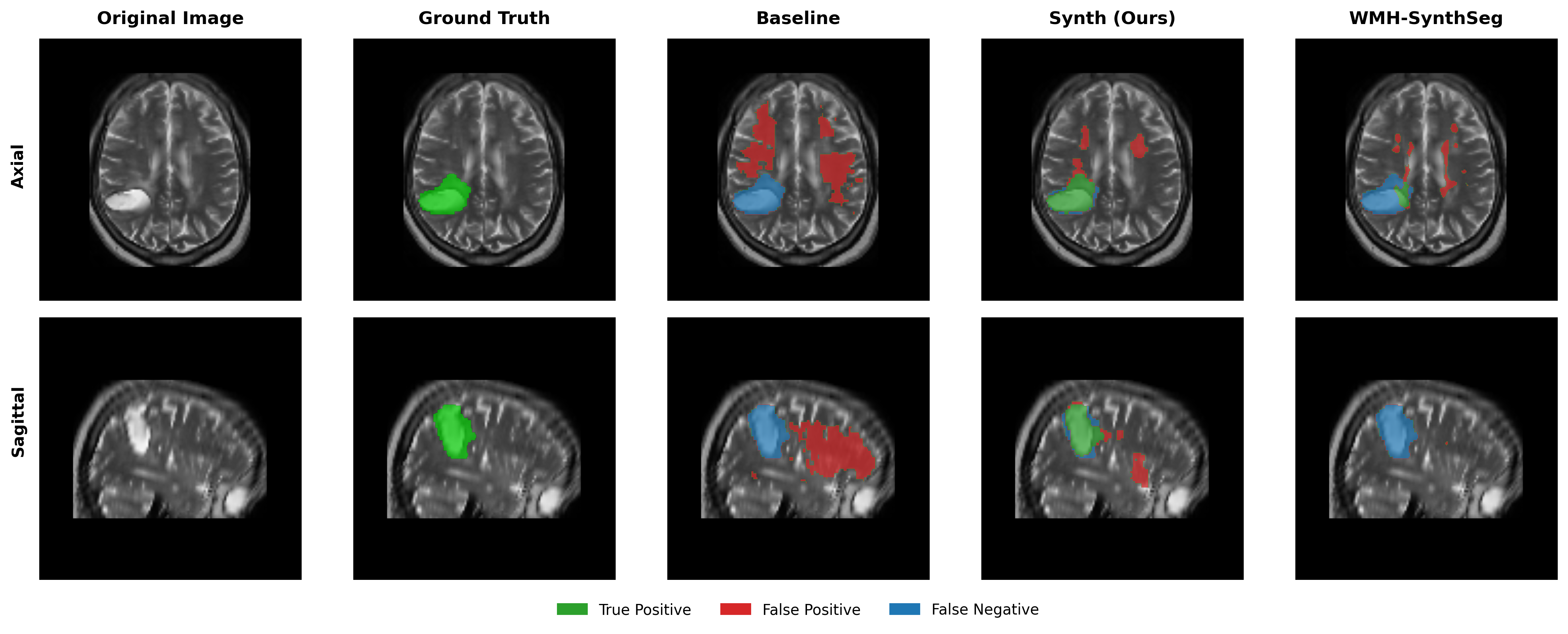}
            \end{subfigure}
            \caption{Sample visualisations of successful cases in the PLORAS T2w dataset. Green indicates a true positive prediction, red a false positive, and blue a false negative.}
            \label{fig:ploras-good-t2}
        \end{figure*}

        \begin{figure*}[h!]
            \centering
            \begin{subfigure}[t]{\textwidth}
                \centering
                \includegraphics[width=\textwidth]{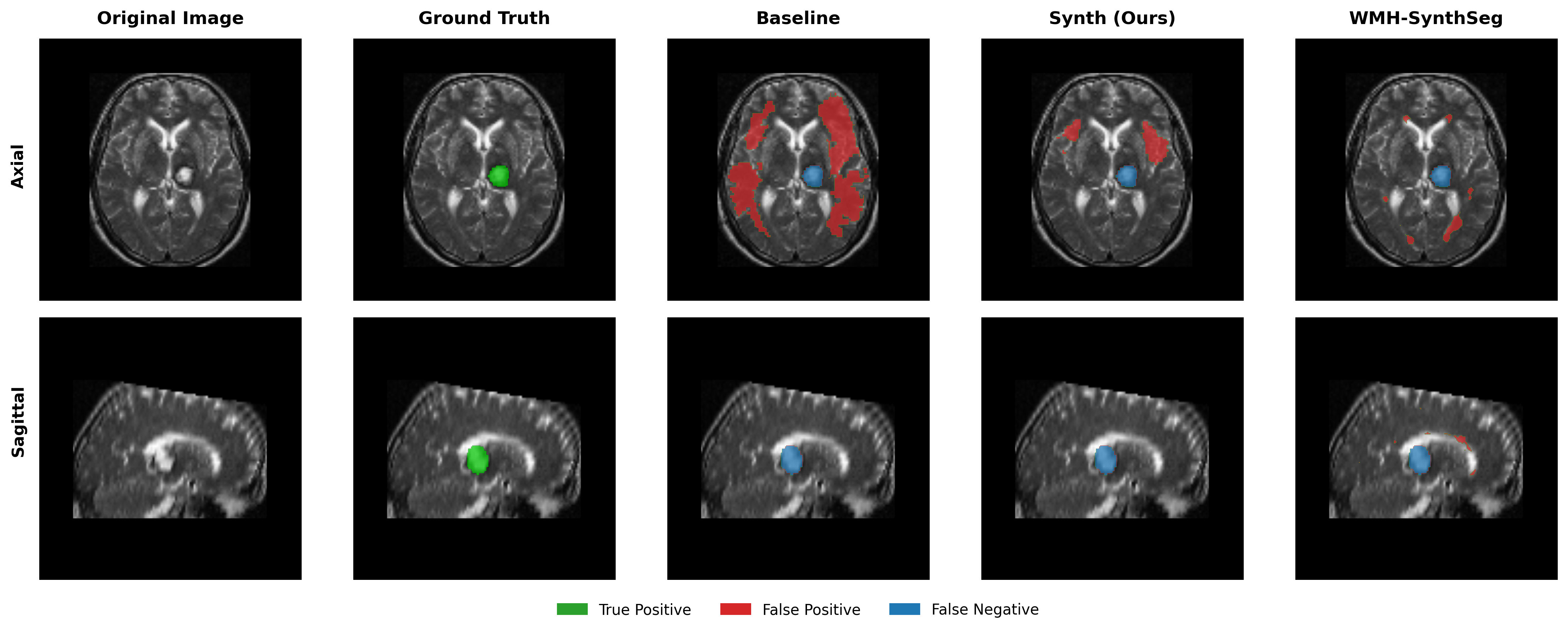}
            \end{subfigure}
            \\
            \begin{subfigure}[t]{\textwidth}
                \centering
                \includegraphics[width=\textwidth]{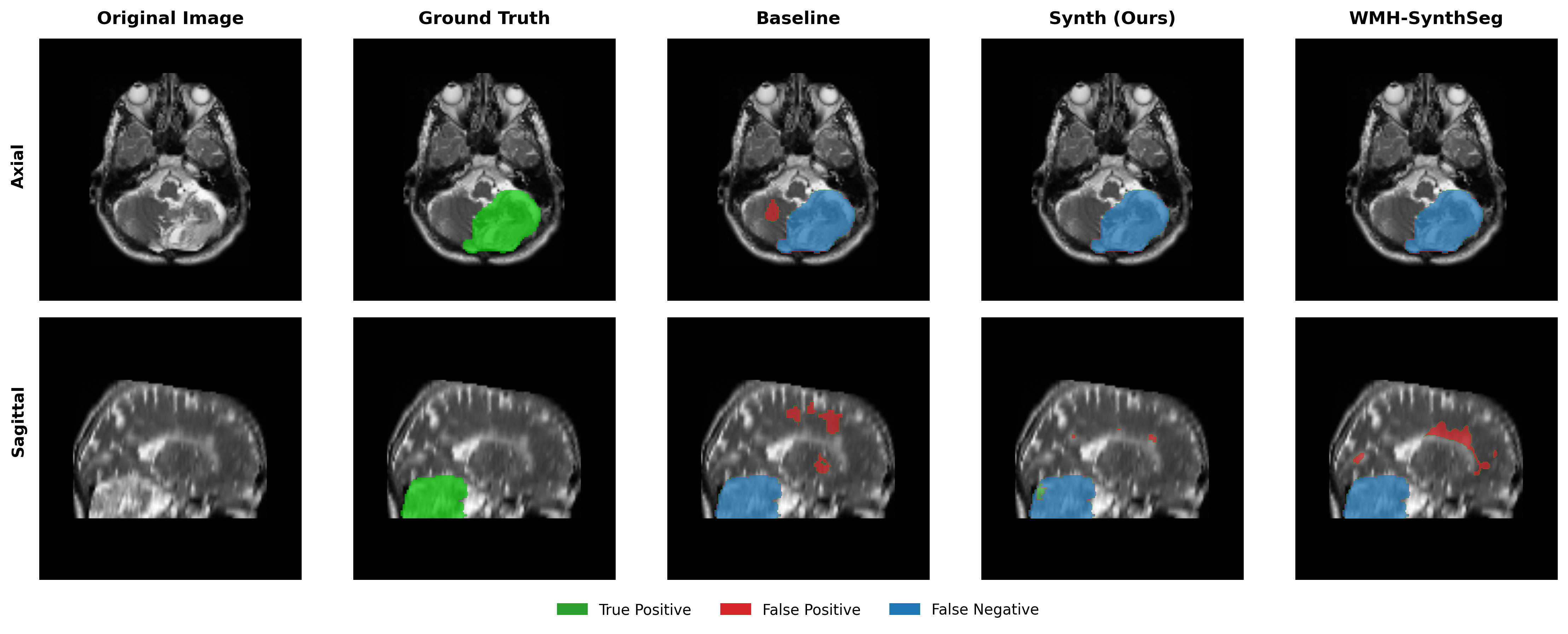}
            \end{subfigure}
            \\
            \begin{subfigure}[t]{\textwidth}
                \centering
                \includegraphics[width=\textwidth]{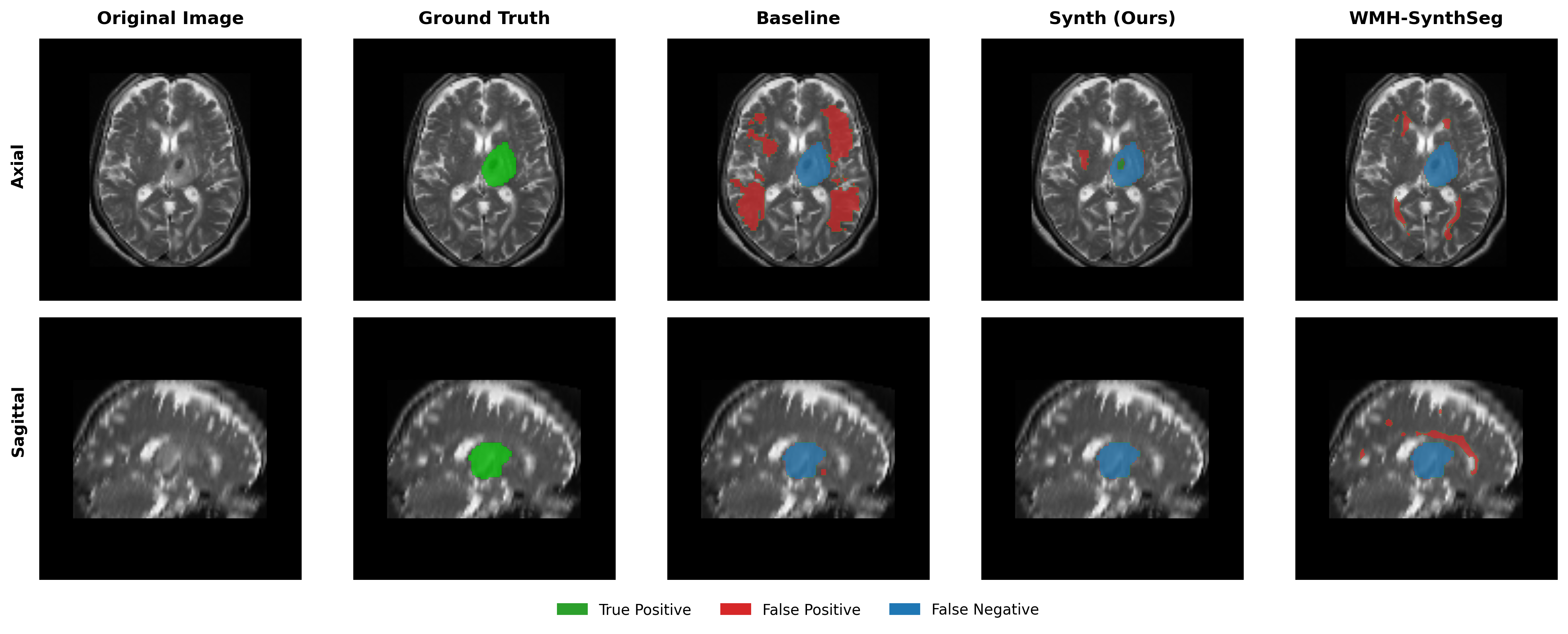}
            \end{subfigure}
            \caption{Sample visualisations of failure cases in the PLORAS T2w dataset. Green indicates a true positive prediction, red a false positive, and blue a false negative.}
            \label{fig:ploras-bad-t2}
        \end{figure*}

        \begin{figure*}[h!]
            \centering
            \begin{subfigure}[t]{\textwidth}
                \centering
                \includegraphics[width=\textwidth]{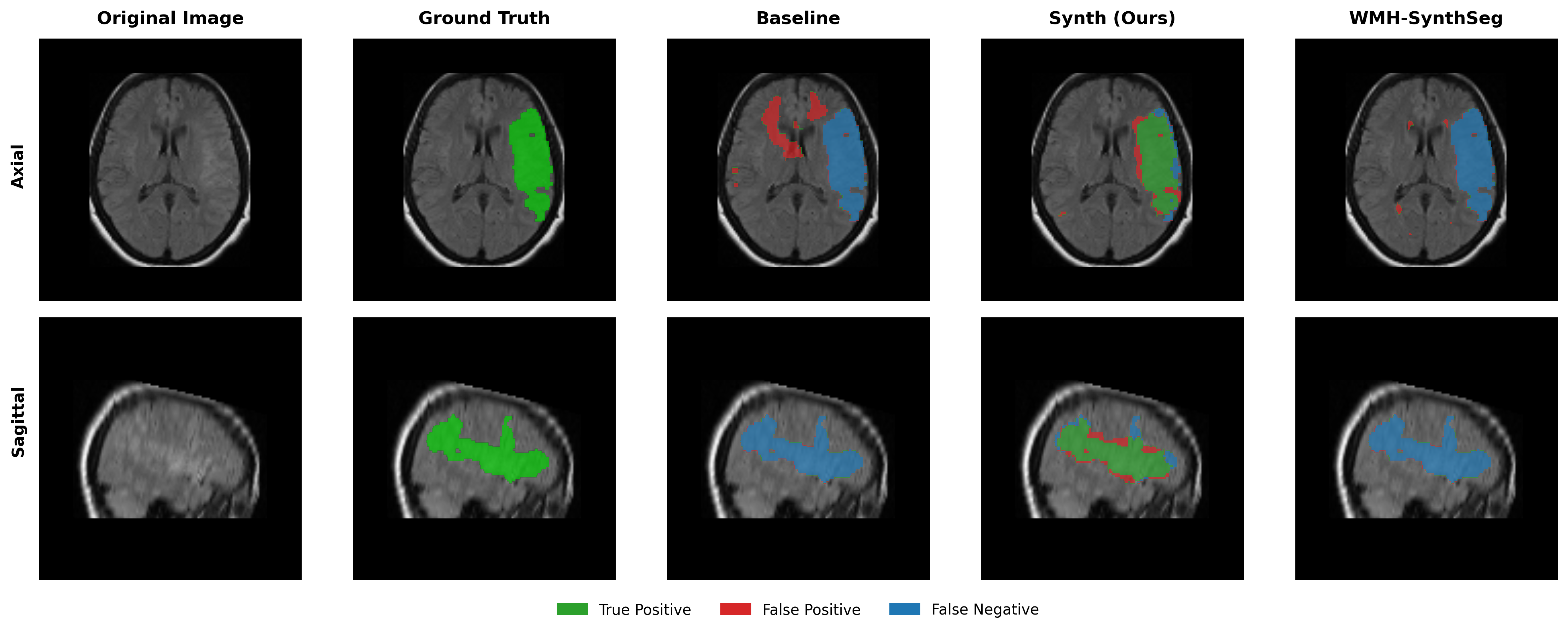}
            \end{subfigure}
            \\
            \begin{subfigure}[t]{\textwidth}
                \centering
                \includegraphics[width=\textwidth]{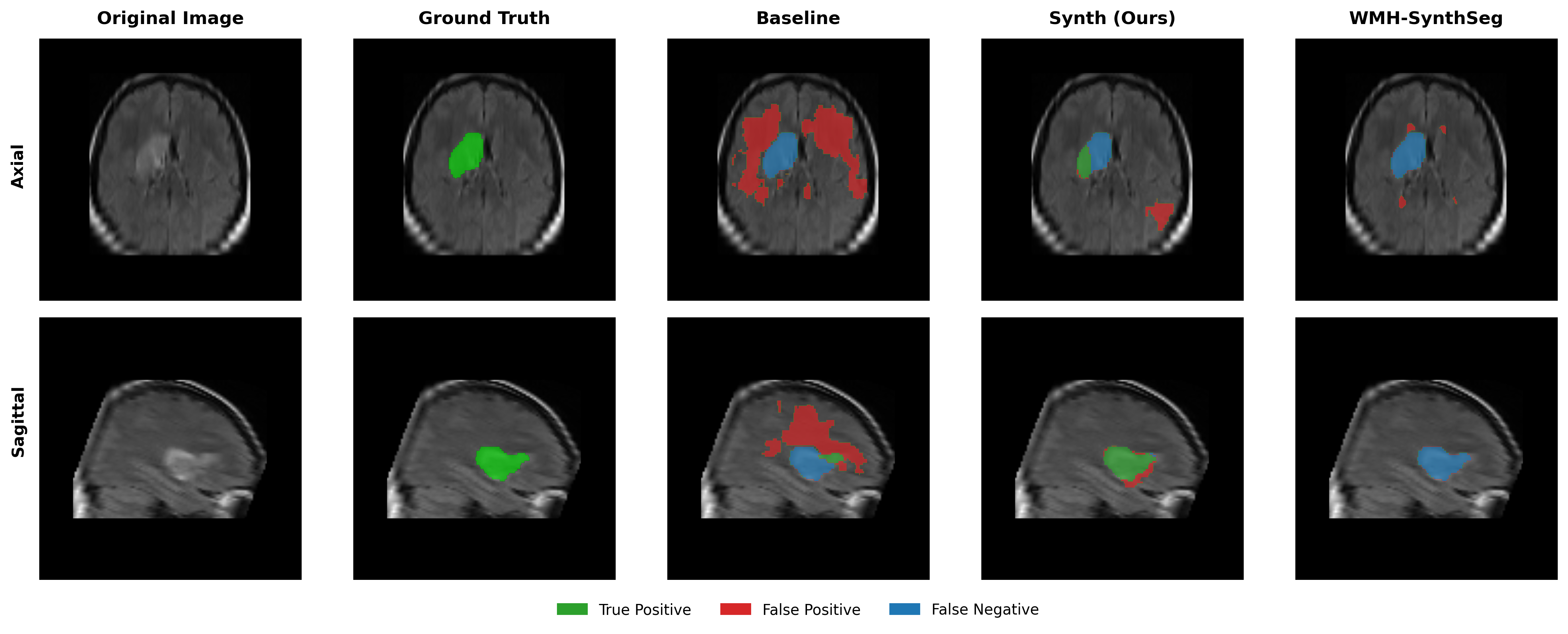}
            \end{subfigure}
            \\
            \begin{subfigure}[t]{\textwidth}
                \centering
                \includegraphics[width=\textwidth]{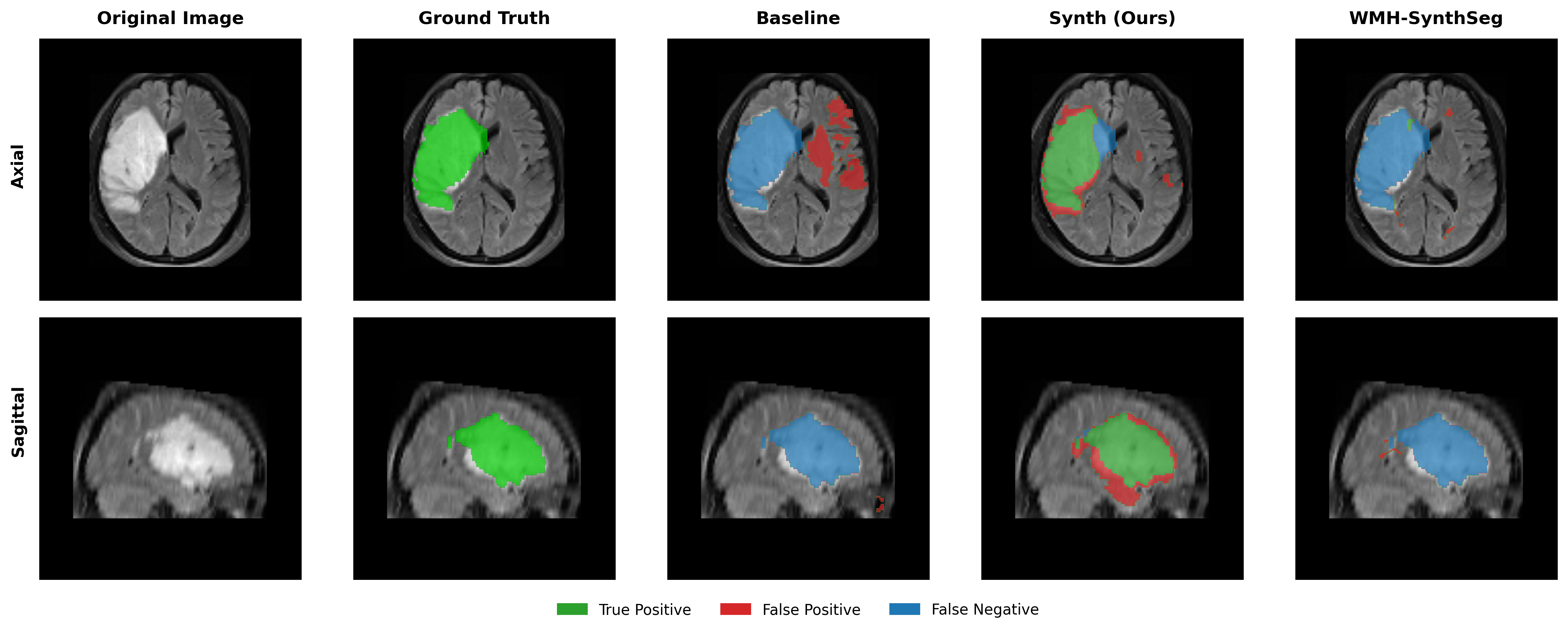}
            \end{subfigure}
            \caption{Sample visualisations of successful cases in the PLORAS FLAIR dataset. Green indicates a true positive prediction, red a false positive, and blue a false negative.}
            \label{fig:ploras-good-flair}
        \end{figure*}

        \begin{figure*}[h!]
            \centering
            \begin{subfigure}[t]{\textwidth}
                \centering
                \includegraphics[width=\textwidth]{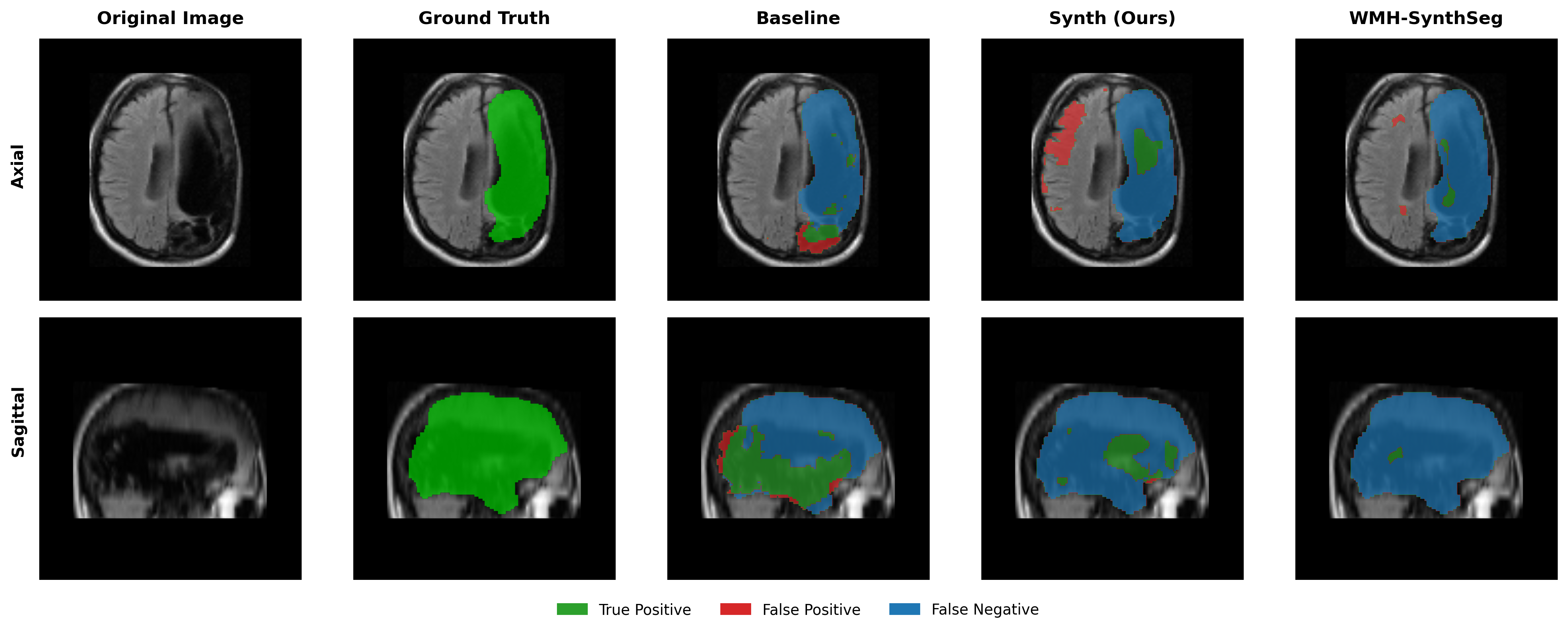}
            \end{subfigure}
            \\
            \begin{subfigure}[t]{\textwidth}
                \centering
                \includegraphics[width=\textwidth]{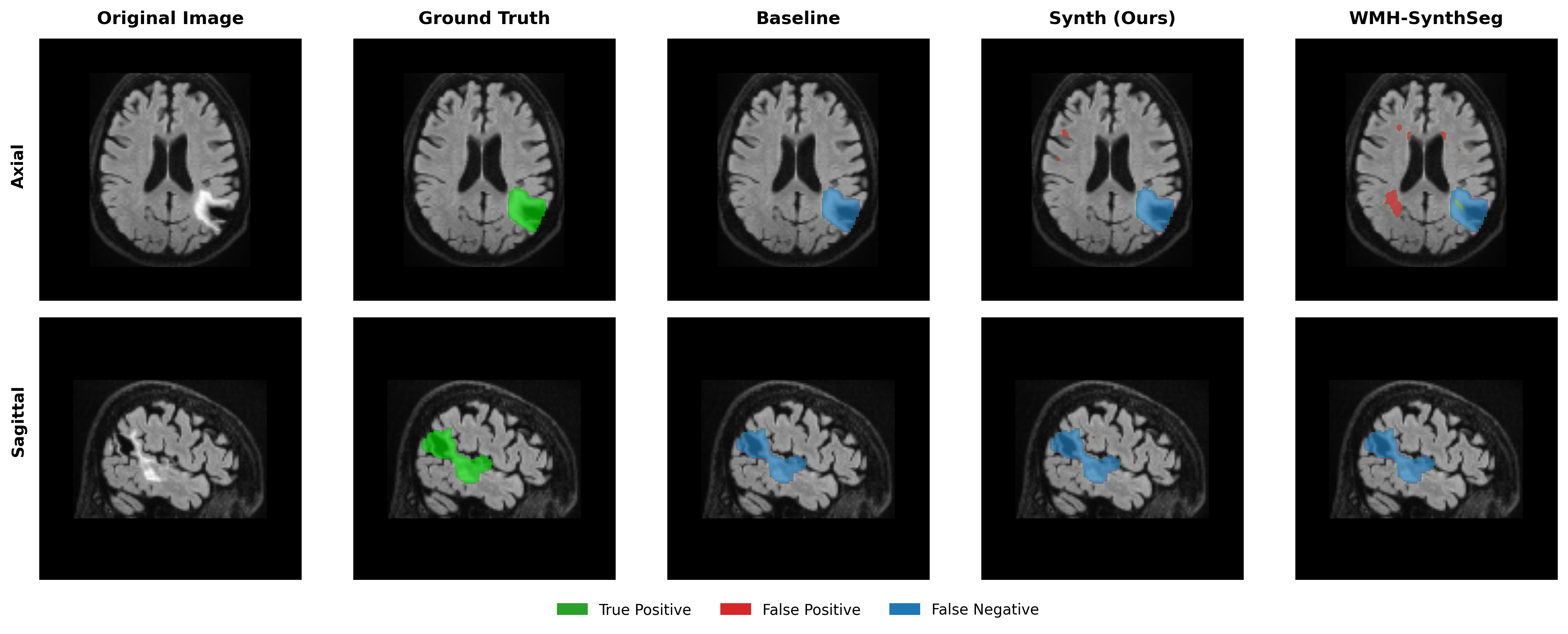}
            \end{subfigure}
            \\
            \begin{subfigure}[t]{\textwidth}
                \centering
                \includegraphics[width=\textwidth]{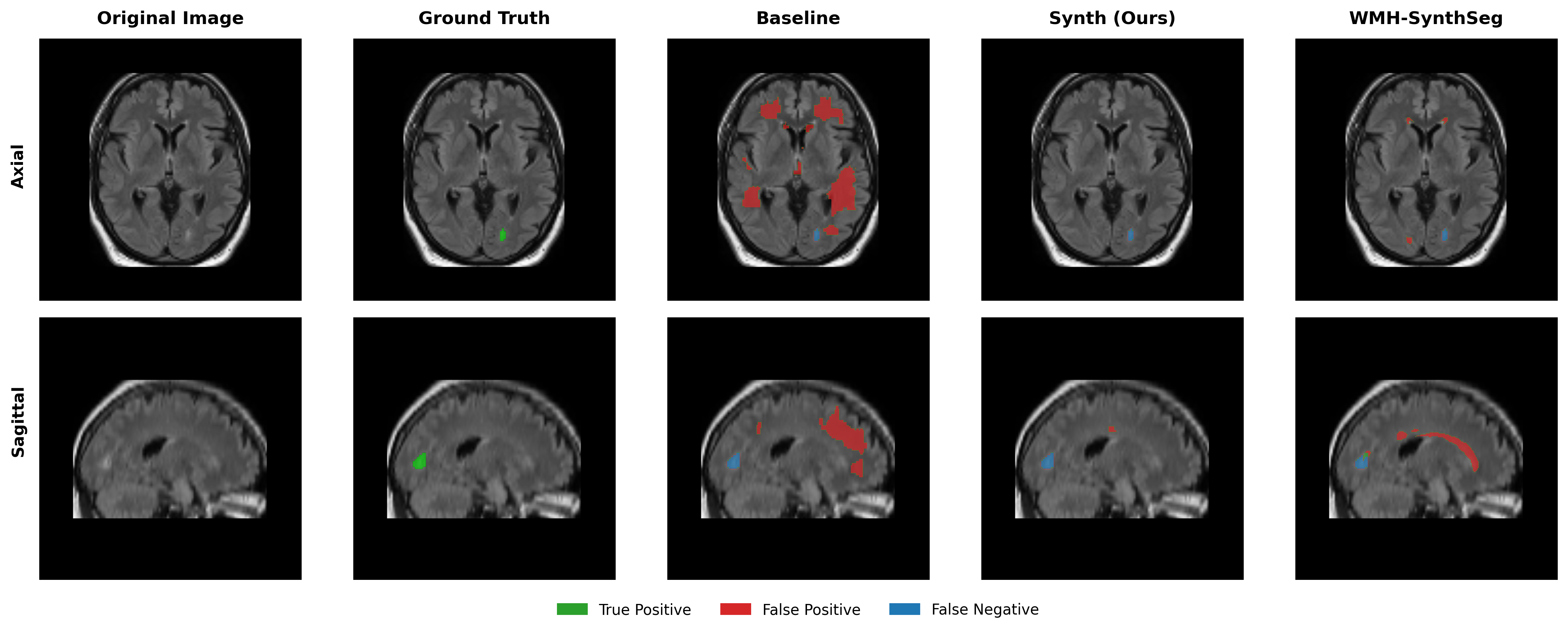}
            \end{subfigure}
            \caption{Sample visualisations of failure cases in the PLORAS FLAIR dataset. Green indicates a true positive prediction, red a false positive, and blue a false negative.}
            \label{fig:ploras-bad-flair}
        \end{figure*}

        \begin{figure*}[h!]
            \centering
            \begin{subfigure}[t]{\textwidth}
                \centering
                \includegraphics[width=\textwidth]{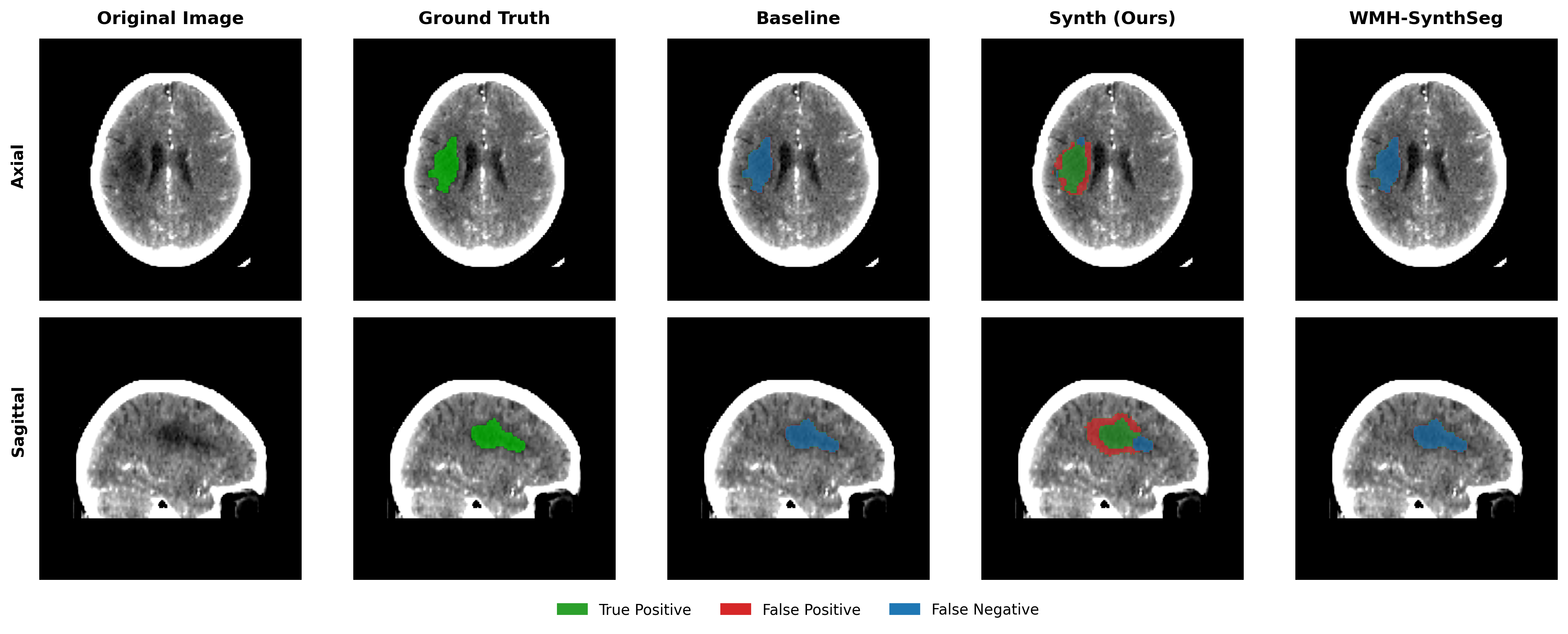}
            \end{subfigure}
            \\
            \begin{subfigure}[t]{\textwidth}
                \centering
                \includegraphics[width=\textwidth]{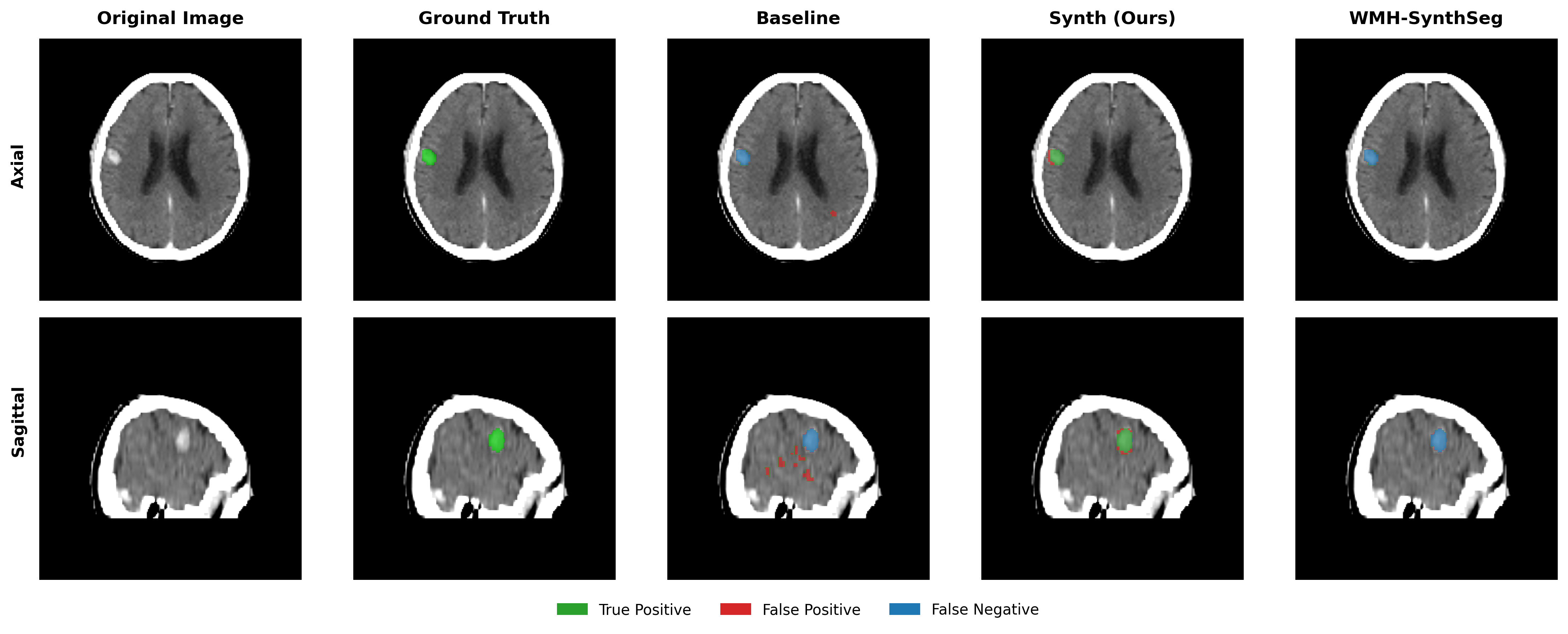}
            \end{subfigure}
            \\
            \begin{subfigure}[t]{\textwidth}
                \centering
                \includegraphics[width=\textwidth]{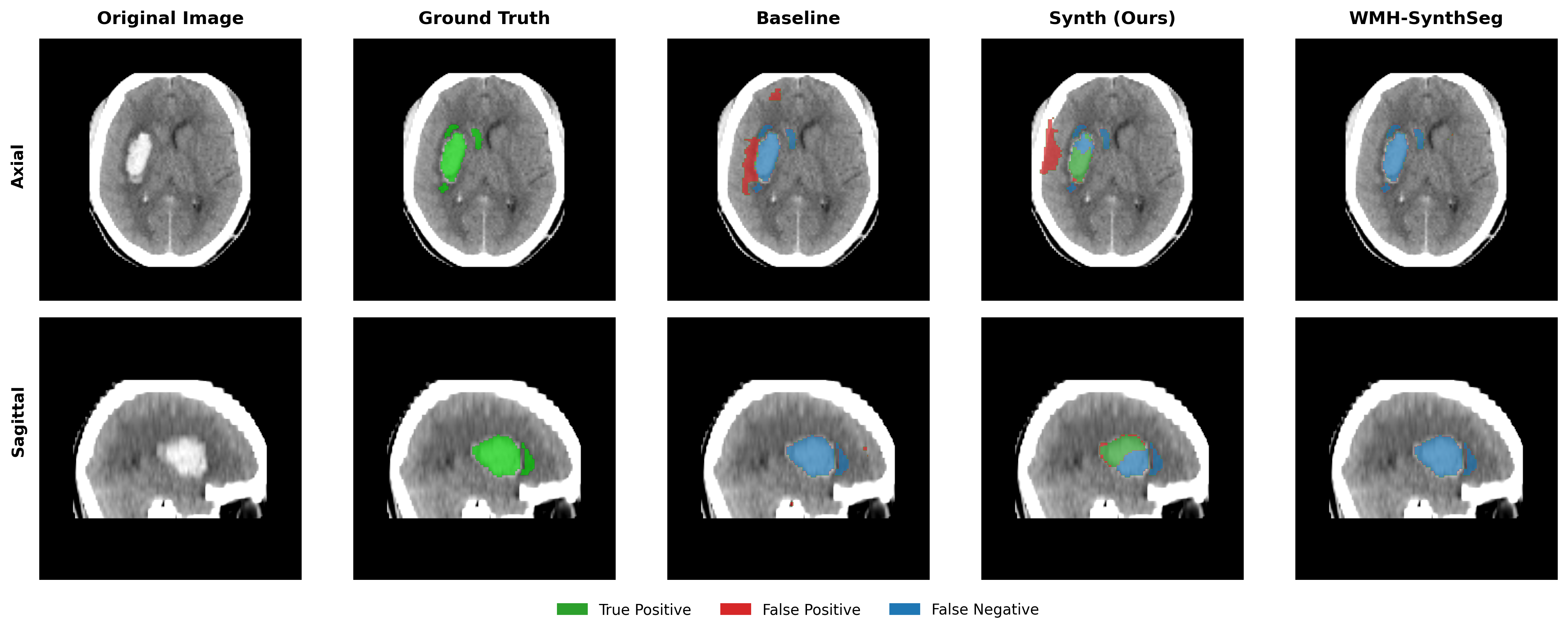}
            \end{subfigure}
            \caption{Sample visualisations of successful cases in the PLORAS CT dataset. Green indicates a true positive prediction, red a false positive, and blue a false negative.}
            \label{fig:ploras-good-ct}
        \end{figure*}

        \begin{figure*}[h!]
            \centering
            \begin{subfigure}[t]{\textwidth}
                \centering
                \includegraphics[width=\textwidth]{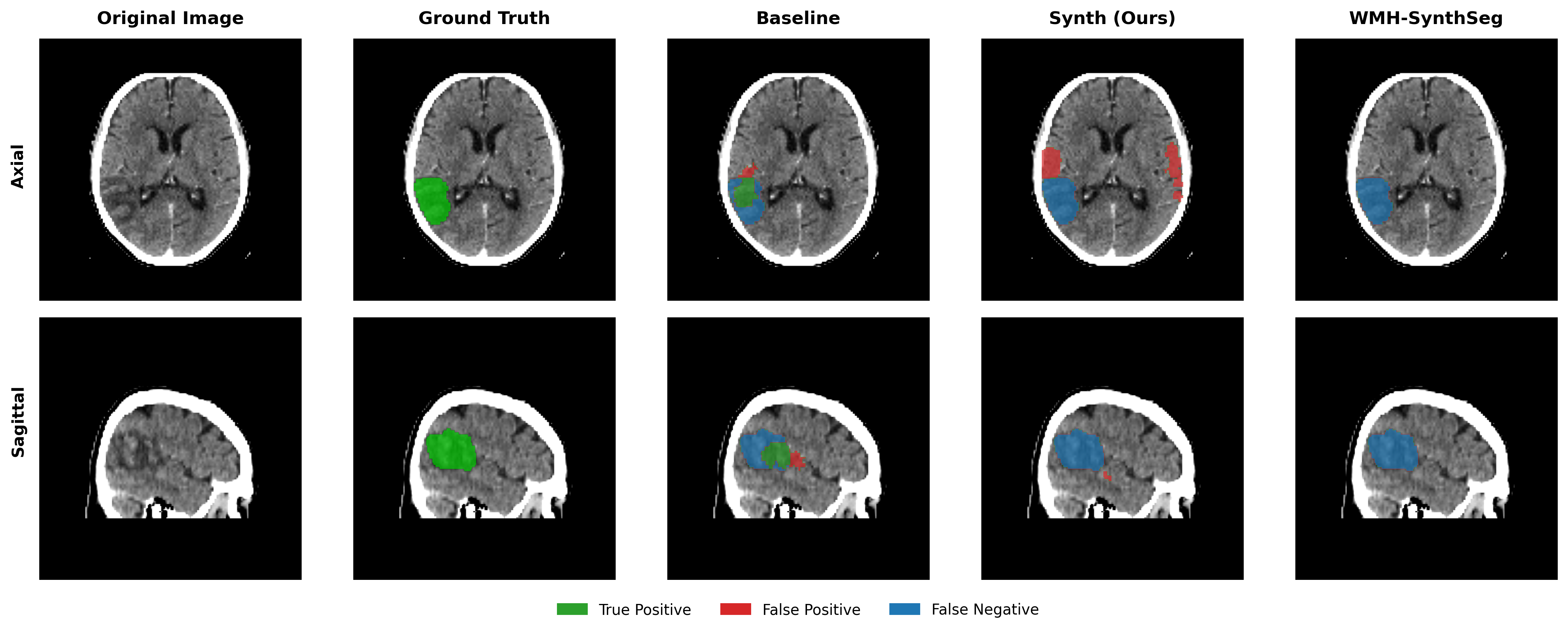}
            \end{subfigure}
            \\
            \begin{subfigure}[t]{\textwidth}
                \centering
                \includegraphics[width=\textwidth]{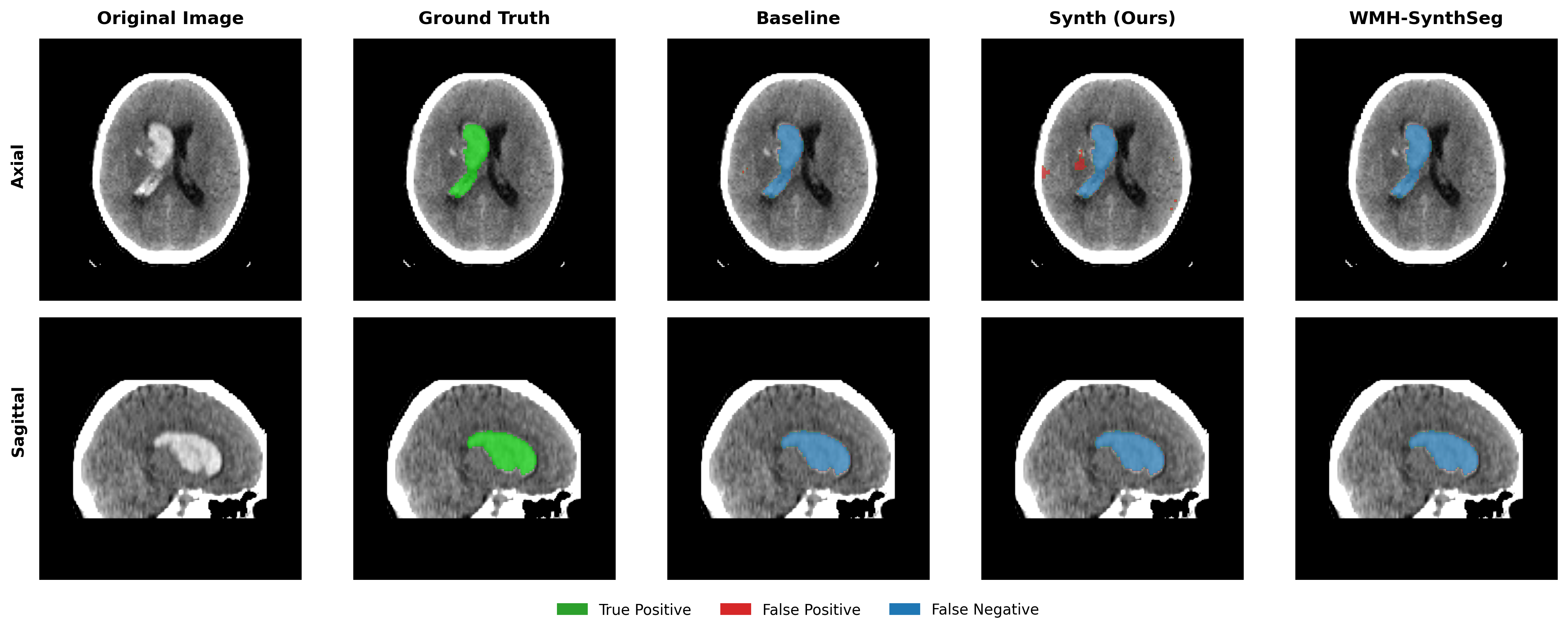}
            \end{subfigure}
            \\
            \begin{subfigure}[t]{\textwidth}
                \centering
                \includegraphics[width=\textwidth]{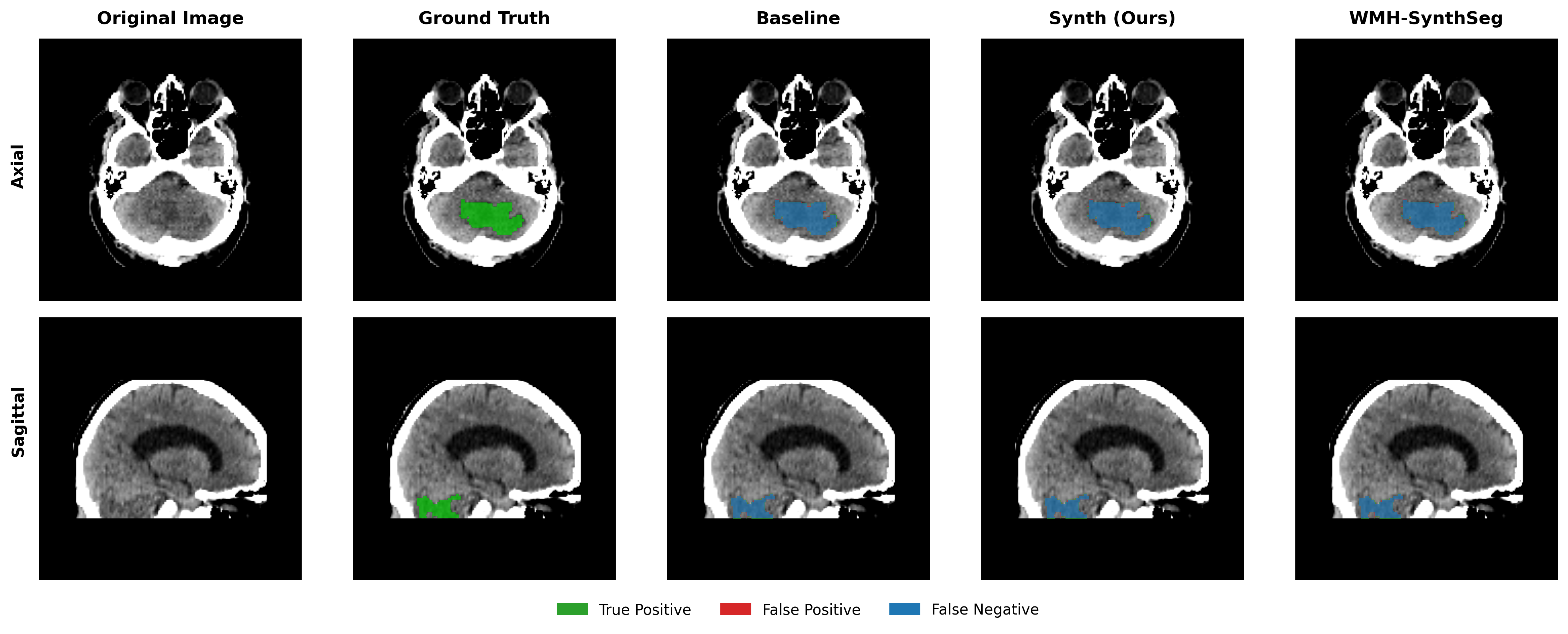}
            \end{subfigure}
            \caption{Sample visualisations of failure cases in the PLORAS CT dataset. Green indicates a true positive prediction, red a false positive, and blue a false negative.}
            \label{fig:ploras-bad-CT}
        \end{figure*}

\end{document}